\documentclass[[aps,prd,a4paper,showpacs,twocolumn]{revtex4-1}
\usepackage{amsmath,amssymb,bm}
\usepackage{gensymb}
\usepackage{nicefrac} 
\usepackage{epsfig}
\usepackage{graphicx}
\usepackage{slashed}
\usepackage{epsfig} \usepackage{graphicx} \usepackage{color}
\usepackage{mathrsfs} \usepackage{amssymb} \usepackage{amsmath} \usepackage{url}

\usepackage{xspace} 
\usepackage[bookmarks, breaklinks, colorlinks,urlcolor=black, citecolor=red, 
linkcolor=blue]{hyperref}

\def\jpsi{\ensuremath{{J\mskip -3mu/\mskip -2mu\psi\mskip 2mu}}}
\def\dsp{\displaystyle}
\def\be {\begin{equation}}
\def\ee {\end{equation}}
\def\bea {\begin{eqnarray}}
\def\eea {\end{eqnarray}}
\def\bc {\begin{center}}
\def\ec {\end{center}}
\def\nn {\nonumber}
\def\sss{\scriptscriptstyle}
\def\gev{\ensuremath{\mathrm{\,Ge\kern -0.1em V}}}
\def\mev{\ensuremath{\mathrm{\,Me\kern -0.1em V}}}

\def \Re{\text{Re}}
\def \Im{\text{Im}}

\def \kstar{{K^{\!*}}}

\def\AFB{A_{\text{FB}}}

\def\hel#1{{\sss{#1}}}

\def\AFB{A_{\text{FB}}}

\def\Gf{\Gamma_{\!\! f}}
\def\qmax{q^2_{\sss\text{max}}}

\def\fb   {\ensuremath{\mbox{\,fb}}\xspace}
\def\invfb   {\ensuremath{\mbox{\,fb}^{-1}}\xspace}
\def\lhcb{LHCb~}

\def\ket|#1>{\left|#1 \right>}

\def\bra<#1|{\left< #1 \right|}

\def\bracket<#1|#2>{\setbox0=\vbox{\hbox{$#1$$#2$}}\left<#1\kern1pt
	\vrule  height\ht0\kern2pt #2\right>} 

\def\dirmat<#1|#2|#3>{\setbox0=\vbox{\hbox{$#1$$#2$$#3$}}\left<#1\kern1pt
	\vrule height\ht0\kern1pt#2\kern1pt \vrule height\ht0\kern1pt
	#3\right>}

\allowdisplaybreaks

\begin{document}

\title{ Signal of right-handed currents using
\texorpdfstring{\boldmath${B\to\kstar\ell^+\ell^-}$}{} observables at the
kinematic endpoint.}

\author{Anirban Karan}\email{kanirban@imsc.res.in}
\affiliation{The Institute of Mathematical Sciences, HBNI, Taramani, Chennai 
600113, India}
\author{Rusa Mandal}\email{rusam@imsc.res.in} 
\affiliation{The Institute of Mathematical Sciences, HBNI, Taramani, Chennai 
600113, India}
\author{Abinash Kumar Nayak}\email{abinashkn@imsc.res.in}
\affiliation{The Institute of Mathematical Sciences, HBNI, Taramani, Chennai 
600113, India}
\author{Rahul Sinha}\email{sinha@imsc.res.in}
\affiliation{The Institute of Mathematical Sciences, HBNI, Taramani, Chennai 
600113, India}
\author{Thomas E. Browder}\email{teb@phys.hawaii.edu}
\affiliation{Department of 
Physics and Astronomy, University of
  Hawaii, Honolulu, HI 96822, USA}


\date{\today}

\begin{abstract}
The decay mode $B\to\kstar\ell^+\ell^-$ is one of the most promising modes to
probe physics beyond the standard model (SM),  since the angular distribution of
the decay products enable measurement of several constraining observables. \lhcb
has recently measured these observables using $3\fb^{-1}$ of data as a binned
function of $q^2$, the dilepton invariant mass squared. We find that LHCb data
implies evidence for right-handed currents, which are absent in
the SM. These conclusions are derived in the maximum $q^2$ limit and are free from
hadronic corrections. Our approach differs from other approaches that probe new
physics at low $q^2$ as it does not require estimates of hadronic parameters but
relies instead on heavy quark symmetries that are reliable at the maximum $q^2$
kinematic endpoint.
\end{abstract}


\maketitle

\section{Introduction}

The rare decay $B\to \kstar\ell^+\ell^-$, which involves a $b\rightarrow s$
flavor changing loop induced quark transition at the quark level, provides an
indirect but very sensitive probe of new physics (NP) beyond the standard model
(SM). The angular distribution of the decay products provides a large number of
observables ~\cite{Kruger:1999xa} and thus can be used to reduce hadronic
uncertainties making the mode a very special tool to probe for NP. Significant
work has been done to probe NP in this mode. Most previous attempts have
focused~\cite{Altmannshofer:2014rta} on the low dilepton invariant mass squared
region $q^2=1-6~\gev^2$. An alternative approach that probes the maximum $q^2$
limit has also been studied in literature \cite{Grinstein:2004,Bobeth:2012vn}.
We show that this limit holds significant promise for clean probes of NP. A
previous study suggested a possible signal of NP in the large $q^2$
region~\cite{Mandal:2015bsa}. In this letter we show that \lhcb data implies a
$5\sigma$ signal for the existence of  NP. While the evidence for right handed
currents is clear, other NP contributions are also possible. Our
conclusions are derived in the maximum $q^2$ limit ($\qmax$) and are free from
hadronic corrections. Our approach differs from  other approaches that probe NP
at low $q^2$ by not requiring estimates of hadronic parameters but relying
instead on heavy quark symmetries that are completely reliable at the kinematic
endpoint $\qmax$~\cite{Grinstein:2004,Hiller:2013}.  While the observables
themselves remain unaltered from their SM values, their derivatives and second
derivatives at the endpoint are sensitive to NP effects. The paper is organized as follows. In Sec.~\ref{sec:Theory}, we discuss the model independent theoretical framework used for the analysis. The numerical procedure for the extraction of right-handed (RH) currents is described in Sec.~\ref{sec:RH}. We illustrate the effect of resonances and the convergence of the polynomial fit in Sec.~\ref{sec:reso} and Sec.~\ref{sec:PolyFit}, respectively. Finally, Sec.~\ref{sec:conclusion} contains concluding remarks.

\section{Theoretical Formalism}
\label{sec:Theory}
In this section we briefly discuss the model independent theoretical framework that has been adopted in this work. The decay $B\to \kstar\ell^+\ell^-$ is described by six transversity amplitudes
that can be written as~\cite{Mandal:2014kma}
\begin{equation}
\mathcal{A}_\lambda^{L,R}=C_{L,R}^{\sss\lambda}\,\mathcal{F}_\lambda 
-\widetilde{\mathcal{G}}_\lambda
\label{eq:amp-def1}
=\big(\widetilde{C}_9^{\sss\lambda}\mp 
C_{10})\mathcal{F}_\lambda -\widetilde{\mathcal{G}}_\lambda
\end{equation}
within the standard model in the massless lepton limit~\cite{leptonmass}. This
parametric form of the amplitude is general  enough to comprehensively include
all short-distance and long-distance effects, factorizable and nonfactorizable
contributions, resonance contributions and complete electromagnetic corrections
to hadronic operators up to all orders. In Eq.~\eqref{eq:amp-def1} $C_9$ and
$C_{10}$ are Wilson coefficients with $\widetilde{C}_9^{\sss\lambda}$ being the
redefined ``effective'' Wilson coefficient
defined~\cite{Mandal:2014kma,BF00,Beneke:2001at} as
\begin{equation}
\label{eq:c9}
\widetilde{C}_9^{\sss\lambda}=C_9+\Delta 
C_9^{\text{(fac)}}(q^2)+\Delta C_9^{{\sss\lambda}{\text{,(non-fac)}}}(q^2)
\end{equation}
where $\Delta C_9^{\text{(fac)}}(q^2)$, $\Delta
C_9^{{\sss\lambda}{\text{,(non-fac)}}}(q^2)$ correspond to factorizable and soft
gluon non-factorizable contributions. The Wilson coefficient $C_{10}$ is
unaffected by strong interaction effects coming from electromagnetic corrections
to hadronic operators \cite{Altmannshofer:2008dz}. The form factors
$\mathcal{F}_\lambda$ and $\widetilde{\mathcal{G}}_\lambda$ introduced in
Eq.~\eqref{eq:amp-def1} can in principle be related to the conventional 
form factors
describing the decay if power corrections are ignored.
However, our approach does not rely on estimates of 
$\mathcal{F}_\lambda$ and $\widetilde{\mathcal{G}}_\lambda$.

In the SM, $\mathcal{F}_\lambda$'s and $C_{10}$ are real, whereas
$\widetilde{C}_9^{\sss\lambda}$ and $\widetilde{\mathcal{G}}_\lambda$ contain
the imaginary contributions of the amplitudes. Defining two variables
$r_\lambda$ and  $\varepsilon_\lambda$, the amplitudes ${\cal A}_\lambda^{L,R}$
in Eq.~\eqref{eq:amp-def1} can be rewritten as,
\begin{equation}
\label{eq:amp-def2}
\mathcal{A}_\lambda^{L,R}=(\mp 
C_{10}-r_\lambda)\mathcal{F}_\lambda+i\varepsilon_\lambda,
\end{equation}
where \vspace*{-0.68cm}
\begin{eqnarray}
\label{eq:rlambda}
r_\lambda&=&\frac{\Re(\widetilde{\mathcal{G}}_\lambda)}{\mathcal{F}_\lambda}
-\Re(\widetilde{C}_9^\hel{\lambda}),\\
\label{eq:epsilon}
\varepsilon_\lambda &=&
\Im(\widetilde{C}_9^\hel{\lambda})\mathcal{F}_\lambda
-\Im(\widetilde{\mathcal{G}}_\lambda).
\end{eqnarray}
The observables $F_\perp$, $F_\|$, $F_L$, $\AFB$ and $A_5$ are defined as,
\begin{align}
  \label{eq:Fperp0}
F_\lambda&= \frac{|\mathcal{A}_\lambda^L|^2+|\mathcal{A}_\lambda^R|^2}
{\dsp\Gamma_{\!f}}~~~  \lambda \in \{\perp,\|,0\},\\
\AFB&=\frac{3}{2}\frac{\Re({\cal A}_\|^L{\cal A}_\perp^{L^*}-{\cal A 
}_\|^R{\cal A}_\perp^{R^*})}{\Gamma_{\!f}}, \\
\label{eq:A50}
A_5&=\frac{3}{2\sqrt{2}}\frac{\Re(\mathcal{A}_0^L\mathcal{A}_\perp^{L^*} 
-\mathcal{A}_0^R\mathcal{A}_\perp^{R^*})}{\Gamma_{\!f}},
\end{align}
where $ \Gamma_{\!\!f}\equiv\sum_\lambda(|{\cal A}_\lambda^L|^2+|{\cal
A}_\lambda^R|^2)$ and are related to the observables measured by \lhcb
\cite{Aaij:2015oid} as follows:
\begin{align}
\label{eq:S4-S5-S9} F_\perp&=\frac{1-F_L+2 S_3}{2}, ~~A_4=-\dsp\frac{2}{\pi}S_4, \nn \\ 
~~A_5&=\dsp\frac{3}{4}S_5, ~~~\AFB\!=\!-\AFB^{\sss \text{LHCb}}.
\end{align}

We neglect the $\varepsilon_\lambda$ contributions to the
amplitude for the time being, but their effect in the numerical analysis is discussed in Appendix \ref{sec:epsilon}. In the presence of
RH currents the transversity amplitudes are given
by~\cite{Altmannshofer:2008dz}
\begin{align}
\label{eq:amp-def1h}
\mathcal{A}^{L,R}_\perp &= \big((\widetilde{C}_9^\hel{\perp}+C_9^\prime) \mp 
(C_{10}+C_{10}^\prime)\big) \mathcal{F}_\perp
    -\widetilde{\mathcal{G}}_\perp \\
\mathcal{A}^{L,R}_\| &= \big((\widetilde{C}_9^\hel{\|}-C_9^\prime) \mp 
(C_{10}-C_{10}^\prime)\big) \mathcal{F}_\|
    -\widetilde{\mathcal{G}}_\| \\
\mathcal{A}^{L,R}_0 &= \big((\widetilde{C}_9^\hel{0}-C_9^\prime) \mp 
(C_{10}-C_{10}^\prime)\big) \mathcal{F}_0
    -\widetilde{\mathcal{G}}_0.
\end{align}
Note that setting the RH contributions $C_9^\prime$ and $C_{10}^\prime$ to zero,
the amplitudes reduce to the SM ones in Eq.~\eqref{eq:amp-def1}.

Introducing new variables 
\begin{align}
\xi = \frac{C_{10}^\prime}{C_{10}} ~~\text{and}~~ \xi^\prime = 
\frac{C_{9}^\prime}{C_{10}} 
\end{align}
the observables $F_{\!\perp}$, $F_{\|}$, $\AFB$, $A_5$ (Eqs.~\eqref{eq:Fperp0} -- \eqref{eq:A50})
can be 
expressed as,
\begin{align}
\label{eq:Fperp} 
F_{\!\perp} &= 2 \zeta\,(1+\xi)^2 (1+R_\perp^2)\\
\label{eq:Fparallel} 
F_{\|}\mathsf{P_1^2} &= 2 \zeta\,(1-\xi)^2 (1+R_\|^2)\\
\label{eq:Flongi} 
F_L \mathsf{P_2^2} &= 2 \zeta\,(1-\xi)^2 (1+R_0^2)\\
\label{eq:AFB} 
\AFB \mathsf{P_1} &=3\zeta\,(1-\xi^2) \big(R_\|+ R_\perp \big)\\
\label{eq:A5} 
\sqrt{2} A_5 \mathsf{P_2} &=3\zeta\,(1-\xi^2) \big(R_0+ R_\perp \big)
\end{align}
where $\mathsf{P_{1}}=\dsp\frac{\mathcal{F}_\perp}{\mathcal{F}_\|}$,~~ 
$\mathsf{P_{2}}=\dsp\frac{\mathcal{F}_\perp}{\mathcal{F}_0},~~\zeta = \frac{\mathcal{F}_\perp^2 C_{10}^2}{\Gf}$,~~
\begin{equation}
\label{eq:Rpdef}
R_\perp=\frac{\dsp\frac{r_\perp}{C_{10}}-\xi^\prime}{1+\xi},~
R_{\|}=\frac{\dsp\frac{r_{\|}}{C_{10}}+\xi^\prime}{1-\xi},~
R_{0}=\frac{\dsp\frac{r_{0}}{C_{10}}+\xi^\prime}{1-\xi}.
\end{equation}

We consider the observables $F_L$, $F_\|$, $F_\perp$, $\AFB$ and $A_5$, with the
constraint $F_L+F_\|+F_\perp=1$. Using Eq.~\eqref{eq:Fperp}--\eqref{eq:A5}, we
obtain expressions for $R_\perp$, $R_\|$, $R_0$ and $\mathsf{P}_2$ in terms of
the observables and $\mathsf{P}_1$:
\begin{align}
 \label{eq:Rperp-2}
R_\perp&=\pm \frac{3}{2} \frac{\Big(\frac{1-\xi}{1+\xi}\Big)F_\perp
+\frac{1}{2}\mathsf{P_1}Z_1}{\mathsf{P_1} \AFB} \\
\label{eq:Rparallel-1}
R_\|&=\pm \frac{3}{2} \frac{\Big(\frac{1+\xi}{1-\xi}\Big)\mathsf{P_1} F_\|
+\frac{1}{2} Z_1}{ \AFB} \\
\label{eq:R0}
R_0&=\pm \frac{3}{2\sqrt{2}} 
\frac{\Big(\frac{1+\xi}{1-\xi}\Big)\mathsf{P_2} F_L
+\frac{1}{2} Z_2}{ A_5} \\
\label{eq:P2-2}
\mathsf{P_2}&=\!\frac{\Big(\frac{1-\xi}{1+\xi}\Big)2\mathsf{P_1}\AFB 
F_\perp}{\sqrt{2}A_5 
\left(\Big(\frac{1-\xi}{1+\xi}\Big)2F_\perp+Z_1\mathsf{P_1}\right)
-Z_2\mathsf{P_1}\AFB}
\end{align}
where $Z_1=(4 F_\|  F_\perp-\frac{16}{9}\AFB^2)^{\nicefrac{1}{2}}$ and $Z_2 =(4
F_L F_\perp-\frac{32}{9} A_5^2)^{\nicefrac{1}{2}}$. Since we have one
extra parameter compared to observables, all of the above expressions depend on
$\mathsf{P_1}$. Fortunately in the large $q^2$ limit, the relations between form
factors enable us to eliminate one parameter.

At the kinematic limit $q^2=\qmax=(m_B-m_\kstar)^2$ the $\kstar$ meson is at
rest and the two leptons travel back to back in the $B$ meson rest frame. There
is no preferred direction in the decay kinematics. Hence, the differential decay
distribution in this kinematic limit must be independent of the angles
$\theta_\ell$  and $\phi$, which can be integrated out. This imposes constraints
on the amplitude $A_\lambda^{L,R}$ and hence the observables. The entire decay,
including the decay $\kstar \to K\pi$ takes place in a single plane, resulting in a
vanishing contribution to the `$\perp$' helicity, or $F_\perp=0$. Since the
$\kstar$ decays at rest, the distribution of $K\pi$ is isotropic and cannot
depend on $\theta_K$. It can easily be seen that this is only possible if
$F_\|=2 F_L$~\cite{Hiller:2013}.

At $q^2\!=\!\qmax$, $\Gf \to 0$ as all the transversity amplitudes vanish in
this limit. The constraints on the amplitudes described above result in unique values
of the helicity fractions and the asymmetries at this kinematical endpoint. The
values of the helicity fractions and asymmetries were derived in
Ref.~\cite{Hiller:2013, Mandal:2014kma} where it is explicitly shown that
\begin{eqnarray}
\label{eq:A4max}
\dsp
F_L(\qmax)=\frac{1}{3}, & F_\|(\qmax)=\dsp\frac{2}{3}, & A_4(\qmax)= 
\dsp\frac{2}{3\pi},~~~~~~\nn \\
F_\perp(\qmax)=0,~ &\AFB(\qmax)=0,~& A_{5,7,8,9}(\qmax)=0.
\end{eqnarray}

The large $q^2$ region where the $\kstar$ has low-recoil energy has also
been studied~\cite{Grinstein:2004,Bobeth:2010wg} in a modified heavy quark
effective theory framework. In the limit $q^2\to \qmax$ the  hadronic form
factors satisfy the conditions
\begin{equation}
\label{eq:qmaxFF}
\frac{\widetilde{\mathcal{G}}_\|}{\mathcal{F}_\|}=
\frac{\widetilde{\mathcal{G}}_\perp}{\mathcal{F}_\perp}=
\frac{\widetilde{\mathcal{G}}_0}{\mathcal{F}_0} 
=-\kappa\frac{2 m_b m_B C_7}{q^2},
\end{equation}
where $\kappa\approx 1$ as shown in~\cite{Bobeth:2010wg}. The helicity 
independence of the ratios 
${\widetilde{\mathcal{G}}_\lambda}/{\mathcal{F}_\lambda}$ at $\qmax$ is easy to 
understand, since both the $B$ and $\kstar$ mesons are at rest, resulting in a 
complete overlap of 
the wave functions of these two mesons and the absence of any preferred 
direction in the $K\pi$ distribution. Due to the 
constraints arising from decay kinematics and Lorentz invariance, on the 
observables at $\qmax$ (in Eq.~\eqref{eq:A4max}), it is shown in 
Ref.~\cite{Hiller:2013} that the non-factorizable contributions are helicity 
independent at the endpoint. Hence from Eq.~\eqref{eq:rlambda} it can 
be seen that, $r_0=r_\|=r_\perp\equiv r$~\cite{Das:2012kz}. Therefore, 
Eq.~\eqref{eq:Rpdef} implies that, by definition of the variables $R_\lambda$, 
in the presence of RH currents, one should expect $R_0=R_\parallel\ne R_\perp$ 
at $q^2=\qmax$ without any approximation. As argued above this relation is 
unaltered by
non-factorizable and resonance contributions at this kinematic endpoint.

We study the values of $R_\lambda$, $\zeta$ and
$P_{1,2}$ in the large $q^2$ region and consider the kinematic limit
$q^2 \to \qmax$. It is easy to see from Eq.~\eqref{eq:Fperp} that $F_\perp(\qmax)=
0$ implies that $\zeta= 0$ in the limit $q^2\to\qmax$. Further, since
$R_\parallel(\qmax)=R_0(\qmax)$, Eqs.~\eqref{eq:Fparallel} and \eqref{eq:Flongi} imply that in the limit $q^2 \to \qmax$, $\mathsf{P}_2=\sqrt{2}\,\mathsf{P}_1$. However, both
$\mathsf{P}_1$ and $\mathsf{P}_2$ go to zero at $\qmax$. It is therefore
imperative that we take into account the limiting values very carefully by 
Taylor expanding all observables around the endpoint $\qmax$ in terms of the
variable $\delta\equiv\qmax-q^2$. The leading power of $\delta$ in the Taylor
expansion must take into account the relative momentum dependence of the
amplitudes ${\cal A}_\lambda^{L,R}$. 
Eq.~\eqref{eq:Fperp0}-\eqref{eq:A50} and \eqref{eq:A4max} together imply that 
$\mathcal{A}_\perp^{L,R}$ must have an expansion at least 
$\mathcal{O}(\sqrt{\delta})$ 
higher compared to $\mathcal{A}_{\|,0}^{L,R}$. This is in agreement with 
Ref.~\cite{Hiller:2013}. Hence
the leading term in
$F_L$ and $F_\|$ must be ${\cal O}(\delta^0)$, whereas the leading term for  
$F_\perp$ is  ${\cal O}(\delta)$. The  leading terms for the asymmetries, 
$A_5$ and $\AFB$, are ${\cal O}(\sqrt{\delta})$.  Thus, we expand
the observables as follows:
\begin{align}
\label{eq:funcFLfit} 
F_L &=\frac{1}{3}+F_L^{(1)}\delta+ F_L^{(2)} \delta^2+F_L^{(3) } \delta^3 \\
\label{eq:funcFPfit} 
F_\perp &= F_\perp^{(1)}\delta+ F_\perp^{(2)} \delta^2+F_\perp^{(3)} \delta^3 \\
\label{eq:funcAFBfit} 
\AFB &= \AFB^{(1)}\delta^{\nicefrac{1}{2}}+ \AFB^{(2)} 
\delta^{\nicefrac{3}{2}} +\AFB^{(3)} \delta^{\nicefrac{5}{2}}\\
\label{eq:funcA5fit} 
A_{5} &= A_5^{(1)}\delta^{\nicefrac{1}{2}}+ A_5^{(2)} 
\delta^{\nicefrac{3}{2}} 
+A_5^{(3)} \delta^{\nicefrac{5}{2}},
\end{align}
where for each observable $O$, $O^{(n)}$ is the coefficient of the $n^\text{th}$
term in the expansion. The polynomial fit to data is not based on Heavy Quark Effective Theory (HQET) or any other theoretical
assumption. A parametric fit to data is performed, so as to obtain the limiting
values of the coefficients to determine the slope and second derivative of the
observables at $\qmax$. It should be noted that the polynomial parameterizations are
inadequate to describe the $q^2$ dependent behavior of resonances. However,
systematics of resonance effects are discussed in Sec.~\ref{sec:reso} in detail
validating the approach followed here.

The relation in Eq.~\eqref{eq:qmaxFF} between form factors is expected to
be satisfied in the large $q^2$ region. Eq.~\eqref{eq:qmaxFF} is
naturally satisfied if it is valid at each order in the Taylor expansion of the
form factors: 
\begin{align}
\label{eq:TaylorFF}
q^2\frac{\widetilde{\mathcal{G}}_\lambda}{\mathcal{F}_\lambda}=
	\qmax 
	\frac{\widetilde{\mathcal{G}}_{\lambda}^{(1)}+\dsp\delta\, 
	(\widetilde{\mathcal{G}}_{\lambda}^{(2)}- 
	\dsp\frac{\widetilde{\mathcal{G}}_{\lambda}^{(1)}}{\qmax})+{\cal 
	O}(\delta^2)}
	{\mathcal{F}_\lambda^{(1)}+ \delta\, \mathcal{F}_\lambda^{(2)}+{\cal 
	O}(\delta^2)}.
\end{align}

We require only that the relation be valid up to order $\delta$.  In order for
Eq.~\eqref{eq:TaylorFF} to have a constant value in the neighborhood of $\qmax$
up to ${\cal O}(\delta)$, we must have $\mathcal{F}_\lambda^{(2)}= c\,
\mathcal{F}_\lambda^{(1)}$ and $(\qmax\,\widetilde{\mathcal{G}}_{\lambda}^{(2)}-
\widetilde{\mathcal{G}}_{\lambda}^{(1)})= c\,
\qmax\,\widetilde{\mathcal{G}}_{\lambda}^{(1)}$ where $c$ is any constant.  As
discussed earlier, $\mathsf{P_{2}}= \sqrt{2} \mathsf{P_{1}}$ at $\qmax$, hence,
we must have $\mathsf{P_{2}^{(1)}}= \sqrt{2} \mathsf{P_{1}^{(1)}}$, where
$\mathsf{P_{1,2}^{(1)}}$ are the coefficients of the leading ${\cal
O}(\sqrt{\delta})$ term in the expansion. However, the above argument implies
that at the next order, we must also have $\mathsf{P_{2}^{(2)}}= \sqrt{2}
\mathsf{P_{1}^{(2)}}$, since 
$\mathcal{F}_\lambda^{(2)}=c\,\mathcal{F}_\lambda^{(1)}$. This provides the
needed input that together with Eq.~\eqref{eq:P2-2} determines
$\mathsf{P_{1}^{(1)}}$ purely in terms of observables.

The expressions for $R_\lambda$ in the limit $q^2\to \qmax$ are
\begin{align}
R_\perp(\qmax) &= \!\dsp \frac{8 \AFB^{(1)}(-2A_5^{(2)}+\AFB^{(2)}) 
+9 (3 F_L^{(1)}+F_\perp^{(1)}) F_\perp^{(1)}}{8\, (2 
A_5^{(2)}-\AFB^{(2)})\,\sqrt{\frac{3}{2} F_\perp^{(1)}-\AFB^{(1)\,2}}}\nn\\
\label{eq:Rp}
&=\dsp\frac{\omega_2-\omega_1}{\omega_2\sqrt{\omega_1-1}},\\
R_\|(\qmax) &= \dsp \frac{3 (3 
F_L^{(1)}+F_\perp^{(1)}) \,\sqrt{\frac{3}{2} F_\perp^{(1)}-\AFB^{(1)\,2}}} 
{\dsp -8 A_5^{(2)}+4\AFB^{(1)}+
3 \AFB^{(1)} (3 F_L^{(1)}+F_\perp^{(1)})}\nn\\
\label{eq:Ra}
&= \dsp\frac{\sqrt{\omega_1-1}}{\omega_2-1}= R_0(\qmax)
\end{align}
where
\begin{align}
\omega_1=&\dsp\frac{3}{2} \frac{F_\perp^{(1)}}{\AFB^{(1) \, 2}} \text{~~and~~}
\omega_2=&\dsp\frac{4\,(2 A_5^{(2)}-\AFB^{(2)})}
{3\,\AFB^{(1)}(3 F_L^{(1)}+F_\perp^{(1)})}.
\label{eq:omega}
\end{align}
It should be noted that Eqs.~\eqref{eq:Rp}--\eqref{eq:omega} are derived 
only at
$\qmax$. However, even at the endpoint, the expressions depend on 
polynomial
coefficients: $F_L^{(1)}$ and $F_\perp^{(1)}$  as well as $\AFB^{(2)}$ 
and
$A_5^{(2)}$ which are not related by HQET. Hence, in our approach, 
corrections
beyond HQET are automatically incorporated through fits to data.

In the absence of RH currents or other NP that treats the ``$\perp$'' amplitude
differently one would expect $R_\perp(\qmax)=R_\|(\qmax)=R_0(\qmax)$. It is 
easily seen that the LHS of Eq.~\eqref{eq:AFB} is 
positive around $\qmax$ and since $\zeta>0$, we must have $R_\perp=R_\|=R_0>0$.
Since very large contributions from RH currents are not possible, as they would
have been seen elsewhere, $R_\lambda(\qmax)>0$ still holds and  restricts $\xi$
and $\xi^\prime$ to reasonably small values.

\begin{table}[thb]
\centering
  \begin{tabular}{ |c| c| c| c |}
    \hline \hline
 &$O^{(1)}(10^{-2})$ & $O^{(2)}(10^{-3})$& 
  $O^{(3)}(10^{-4})$\\ \hline
    $F_L$     & $-2.85\pm1.26$& $~12.13\pm1.90$ & 
    $-5.68\pm0.67 $ \\ 
    $F_\perp$   & $~6.89\pm1.65$ & 
    $-9.79\pm2.47$& $~3.83\pm0.86$\\ 
    $\AFB$    & $-30.58\pm1.95$& $~26.96\pm3.58$ & 
    $-4.15\pm1.47$\\ 
    $A_5$     & $-15.85\pm1.87$& $~5.38\pm3.33$ & 
    $2.46\pm1.29$\\ 
    \hline 
  \end{tabular}
  \caption{Best fit and 1$\sigma$ uncertainties for the coefficients of observables (in 
  Eqs.~\eqref{eq:funcFLfit}-- \eqref{eq:funcA5fit}) obtained by fitting recent 
  LHCb's 14- bin measurements \cite{Aaij:2015oid} as a function of $q^2$ for 
  the entire region.}
\label{Table-1}
\end{table}

\begin{center}
\begin{figure}[htb]		
 \begin{center}
	\includegraphics*[width=1.68in]{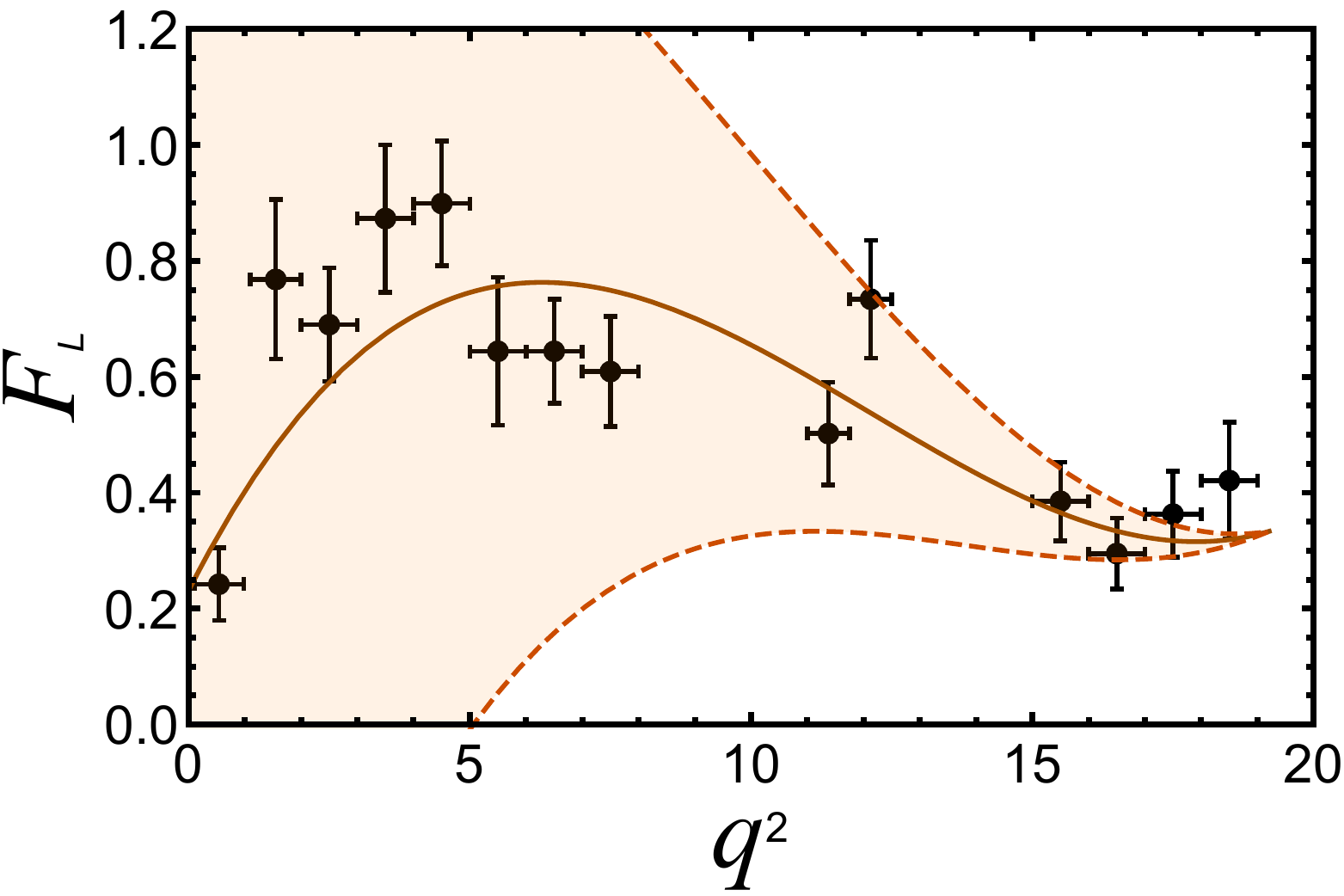}%
	\includegraphics*[width=1.68in]{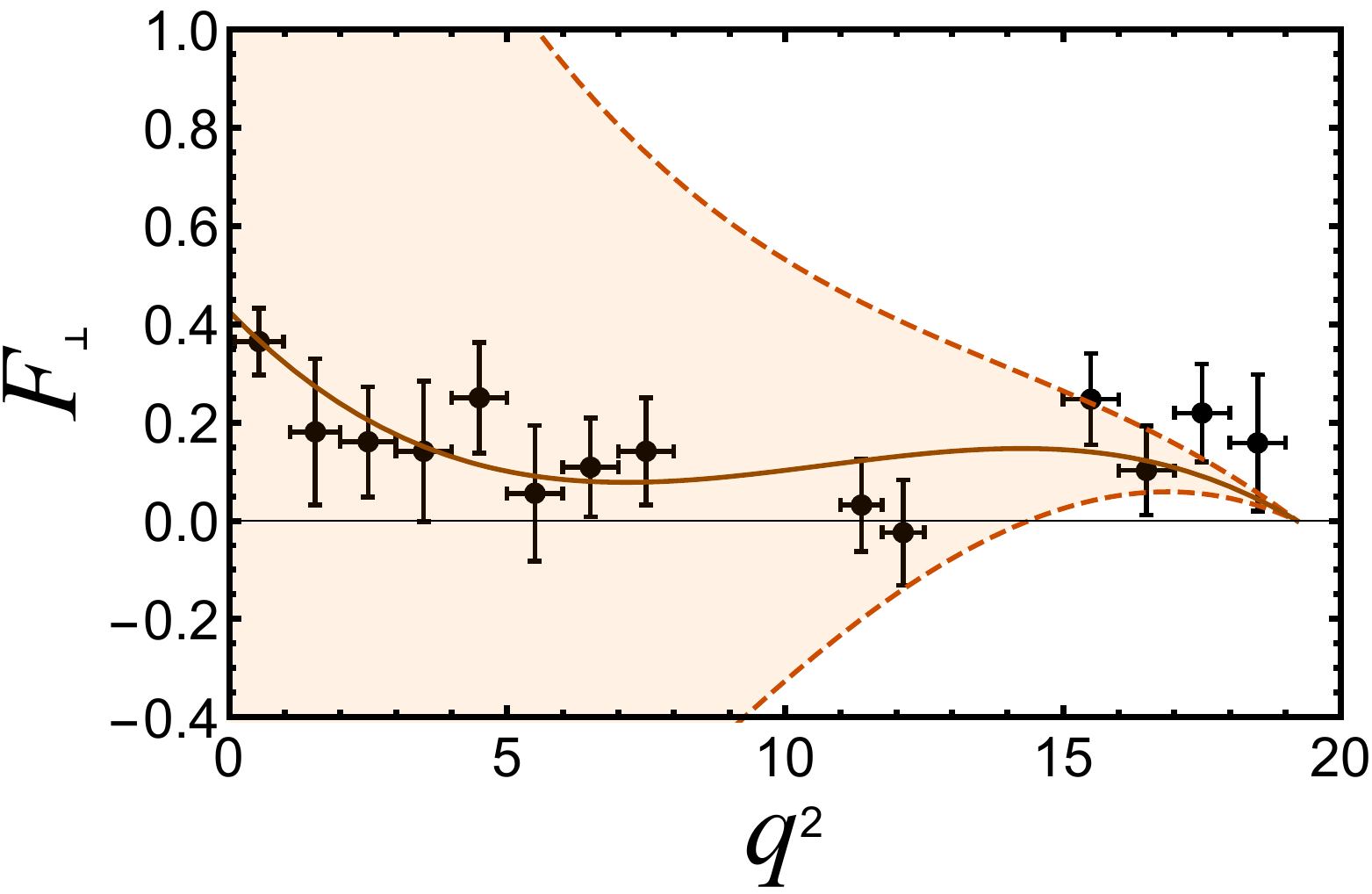}
	\includegraphics*[width=1.68in]{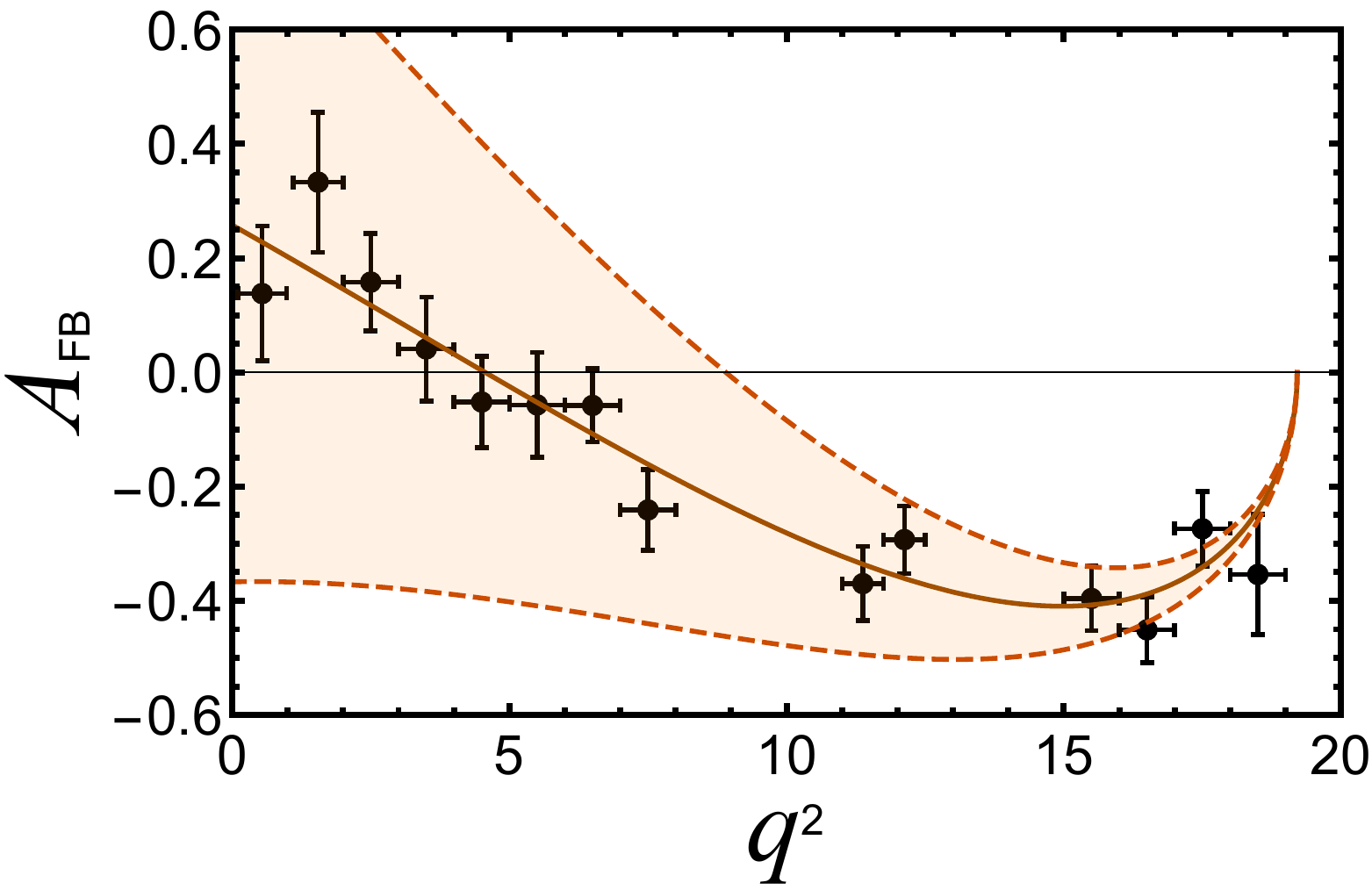}%
	\includegraphics*[width=1.68in]{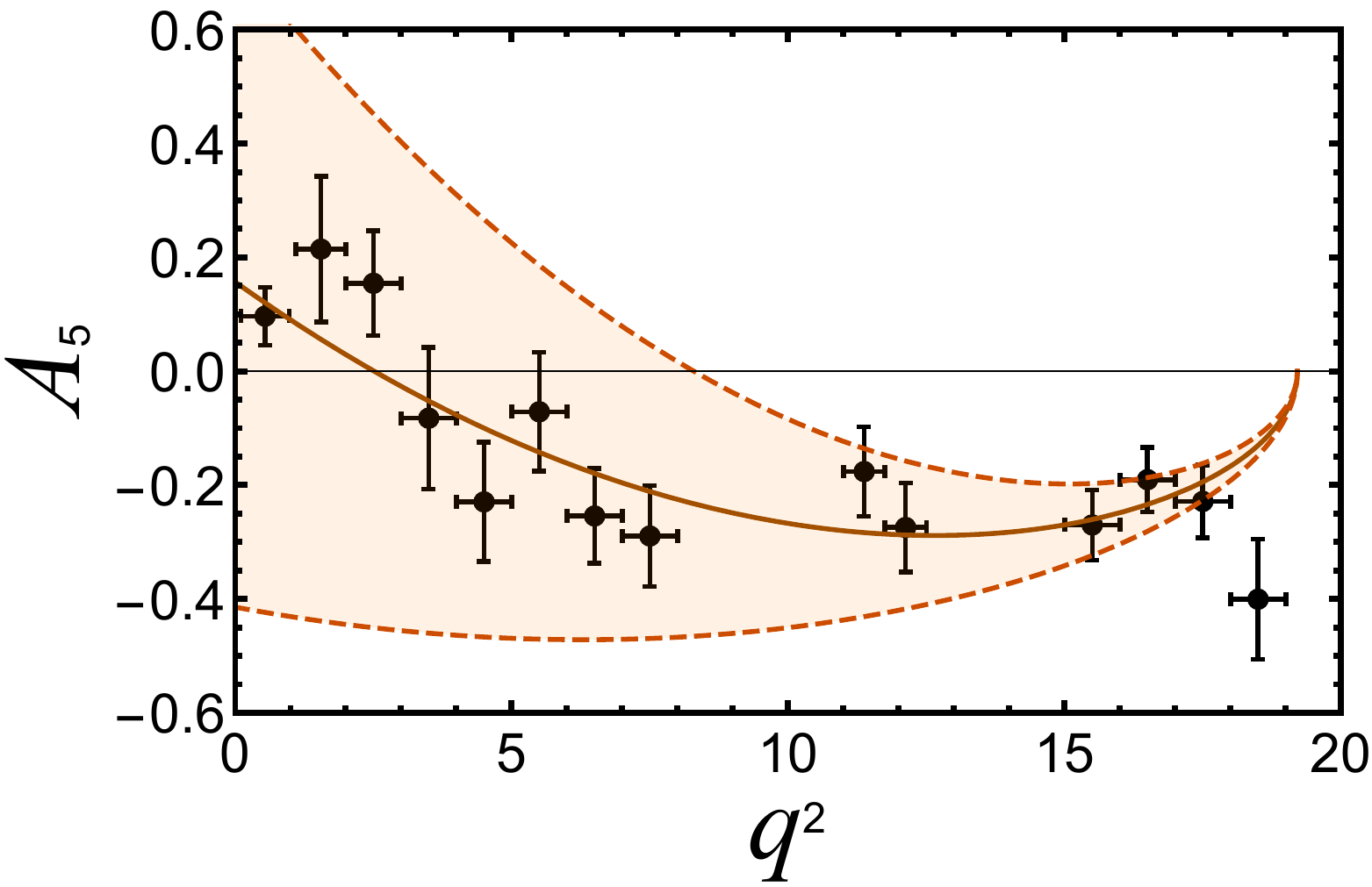}
	\caption{An analytic fit to 14-bin \lhcb data using Taylor expansion at 
	$\qmax$ for the observables
	$F_L$, $F_\perp$, $\AFB$ and $A_5$ are shown as the brown curves. The 
	$\pm1\sigma$ error bands are indicated by the light brown shaded regions, derived including correlation among all observables. The points with the black error bars are \lhcb 14-bin measurements \cite{Aaij:2015oid}.} \label{fig:1}
\end{center}
\end{figure}
\end{center}

\section{Right-Handed current Analysis}
\label{sec:RH}
In this section we describe the numerical analysis based on the theoretical formalism derived in the previous section. We start by fitting the latest \lhcb measurements \cite{Aaij:2015oid} of the observables
$F_L$, $F_\perp$, $\AFB$ and $A_5$  as functions of $q^2$ using the Taylor expansion
at $\qmax$ as given in Eqs.~\eqref{eq:funcFLfit}-- \eqref{eq:funcA5fit}. The fits were performed by minimizing the $\chi^2$ function, which
compares the bin integrated values of $q^2$ functions of the observables with
their measured experimental values for all 14 bins. The correlations reported by \lhcb among all observables have also been considered. The bin integration for
the polynomial fit is weighted with  the recent measurements of differential
decay rate~\cite{Aaij:2016flj}. A polynomial is fitted for $d\Gamma/dq^2$ data for the entire $q^2$ region. This fitted polynomial for $d\Gamma/dq^2$ (say denoted by $\Gamma(q^2)$) is then used in weighted average for all the observables. For an observable
$\mathcal{O}$ the bin averaged value within the $q^2$ interval $[b_i,b_f]$ is
obtained by,
$\int_{b_i}^{b_f} \mathcal{O}(q^2) \Gamma(q^2)\, dq^2 \Big/\int_{b_i}^{b_f} \Gamma(q^2)\,dq^2$.
%
We use the 14 bin data set based on the method
of moments~\cite{Beaujean:2015xea} from \lhcb rather than the  8 bin data set as
it enables better constraints near $\qmax$. The best fit values for each
coefficient of the observables $F_L$, $F_\perp$, $\AFB$ and $A_5$
(Eqs.~\eqref{eq:funcFLfit}-- \eqref{eq:funcA5fit}) are given in
Table~\ref{Table-1}. The errors in each coefficient are evaluated using a
covariance matrix technique. A detailed study of the systematics  in
fitting the polynomial is described in Sec.~\ref{sec:PolyFit}. Variations in  the
order of the polynomial from two to four and the number of bins used in fitting
(from the last four to all fourteen), demonstrate good convergence when larger numbers
of bins are considered.

In Fig.~\ref{fig:1} the results of the fits for the observables $F_L$,
$F_\perp$, $\AFB$ and $A_5$, respectively,  are compared with the measured \lhcb
data \cite{Aaij:2015oid}. We notice that the factorization requirement
$\AFB^{(1)}=2A_5^{(1)}$ holds to within $\pm 1\sigma$. We treat $\AFB^{(1)}$ and
$2A_5^{(1)}$ as two independent measurements of the same quantity as we have 
neglected correlation between observables. We obtain $\omega_1=1.10\pm
0.30~(1.03\pm 0.34)$ and $\omega_2=-4.19\pm 10.48~(-4.04\pm 10.12)$,
where the first values are determined using $\AFB^{(1)}$ and the values in
the round brackets use $2A_5^{(1)}$. 
\begin{center}
\begin{figure}[tb]
\begin{center}
	\includegraphics*[width=2.in]{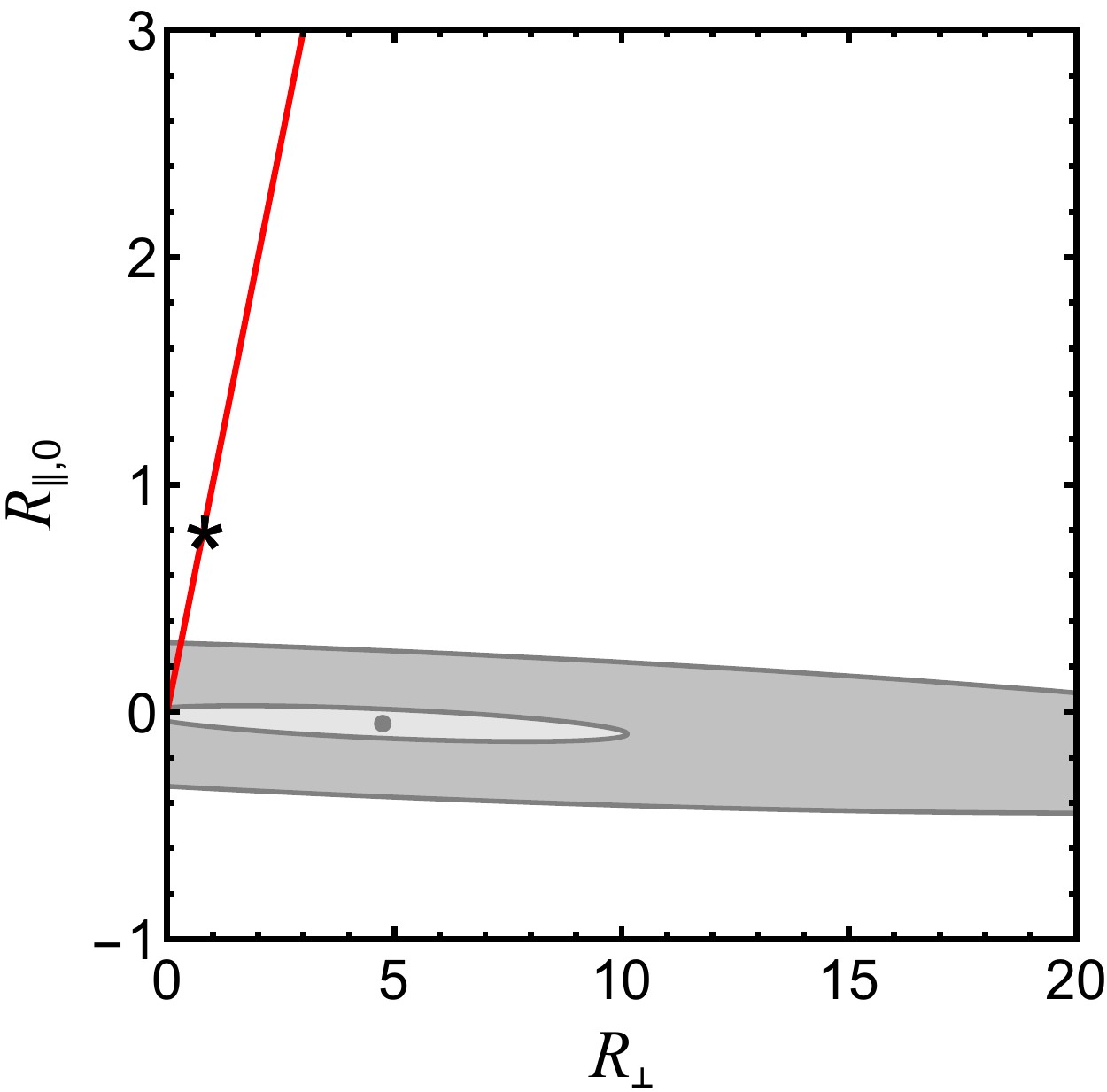}%
	\caption{Allowed regions in $R_\perp$ -- $R_{\|,0}$ plane are shown. The solid red straight line on the 
	far left  corresponds to the case $R_\perp=R_{\|,0}$. The SM value is indicated 
	by the star. The gray point 
	corresponds to best fit central values. The light and dark gray contours correspond to $1\sigma$ and $5\sigma$ confidence level regions, respectively.} 
	\label{fig:2}
\end{center}
\end{figure}
\end{center}

We estimate the range of values for $R_\perp$ and $R_{\|,0}$ in two 
different
ways. One approach estimates $R_\perp$ and $R_{\|,0}$ using randomly 
chosen
values of $F_L^{(1)}$, $F_P^{(1)}$, $\AFB^{(1)}$, $A_5^{(1)}$, 
$\AFB^{(2)}$ and
$A_5^{(2)}$, from a Gaussian distribution with the central value as the 
mean and
errors from Table~\ref{Table-1}. If RH currents are absent the values 
would lie
along a straight line with a $45^\text{o}$ slope in the 
$R_\perp-R_{\|,0}$
plane. However, we find a slope that is nearly horizontal, indicating 
that
$R_\perp \gg R_{\|,0}$. The deviation of slope from $45^\text{o}$ 
provides
evidence of contributions from RH currents.

In an alternate approach we fit the values of $R_\perp$ and $R_{\|,0}$ with the
two estimated values of $\omega_1$ and $\omega_2$  by minimizing a $\chi^2$
function. The allowed regions in the $R_\perp$ -- $R_{\|,0}$  plane are shown in
Fig.~\ref{fig:2}.  The solid red straight line on the far left  corresponds to
the case $R_\perp=R_{\|,0}$. The SM value is indicated by the star on the red
line. The light gray and dark gray contours indicate the $1\sigma$ and $5 \sigma$
permitted regions. We emphasize that for the SM, even in the presence of
resonances the contours should be aligned along the $45^\text{o}$ line, since
resonances contribute equally to all helicities through $\Delta C_9$ in
Eq.~\eqref{eq:c9}. Hence the deviation of the contours from the SM expectation
is a signal for RH currents. As it will be discussed in Sec.~\ref{sec:reso}, charmonium resonance contributions in bin averaged data always raise the values of $\omega_1$,
whereas we find that the values of $\omega_1$ are  close to the lowest possible
physical value allowed.

\begin{center}
	\begin{figure}[!t]
		\begin{center}
			\includegraphics*[width=1.6in]{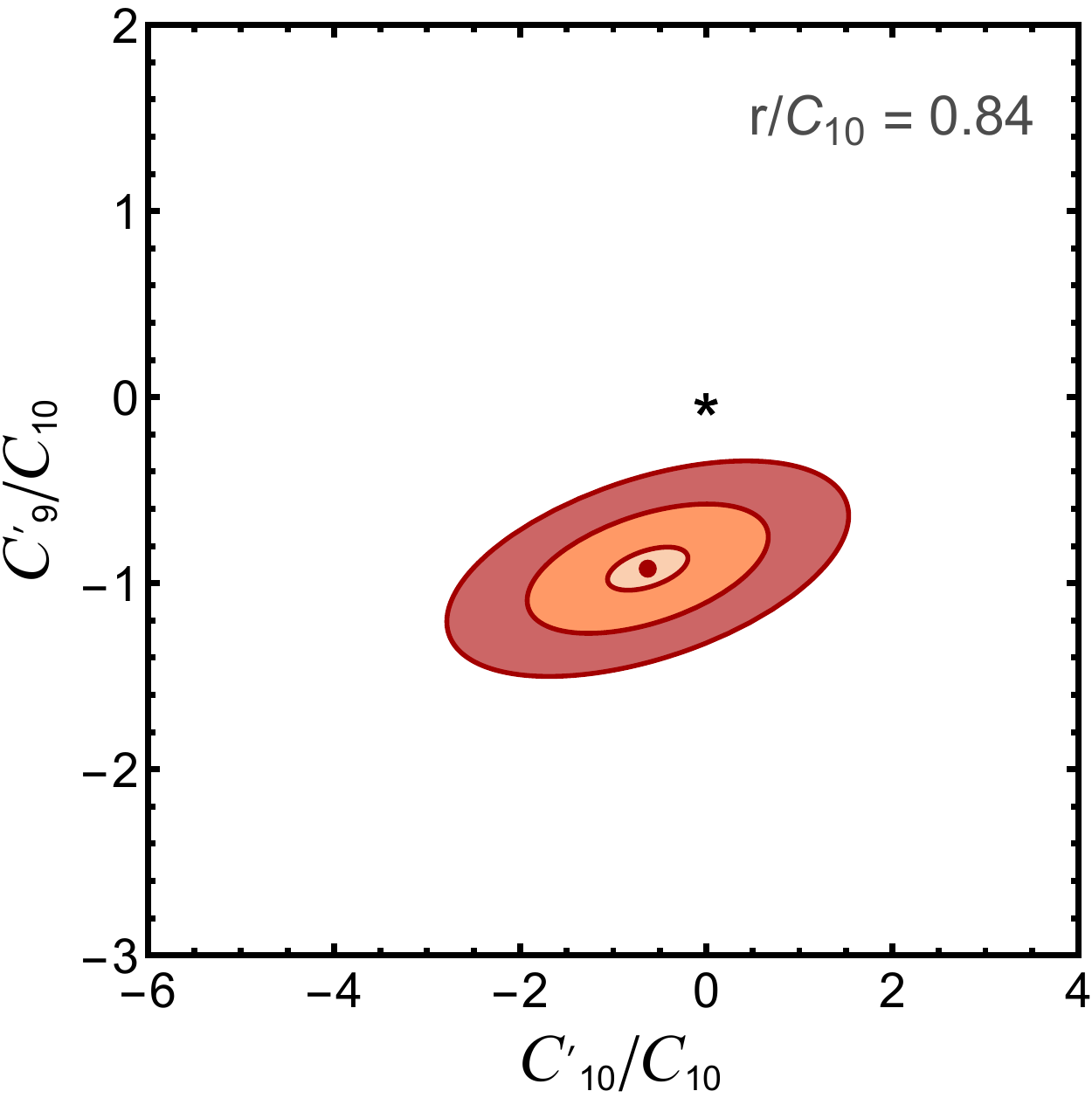} %
			\includegraphics*[width=1.6in]{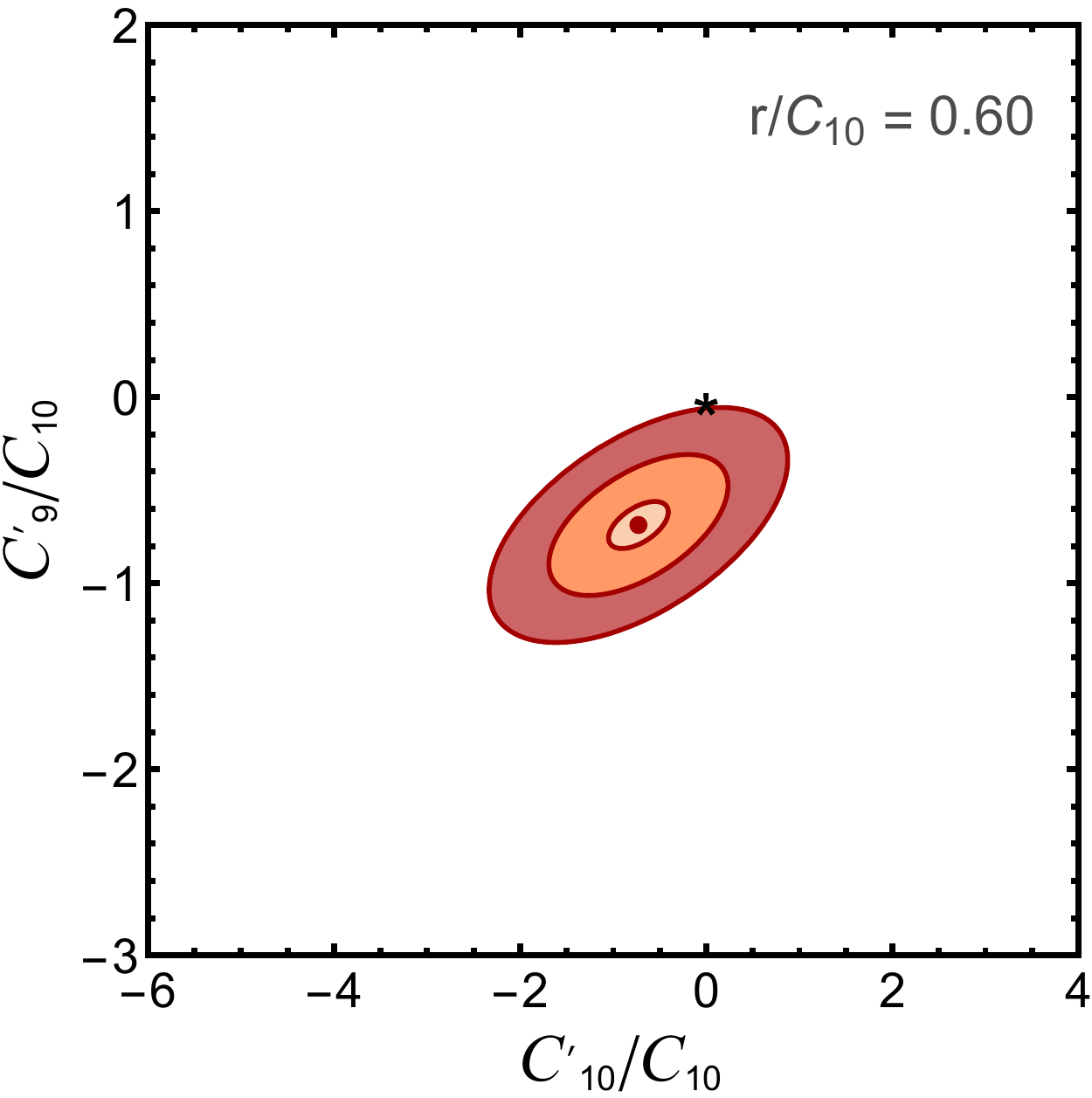}
			\includegraphics*[width=1.6in]{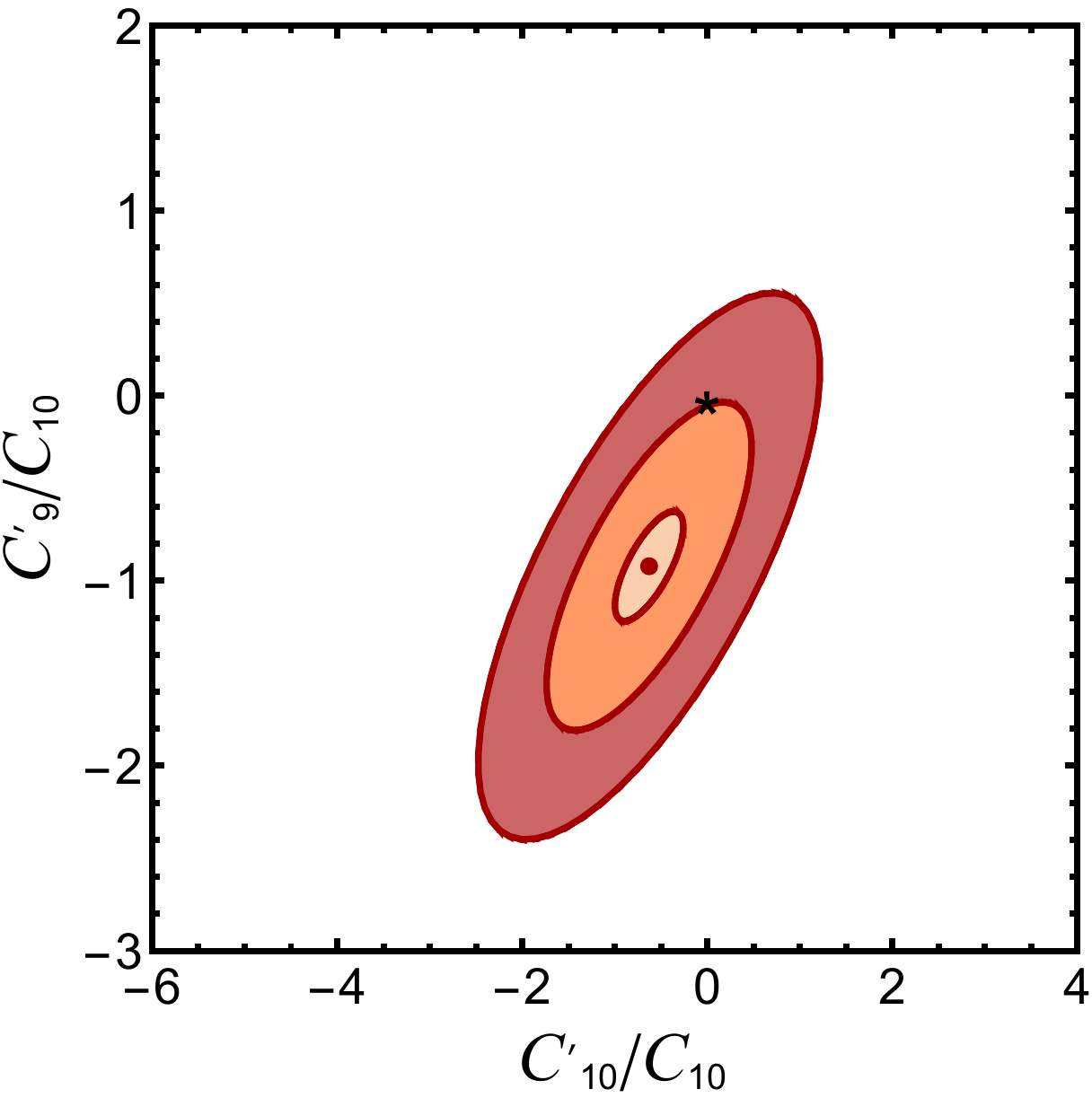}
			\caption{Allowed regions in $C_{10}^\prime/C_{10}$ -- $C_{9}^\prime/C_{10}$
				plane are shown in three different panels. The yellow, orange and red 
				bands are the 1$\sigma$, 3$\sigma$ and 5$\sigma$ confidence level regions, 
				respectively. The center red 
				dot denotes best fit point; the SM values of
				$C_{10}^\prime/C_{10}$ and $C_{9}^\prime/C_{10}$ are indicated by a `star', 
				which sits more than 5$\sigma$ confidence level contour in the upper left panel, at 5$\sigma$ contour in the upper right panel and at 3$\sigma$ contour in the bottom panel. The plots illustrates the sensitivity to $r/C_{10}$. The upper left panel shows the SM value, the upper right panel includes an additional NP contribution $C_9^{\text{NP}}\approx -1$ \cite{Altmannshofer:2014rta} while the bottom panel highlights the case where $r/C_{10}$ is considered as a nuisance parameter (see text for details).}
			\label{fig:3}
		\end{center}
	\end{figure}
\end{center}

Having established the existence of RH contributions, we  perform a $\chi^2$ fit
to the parameters $\xi$ and $\xi^\prime$ which indicate the size of the new
Wilson coefficients. This is easily done using Eqs.~\eqref{eq:Rpdef},
\eqref{eq:Rp} and \eqref{eq:Ra}. However, this requires as an input the estimate
of $r/C_{10}$ from Eq.~\eqref{eq:qmaxFF} at $\qmax$. The allowed regions in the
$\xi$ -- $\xi^\prime$ plane are shown in Fig.~\ref{fig:3}.  The left panel shows
the region obtained using SM estimate of $r/C_{10}=0.84$ \cite{Bobeth:2010wg}. The best fit values of
$\xi$ and $\xi^\prime$, with $\pm1\sigma$ errors are $-0.63\pm0.43$ and
$-0.92\pm0.10$, respectively. The yellow, orange and red bands denote 1$\sigma$,
3$\sigma$ and 5$\sigma$ confidence level regions, respectively. The SM value of
$C_{10}^\prime/C_{10}$ and $C_{9}^\prime/C_{10}$ is indicated by the star,
beyond the 5$\sigma$ confidence level contour, which is in an agreement with the result shown in Fig.~\ref{fig:2}. The SM estimate of $r/C_{10}$ can
have uncertainties that cannot easily be accounted for. These could range from
errors in Wilson coefficients, contributions from other kinds of new physics or
even the contributions from resonances. In order to ascertain the accuracy of
our conclusion to these uncertainties, we have scanned $r/C_{10}$
over a range of values. While the evidence for right handed currents is
clear, the central values of $\xi$ and $\xi^\prime$ obtained from the fit
can be reduced somewhat if $r/C_{10}$ is smaller due to NP contributions that
alter the Wilson coefficient $C_9$ and the significance of discrepancy can also
be reduced $\sim 5\sigma$ as can be seen from  Fig.~\ref{fig:3} upper right panel
plot. The value $r/C_{10}=0.6$ corresponds to the scenario in which NP contribution
to the Wilson coefficient $C_9$ is $C_9^{\text{NP}}\approx -1$ as indicated by a
global fit analysis for $b\to s $ transition \cite{Altmannshofer:2014rta}. In
this case, best fit values of $\xi$ and $\xi^\prime$ with $\pm1\sigma$ errors
are $-0.73\pm0.32$ and $-0.69\pm0.10$. We have performed  another analysis where the input $r/C_{10}$ is considered as nuisance parameter and the result is shown in the bottom panel of Fig.~\ref{fig:3}. In this case the best fit value with $\pm 1 \sigma$ error for the parameters $\xi$, $\xi^\prime$ and $r/C_{10}$ are $-0.63\pm 0.43$, $-0.92 \pm 0.14$ and $0.84\pm 0.10$, respectively. It can be seen that the uncertainties in $C_{10}^\prime/C_{10}$ and $C_{9}^\prime/C_{10}$ parameters have increased due to the variation of $r/C_{10}$ and the SM prediction still remains on a $3\sigma$ level contour providing evidence of RH currents. 
We note that if $\xi\ne 0$ is confirmed
by further measurements, additional scalar and or pseudoscalar contributions
would be needed in order to have consistency with $B_s\to \mu^+\mu^-$
data~\cite{DeBruyn:2012wk}.

We now discuss the  effect of complex part of the transversity amplitudes i.e
$\varepsilon_\lambda$ contributions (in Eq.~\eqref{eq:amp-def2}), which was not
considered so far. In our approach $\varepsilon_\lambda$ can be estimated at the
endpoint purely from data. The $\varepsilon_\lambda$ corrections do not
contribute to the asymmetries $\AFB$ and $A_5$, however, they do contribute to
the helicity fractions $F_L$ and $F_\perp$ \cite{Mandal:2014kma}. 
Interestingly, in  a Taylor
expansion of $\widehat{\varepsilon}_\lambda\equiv
2|\varepsilon_\lambda|^2/\Gamma_f$, the coefficient of the leading term must be
positive. We have used \lhcb data to estimate $\widehat{\varepsilon}_0^{(0)}$,
$\widehat{\varepsilon}_0^{(1)}$,
$\widehat{\varepsilon}_\|^{(1)}$ and $\widehat{\varepsilon}_\perp^{(1)}$ that
modify the estimates of $\omega_1$ and $\omega_2$. The detailed expressions and discussions are given in Appendix \ref{sec:epsilon}. We have also studied the 
effects of non-zero $\kstar$, width in Appendix \ref{sec:finite width}. Including these effects we
find that our  conclusions are slightly strengthened.

\section{Effect of Resonances}
\label{sec:reso}

In this section we examine if resonances can alter the results that are obtained
using a polynomial fit to the observables in Eqs.~\eqref{eq:funcFLfit}--
\eqref{eq:funcA5fit}, where it is assumed that resonances are absent. The
data includes resonance contributions in the bin averaged observables and these
averages may not fit well to a polynomial if resonance contributions are sizable.
It may be noted that in our approach the polynomial function is used only for a
parametric fit to data. In principle, the data could have been fitted to any
chosen function. An inappropriate function will result in a poor fit with large
errors. We have estimated all errors and the fits reflect the errors caused by
the assumption of ignoring resonances. It should be noted that the fit itself is
not invalidated, however the errors estimated in Table.~\ref{Table-1} will decrease if a better function or a better estimate of systematics of resonances are
accounted for. Thus, our uncertainties are an overestimate. %
We also emphasize that the real part of resonance contributions are notionally
included in the amplitudes and the imaginary parts are also accounted for as discussed in Appendix~\ref{sec:epsilon}. Since, both our theory and
experimental data include resonance contributions, the observed discrepancy cannot 
arise due to resonances. Below we discuss the
differences between $q^2$ distributions with and without resonances   as systematic
uncertainties.

This study is
performed on observables, evaluated using 
theoretical estimates of
form factors and Wilson  coefficients. We assume the values of the form factors
evaluated using LCSR~\cite{Straub:2015ica}
for $q^2\le 15\gev^2$ and from Lattice QCD~\cite{Horgan:2013pva} for 
$q^2\ge
15\gev^2$ region.
The effects of resonances
are incorporated as in \cite{Kruger:1996cv}. 
The procedure defines the function $g(m_c,q^2)$, in $C_9^{\rm eff}$,  as
\begin{align}
g(m_c,q^2)&=-\frac{8}{9} \ln\frac{m_c}{m_b} -\frac{4}{9} \nn \\
	&+ \frac{q^2}{3} 
\mathcal{P}  
\int_{4 \hat{m}_D^2}^{\infty} \frac{R_{\rm had}^{c\bar{c}}(x)}{x(x-q^2)} dx + i 
\frac{\pi}{3}  R_{\rm had}^{c\bar{c}}(q^2).
\end{align}
where $\mathcal{P}$ is the Principal Value of the integral 
and $\hat{m}_D=m_D/m_b$. The cross-section ratio 
$R_{\rm had}^{c\bar{c}}(q^2)$ is given by,
\begin{align}
R_{\rm had}^{c\bar{c}}(q^2)=R_{\rm cont}^{c\bar{c}}(q^2)+R_{\rm 
res}^{c\bar{c}}(q^2) .
\end{align} 
Here, $R_{\rm cont}^{c\bar{c}}$ and $R_{\rm res}^{c\bar{c}}$ denote the 
contributions from the continuum and the narrow resonances, respectively. The 
latter is given by the Breit-Wigner formula
\begin{align}
R_{\rm res}^{c\bar{c}}(q^2)\!=\! N_r\!\!\!\!\!\! \sum_{V=\jpsi,\psi'..}\hspace*{-0.5cm} \frac{9\,q^2}{\alpha} 
\frac{\text{Br}(V \! \rightarrow \ell^+ 
\ell^-)\Gamma^V_{\rm tot}\Gamma^V_{\rm had}}{(q^2\!-\!m_V^2)^2+m_V^2\Gamma^{V2}_{\rm 
tot}}  e^{i\delta_V}
\end{align}
where $\Gamma^V_{\rm tot}$ is the total width of the vector meson `$V$', $\delta_V$
is an arbitrary relative strong phase associated with each of the resonances 
and  $N_r$ is a normalization factor that fixes the size of the
resonance contributions compared to the non-resonant ones correctly. We
include the $\jpsi(1S)$, $\psi(2S)$, $\psi(3770)$,
$\psi(4040)$, $\psi(4160)$ and $\psi(4415)$  resonances in our study. The masses and widths
of these vector mesons are taken from the PDG compilation~\cite{Agashe:2014kda}.

The continuum term $R_{\rm cont}^{c\bar{c}}(q^2)$ is parametrized differently in
Refs. \cite{Kruger:1996cv} and \cite{Lyon:2014hpa}, but we have verified that
both of these parameterization gives indistinguishable results for our
analysis.  We introduce yet another overall normalization factor $N_b$ that
normalizes the value of $d\Gamma/dq^2$ so as to match it with its experimentally
measured value.

We numerically integrate the theoretical differential decay rate 
including all the resonances, in $q^2$, for eight bin intervals given in
\cite{Aaij:2016flj}. 
 We add all these eight bin averaged values to obtain a quantity, which we refer to here as $d\Gamma_{\rm th}^{\rm 
 	tot}/dq^2$. However, $d\Gamma_{\rm th}^{\rm 
 	tot}/dq^2$ depends on the two unknown quantities $N_b$ and $N_r$. We integrate the same theoretical  
differential decay rate again including all the resonances, in the $q^2$ region 
$[2.97^2,3.21^2]$  to match the cuts used in the LHCb
experiment (Ref.~\cite{Aaij:2012nh})  and
denote the result as $d\Gamma_{\rm th}^{\jpsi}/dq^2$, which is once again also a function 
of the same two quantities $N_b$ and $N_r$. These two theoretical quantities,
$d\Gamma_{\rm th}^{\rm tot}/dq^2$ and $d\Gamma_{\rm th}^{\jpsi}/dq^2$, are then compared
with the central values of the experimentally measured differential  decay rates
$4.379\times 10^{-7}$ (total bin average value for eight bins) and $1.29\times
10^{-3}$, respectively. %
The solution for $N_b$ and $N_r$ are obtained by solving the two equations,
\begin{align*}
\frac{d\Gamma_{\rm th}^{\rm tot}(N_b,N_r)}{dq^2}= 4.379\times 10^{-7} 
\\
\frac{d\Gamma_{\rm th}^{\jpsi}(N_b,N_r)}{dq^2}= 1.29\times 10^{-3}.
\end{align*}
Two solutions for normalizations are obtained from the resultant quadratic
equations. 
For every set of
$\delta_V$ chosen, two sets of $N_b$ and $N_r$ are calculated. We have also
verified that our results are insensitive to the variation in $q^2$ cuts for the 
$\jpsi$ resonance. This implies that if the $q^2$ cut is changed to
$[3.05^2,3.15^2]$~\cite{Abe:2002haa}, the normalization factors are modified only 
by a few percent. 

We have varied $\delta_V$ from 0 to $2\pi$ through $15^{\text o}$ intervals
for each resonance. In order to keep the size of data limited we present only a 
sample of some of the plots obtained by varying 
$\delta_V$ for  the $\jpsi(1S)$, 
$\psi(4040)$ and  $\psi(4160)$ resonances. The plots are given in 
link~\cite{link} as movies. The 
movies were created using more than 22000 plots.

It may be noted from these plots that when resonances are included, the helicity fractions do not vary significantly due to resonance contributions. The asymmetries  $\AFB$ and
$A_5$ always decrease in magnitude for the $15\gev^2\le q^2\le 19\gev^2$ region.
Hence if the effect of resonances could somehow be removed from the data, the
values of $\AFB$ and $A_5$ would be larger in magnitude.
This observation is also valid for the slope of the fitted polynomial for $\AFB$
and $A_5$ at the endpoint. The value of $\omega_1$ in this case would be smaller
compared to the values obtained from fits to experimental data in which resonances
are automatically present. In other words including resonance effects in
$15\gev^2\le q^2\le 19\gev^2$ region always increases $\omega_1$.  It should be 
noted that
the values of $\omega_1$, obtained by fitting to experimentally observed data, are
already close to unity and any further reduction will force $\omega_1$ into the
un-physical domain.

In a Toy Monte Carlo study, values of $\delta_V$ were randomly
chosen one million times and  the values of observables obtained  without
resonances  were compared to those where resonances were included. This
enabled us to verify that the conclusions drawn for $\delta_V\in n\pi/12$
($\forall n=1,...,12$) are valid in general.

It is also easy to see analytically that adding resonances would strengthen the case for NP rather than weaken it.
Consider the observable $Z_1=\sqrt{4 F_\|F_\perp-\frac{16}{9}\AFB^2}$ from Refs.~\cite{Das:2012kz,Mandal:2014kma}, which can 
be cast as 
\begin{equation}
Z_1=\frac{4}{3}|\AFB|\sqrt{\frac{9 F_\parallel F_\perp}{4 
\AFB^2}-1}=\frac{4}{3}|\AFB|\sqrt{ \Omega_1-1}
\end{equation}
where $\Omega_1=\dsp \frac{9 F_\parallel F_\perp}{4\AFB^2}$. Since $Z_1$ is 
real it is obvious that $\Omega_1>1$. Experimental data indicates that 
$\Omega_1$ is very close to unity for the entire range above 
$q^2>15\gev^2$. If resonance contributions are explicitly included $Z_1$ becomes,
\begin{align}
Z_1&=\frac{4}{3}|\AFB|\sqrt{\frac{9 (F_\parallel-\frac{2\varepsilon_\|^2}{\Gamma_f}) 
(F_\perp-\frac{2\varepsilon_\perp^2}{\Gamma_f})}{4 
\AFB^2}-1} \nn \\
&=\frac{4}{3}|\AFB|\sqrt{ \Omega_1-{\cal 
O}\Big(\frac{2\varepsilon_{\|,\perp}^2 \Omega_1}{F_{\|,\perp}\Gamma_f}\Big)-1}
\end{align}
where $\varepsilon_\lambda$ is defined in Eq.~\eqref{eq:epsilon}. Note that ${\cal 
O}\Big(\frac{2\varepsilon_{\|,\perp}^2 \Omega_1}{F_{\|,\perp}\Gamma_f}\Big)$
is always positive, decreasing the radical.
Hence, one can conclude that resonance contributions cannot be significant in
data or else the value of $\Omega_1$ would become unphysical. It should be noted
that $\omega_1\equiv\Omega_1(\qmax)$, implying that the value of $\omega_1$
which we find very close to unity is consistent and would only decrease and
become unphysical if resonances were included. The same arguments hold for the
observables $Z_2$ and $\Omega_2$ or $\omega_2$.

 It may be noted that in a previous study of resonance effects in $B\to
K\ell^+\ell^-$~\cite{Lyon:2014hpa}, the difficulty in accommodating the
LHCb-result in the standard treatment of the SM or QCD was noted and  possible
right-handed current contributions were suggested.

\begin{widetext}
\begin{center}
\begin{figure}[!h]
 \begin{center}
	\includegraphics*[width=2.2in]{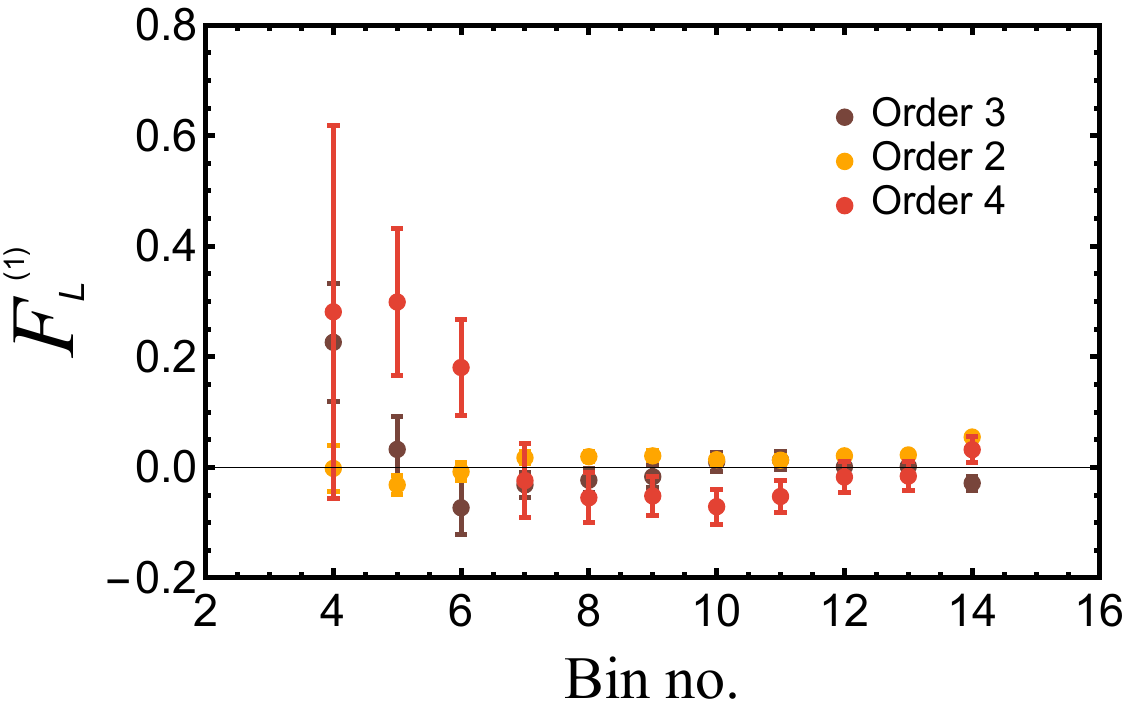}%
	\includegraphics*[width=2.2in]{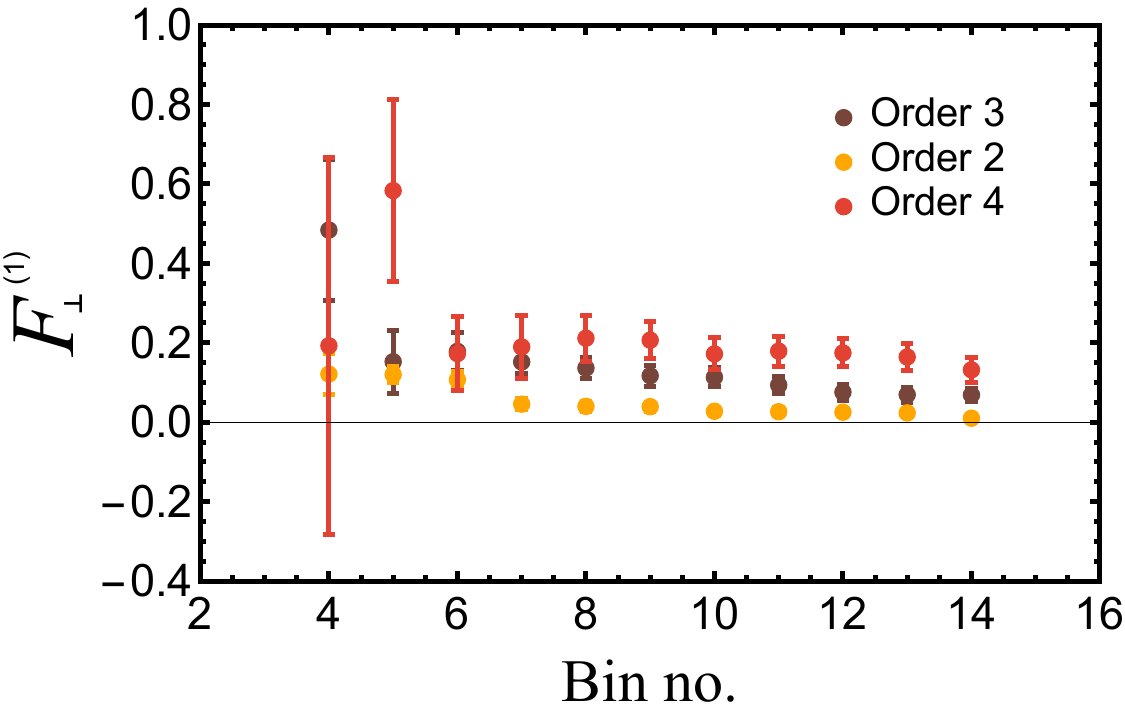}
	\includegraphics*[width=2.2in]{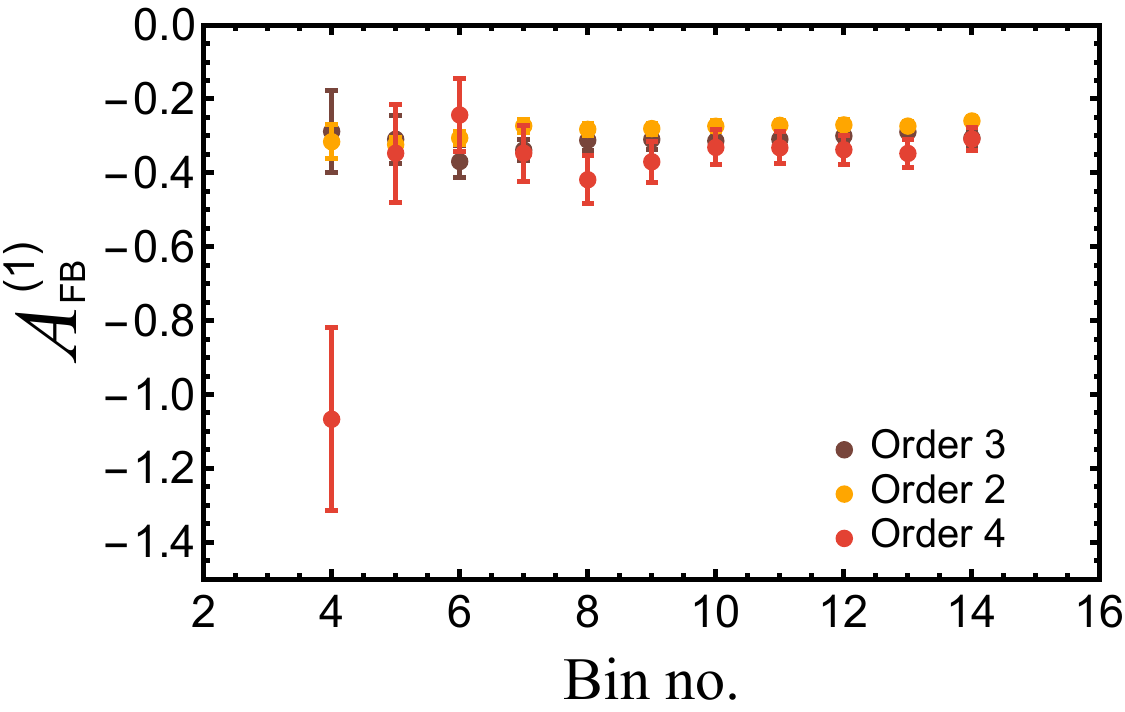}
	\includegraphics*[width=2.2in]{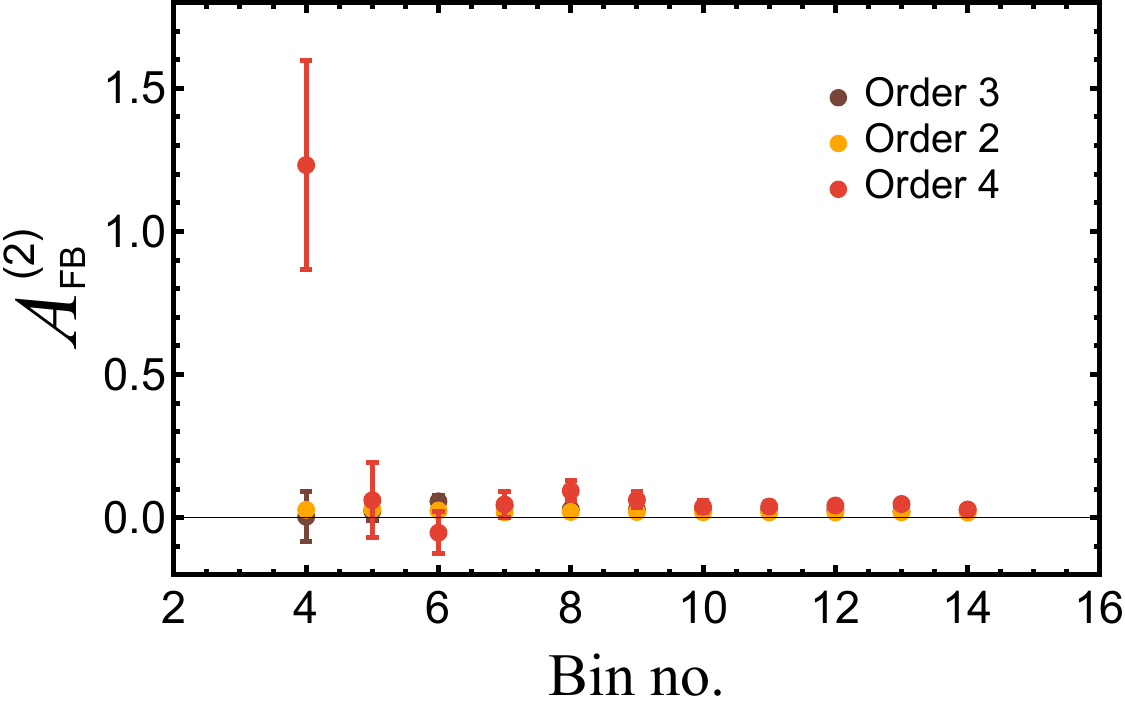} %
	\includegraphics*[width=2.2in]{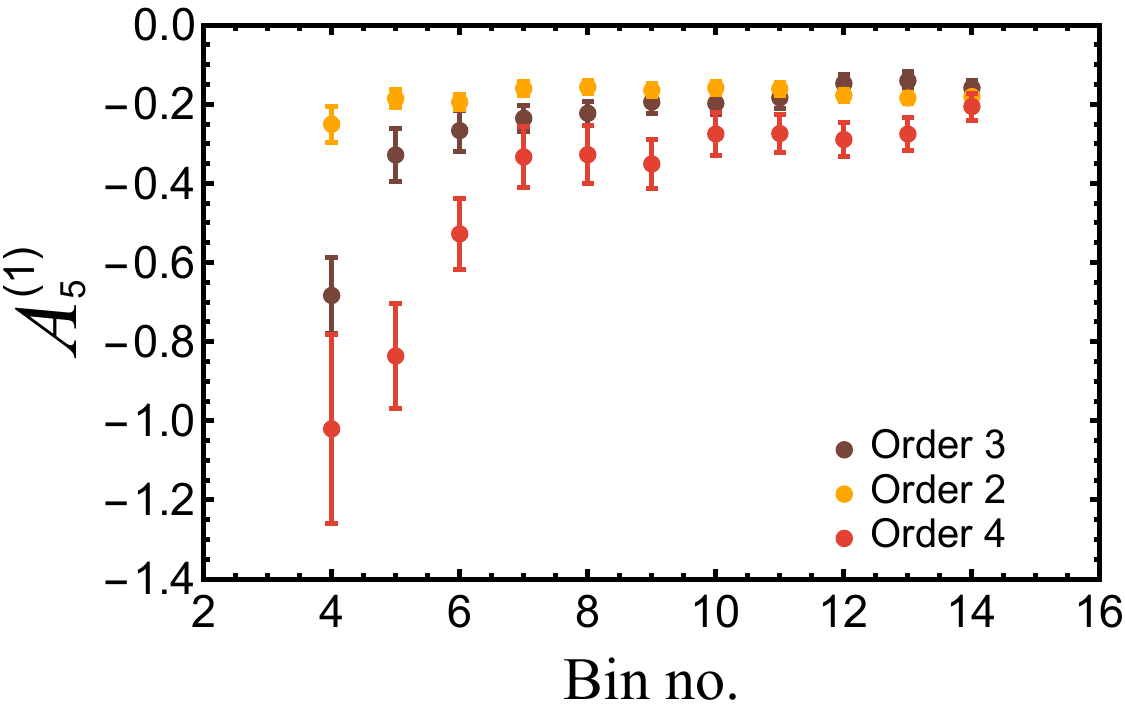}
	\includegraphics*[width=2.2in]{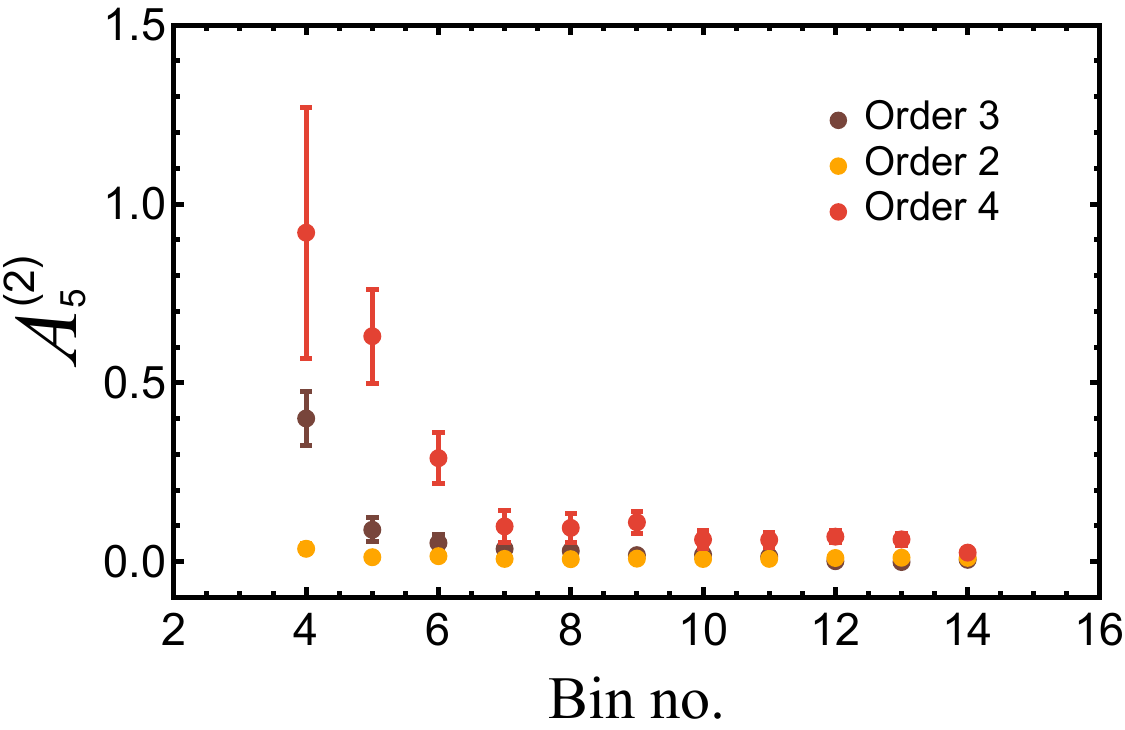}
	\caption{Systematic study of the coefficients of observables with the variation of polynomial order and the number of bins used for the fit. The color code for the different orders of the fitted polynomial is depicted in the panel. The $x$-axis denotes the number of bins used for the fit from last 4 to 14 bins. Coefficient values show good convergence within the $\pm1\sigma$ error bars except for few bins in the $F_\perp^{(1)}$ and $A_5^{(1)}$ distributions.}
	\label{fig:5}
\end{center}
\end{figure}
\end{center}

\end{widetext}

\begin{center}
	\begin{figure}[!h]		
		\begin{center}
			\includegraphics*[width=1.68in]{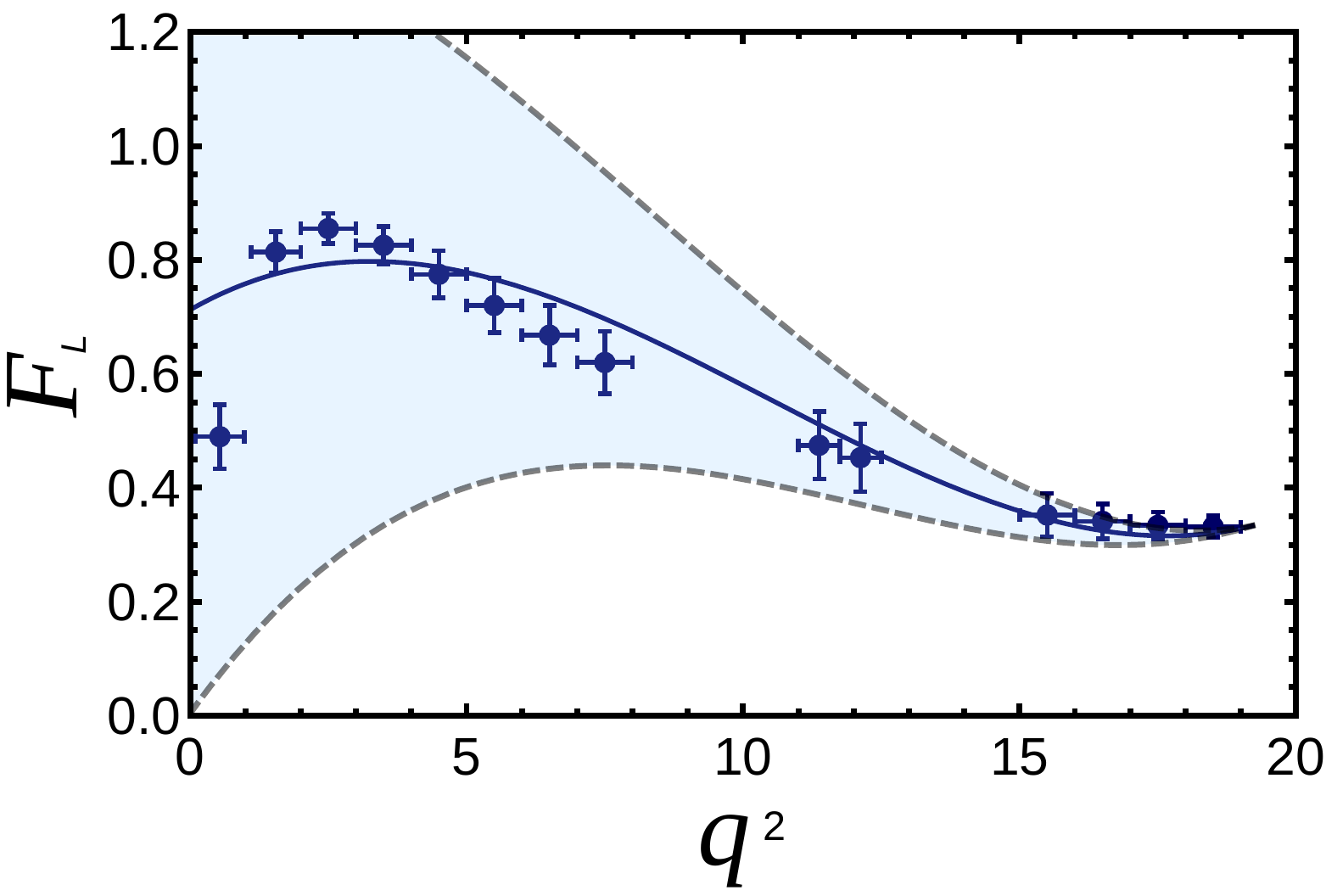}%
			\includegraphics*[width=1.68in]{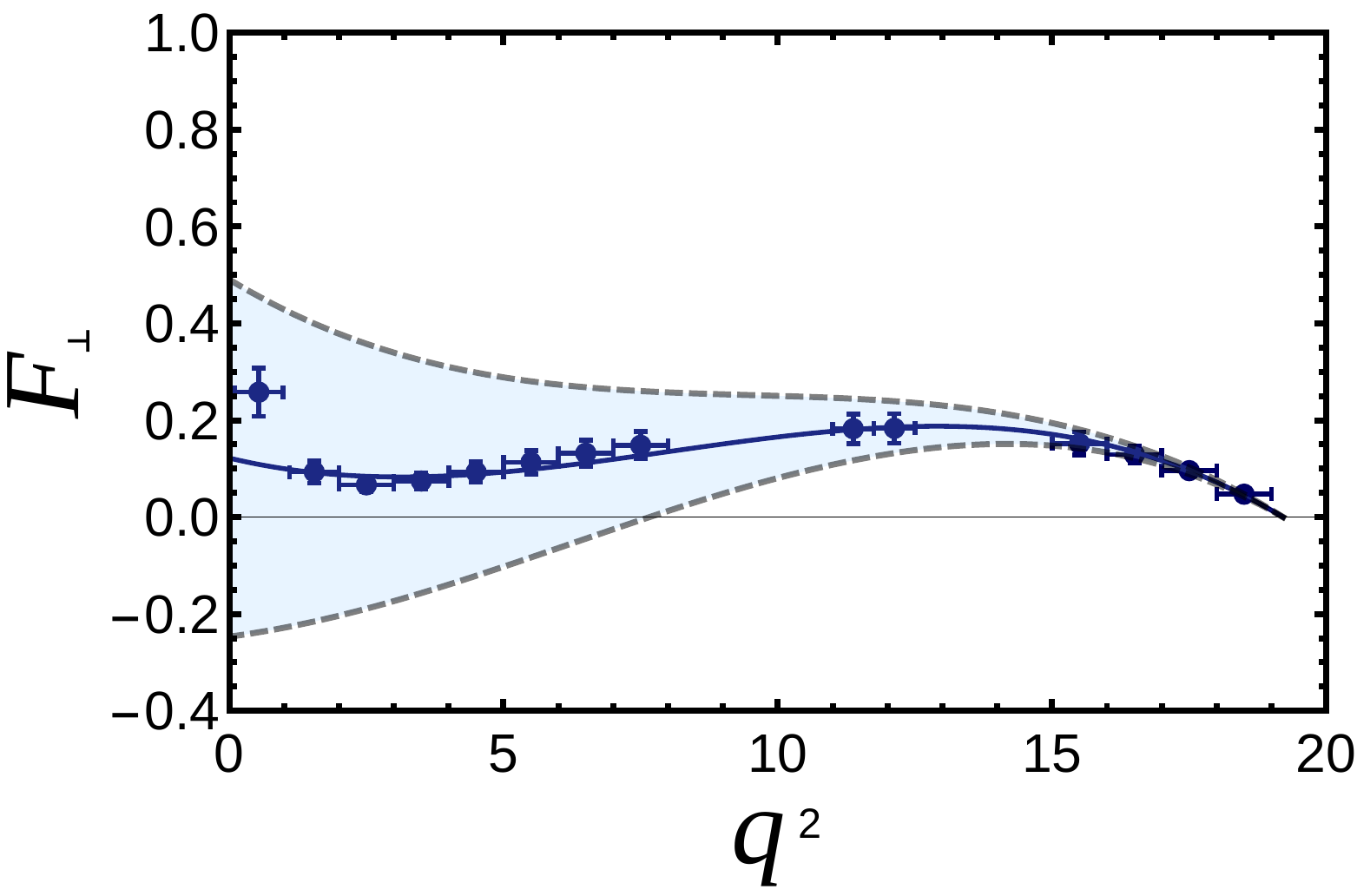}
			\includegraphics*[width=1.68in]{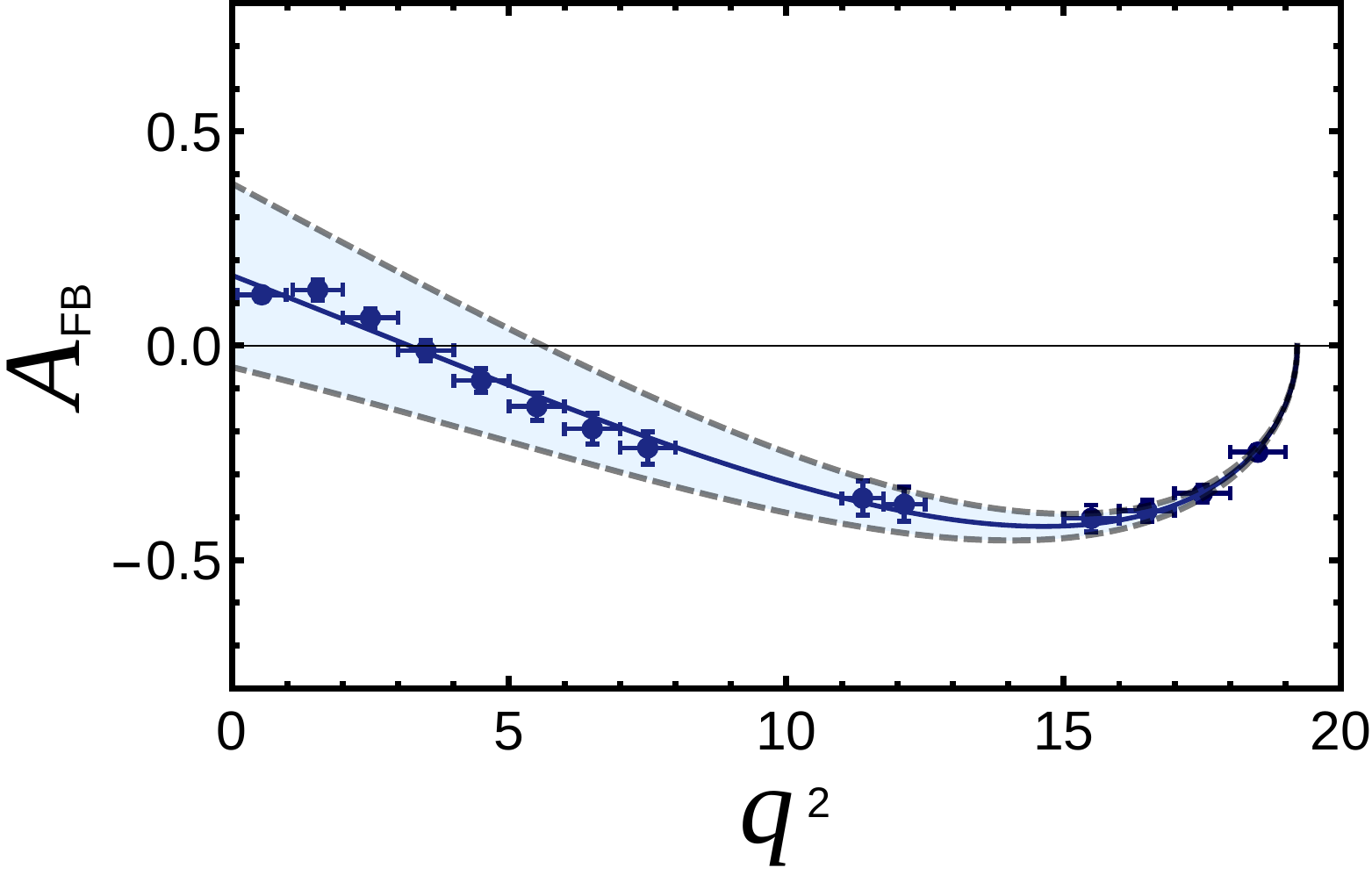}%
			\includegraphics*[width=1.68in]{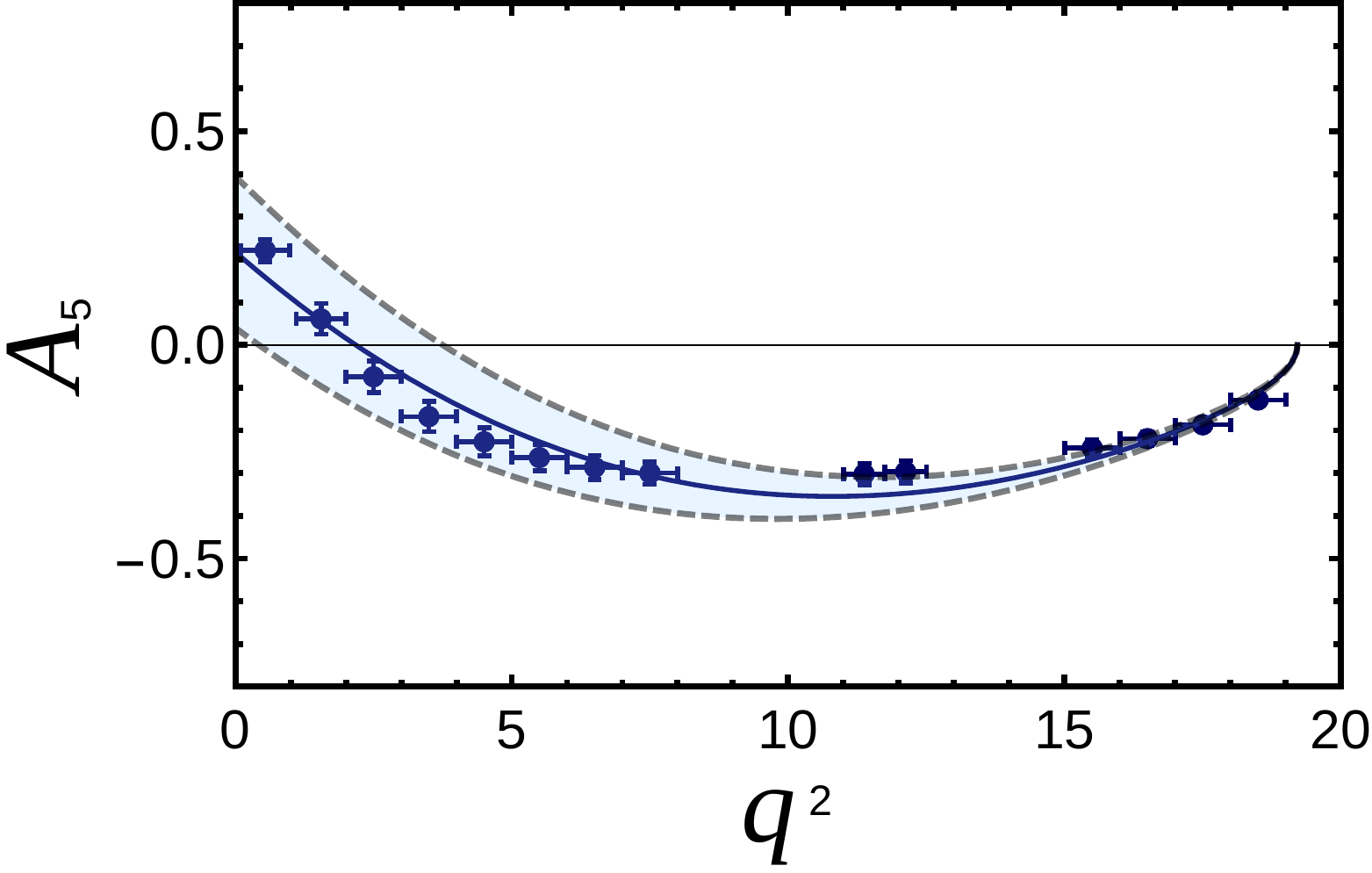}
			\caption{Fits with third order polynomials to the theoretical SM observables, generated using LCSR form factors for $q^2\le 15\gev^2$ and Lattice QCD form factors for $q^2\ge 15\gev^2$.} \label{fig:4}
		\end{center}
	\end{figure}
\end{center}

\section{Convergence of Polynomial Fit}
\label{sec:PolyFit}

It is  discussed in Sec.~\ref{sec:Theory} the observables are Taylor expanded around the endpoint $\qmax$ in Eqs.~\eqref{eq:funcFLfit}-- \eqref{eq:funcA5fit}. In this section, we study the systematics of the fits to coefficients $F_L^{(1)}$, $F_P^{(1)}$, $\AFB^{(1)}$,
$\AFB^{(2)}$, $A_5^{(1)}$ and $A_5^{(1)}$, which appear in the expressions of
$\omega_1$ and $\omega_2$ given in Eq.~\eqref{eq:omega}.  We vary the order the
polynomial fitted from $2$ to $4$. Fits are also performed by varying the number
of bins from the last $4$ to $14$ bins. The plots for the observables $\AFB$, $A_5$ $F_L$ and $F_\perp$ are shown in Appendix.~\ref{sec:Poly_Var}. 
The summary of the variation of fits with respect to
the order of the polynomial and number of bins are given in Fig.~\ref{fig:5} for all observable coefficients $F_L^{(1)}$, $F_P^{(1)}$, $\AFB^{(1)}$, $\AFB^{(2)}$, $A_5^{(1)}$ and $A_5^{(2)}$,  respectively. The color code for the order of the polynomial used to fit is given in the panel.
The $x-$ axis denotes the number of bins from last 4 to last 10 bins. We find that all the fitted coefficients show a good degree of convergence even when larger number of bins are added. The values obtained for the coefficients are consistent within $\pm 1 \sigma$ regions apart from some small mismatches in $F_P^{(1)}$, and $A_5^{(1)}$.
We choose as a benchmark the third order polynomial fit to all 14 bins.

To validate this choice of a third order polynomial fit to all 14 bins, we
also perform an identical fit for observables generated using form factor values
from LCSR~\cite{Straub:2015ica} for the $q^2\le 15\gev^2$
and from Lattice QCD~\cite{Horgan:2013pva} for $q^2\ge
15\gev^2$ region. The fits are shown in Fig.~\ref{fig:4}, where the blue error 
bars are bin integrated SM estimates and the solid blue curve with the shaded
region represents the best fit polynomial with $\pm 1\sigma$ errors. The fits to
SM observables are satisfactory for the entire $q^2$ region.

\section{Conclusion}
\label{sec:conclusion}

In conclusion, we have shown how RH currents can be uniquely probed without any
hadronic approximations at $\qmax$. Our approach adopted in Sec.~\ref{sec:Theory} differs from other approaches \cite{Altmannshofer:2014rta}
that probe new physics at low $q^2$, as it does not require estimates of
hadronic parameters but relies instead on heavy quark symmetry based arguments
that are reliable at $\qmax$~\cite{Grinstein:2004,Bobeth:2012vn}. Our parameters are defined so as to notionally include
non-factorizable loop correction and power-corrections and must differ from
those of others. It should be noted that we  use data directly, instead of
theory estimates, to derive our conclusions. We understand that experimental
measurements cannot alone result in discovery of NP as re-parameterization
invariance suggests and to that end we rely on theoretical understanding of
symmetries at the endpoint.
While the
observables themselves remain unaltered from their SM values, their derivatives
and second derivatives at the endpoint are sensitive to NP effects. Large
values of $\AFB$ and $A_5$, which do not rapidly approach zero in the
neighborhood of $\qmax$, are indicative of NP effects. In Sec.~\ref{sec:RH} we show that \lhcb data implies $5 \sigma$ evidence of NP at $\qmax$. While the signal for right
handed currents is clear, the large central values of $\xi$ and $\xi^{\prime}$ obtained will be reduced if other NP contributions are present. Allowing variation in the only input parameter i.e $r/C_{10}$, we obtain $3\sigma$ evidence of RH currents from the latest \lhcb measurements. A detailed study of resonance effects has been carried out in
Sec.~\ref{sec:reso}, which provides more significant evidence for RH currents. The systematics of polynomial fit has been discussed in Sec.\ref{sec:PolyFit} where a good convergence has been observed when a large number of bins are considered. The choice of a particular polynomial parametrization has been justified with a fit to SM observables. The effect of complex contributions in the amplitudes (in Appendix \ref{sec:epsilon}) and the finite $\kstar$ width (in Appendix \ref{sec:finite width}) leaves the conclusions unchanged. In view of these, we speculate that if the current features of data persist with higher statistics the existence of RH currents can be established in the near future.

\acknowledgments
We thank Enrico Lunghi and Joaquim Matias for useful discussions and also encouraging us to incorporate correlations among experimental observables in our analysis.
We thank Marcin Chrzaszcz, Nicolla Serra and other members of the LHCb
collaboration for extremely valuable comments and suggestions.  We also 
thank B.
Grinstein, N. G. Deshpande and S. Pakvasa for useful discussions. T. E. 
Browder
thanks the US DOE grant no. DE-SC0010504 for support.

\appendix

\section{Effect of Complex Contributions of amplitude}
\label{sec:epsilon}

We show that the contributions arising from the  complex part 
$(\varepsilon_\lambda)$ of the amplitudes, in Eq.~\eqref{eq:amp-def2}, can be 
incorporated in the following way.

Defining a new notation $\widehat{\varepsilon}_\lambda\equiv
2|\varepsilon_\lambda|^2/\Gamma_f$, the Taylor expansions for each  
$\widehat{\varepsilon}_\lambda$ around $q^2=\qmax$ are given by,

\begin{align*}
\widehat{\varepsilon}_\perp&= \widehat{\varepsilon}_\perp^{(1)} \delta + \widehat{\varepsilon}_\perp^{(2)} \delta^2+\widehat{\varepsilon}_\perp^{(3)} \delta^3 \\
\widehat{\varepsilon}_0&= \widehat{\varepsilon}_0^{(0)} + \widehat{\varepsilon}_0^{(1)} \delta + \widehat{\varepsilon}_0^{(2)} \delta^2 \\
\widehat{\varepsilon}_\|&= \widehat{\varepsilon}_\|^{(0)} + \widehat{\varepsilon}_\|^{(1)} \delta + \widehat{\varepsilon}_\|^{(2)} \delta^2 
\end{align*}
where $\delta\equiv\qmax-q^2$ and the limiting values of helicity fractions, $F_L(\qmax)=1/3 \text{~and~} F_\|(\qmax)=2/3$, constrain the coefficients i.e. $\widehat{\varepsilon}_\|^{(0)}=2 \,\widehat{\varepsilon}_0^{(0)}$. The presence of complex amplitudes leads to a modification of the expressions of $\omega_1$ and $\omega_2$ (Eq.~\eqref{eq:omega}) in the following way,
\begin{align}
\label{eq: w1}
\omega_1=\,&\dsp\frac{9}{4} \frac{\left(\frac{2}{3}-2\,\widehat{\varepsilon}_0^{(0)} \right)\left(F_\perp^{(1)}-\widehat{\varepsilon}_\perp^{(1)}\right)}{\AFB^{(1) \, 2}} \\
\label{eq: w2}
\omega_2=\,&\dsp\frac{4\,\Big(2 A_5^{(2)}-\AFB^{(2)}\Big) \Big(1-3\, \widehat{\varepsilon}_0^{(0)}\Big)}
{3\,\AFB^{(1)}\Big(3 F_L^{(1)}+F_\perp^{(1)}+\widehat{\varepsilon}_\|^{(1)}-2\,\widehat{\varepsilon}_0^{(1)}\Big)}.
\end{align}

The procedure to incorporate the complex part of the amplitudes
$\varepsilon_\lambda$ is described in Ref.~\cite{Mandal:2014kma}, where it was
shown  that the complex part of the amplitudes $\varepsilon_\lambda$ are
proportional to the asymmetries $A_7$, $A_8$ and $A_9$. Using $3\invfb$ of \lhcb
data~\cite{Aaij:2015oid}, we simulated values of the coefficients
$\widehat{\varepsilon}_0^{(0)}$, $\widehat{\varepsilon}_0^{(1)}$,
$\widehat{\varepsilon}_\|^{(1)}$ and $\widehat{\varepsilon}_\perp^{(1)}$; these
turn out to be very small at the kinematic endpoint. These estimated coefficients are
used to evaluate $\omega_1=1.03\pm 0.31~(0.98\pm 0.29)$ and $\omega_2=-4.52\pm
17.40~(-3.94\pm 9.86)$ (Eq.~\eqref{eq: w1} and \eqref{eq: w2}), where the
first values are determined using $\AFB^{(1)}$ and $A_9^{(1)}$ whereas the values
in the round brackets use $2A_5^{(1)}$ and $-\frac{2}{3}A_8^{(1)}$. The
factorization assumption is needed only at leading order in the expansions of observables,
which requires $\AFB^{(1)}= 2A_5^{(1)}$ and $A_9^{(1)}=-\frac{2}{3}A_8^{(1)} $.

It should be noted that the inclusion of 
$\widehat{\varepsilon}_\lambda$'s change the values of $\omega_1$ and 
$\omega_2$ insignificantly, with corresponding estimates for the real case being 
well within the $\pm 1\sigma$ errors.
Hence, 
the conclusions derived in the paper are robust against the inclusion of complex contributions in the amplitudes.

\section{Finite \boldmath$\kstar$ width effect}
\label{sec:finite width}

The finite width of the $\kstar$ can alter the position of the kinematic 
endpoint i.e 
$\qmax$ value. As LHCb considers a much wider range for the width of $K^*$, compared to the observed width (which is $\sim 50$ MeV), we have varied the $\qmax$ value in the Taylor expansion of 
observables (Eqs.~\eqref{eq:funcFLfit}--\eqref{eq:funcA5fit}) within an interval $18.34 - 20.10$ GeV$^2$. 
The observables $\omega_1$ and $\omega_2$ are evaluated for each case and a 
weighted average over the Breit-Wigner shape for a $\kstar$ gives 
$\omega_1=1.11\pm 0.30~(1.03\pm 
0.35)$ and $\omega_2=-3.56\pm 28.34~(-3.50\pm 27.44)$. The change in the values 
of $\omega_1$ and $\omega_2$ have an insignificant effect in Fig.~\ref{fig:1} and \ref{fig:2} and the results derived in this work.

\section{Polynomial fit variation }
\label{sec:Poly_Var}
The variation of fits with respect to
the order of the polynomial and number of bins are shown in Fig.~\ref{fig:0a},
\ref{fig:0b}, \ref{fig:0c} and \ref{fig:0} for observables $\AFB$, $A_5$, $F_L$
and $F_\perp$, respectively The color code is the same as in Fig.~\ref{fig:1}. The panel in each plot depicts the number of bins (from kinematic endpoint) and the order of polynomial is used for the fit. The extracted coefficient values of the observables from these plots are summarized in Fig.~\ref{fig:4}.

\begin{widetext}
	
	\begin{center}
		\begin{figure}[!th]
			\begin{center}
				\includegraphics*[width=1.5in]{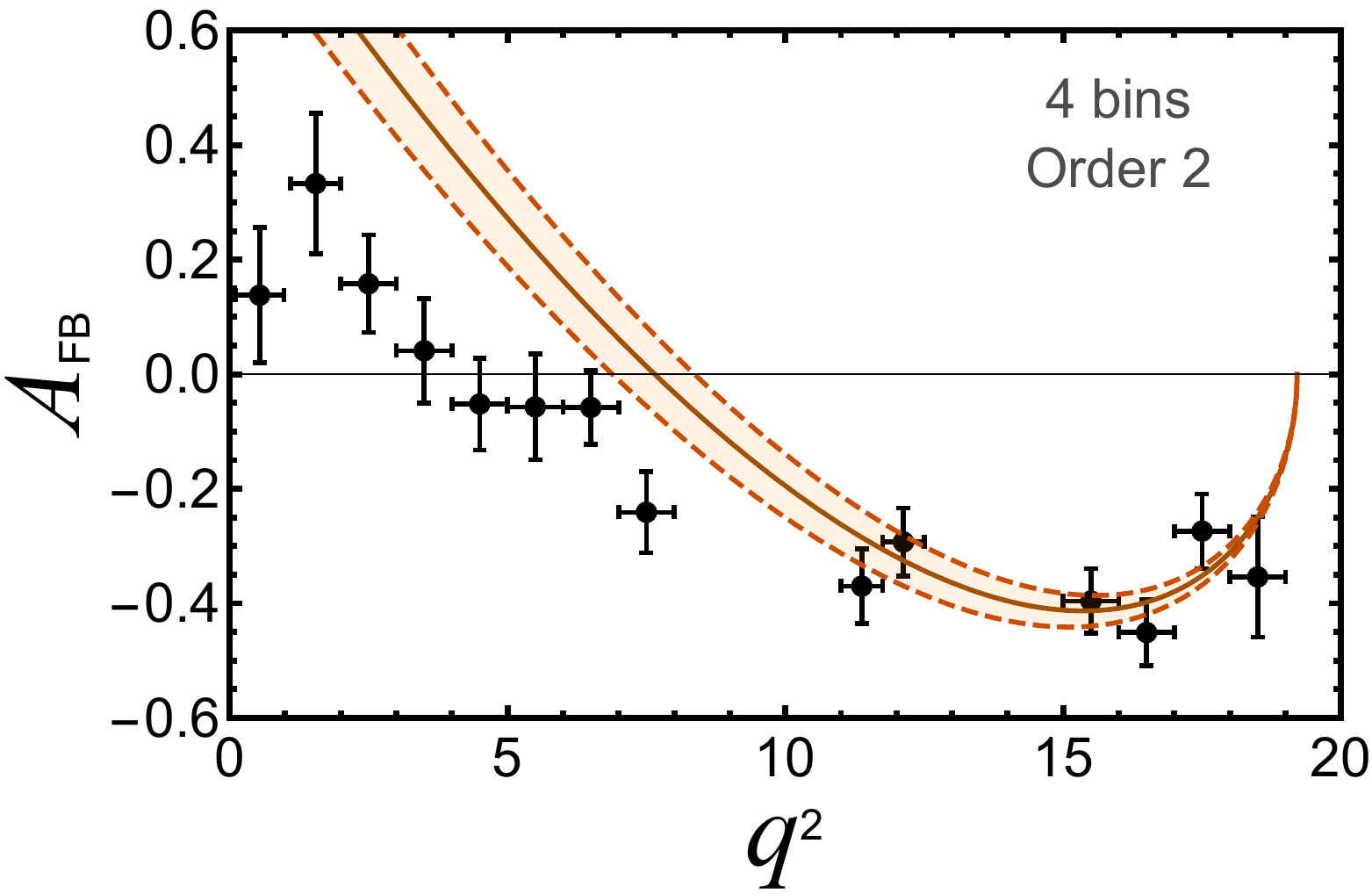}%
				\includegraphics*[width=1.5in]{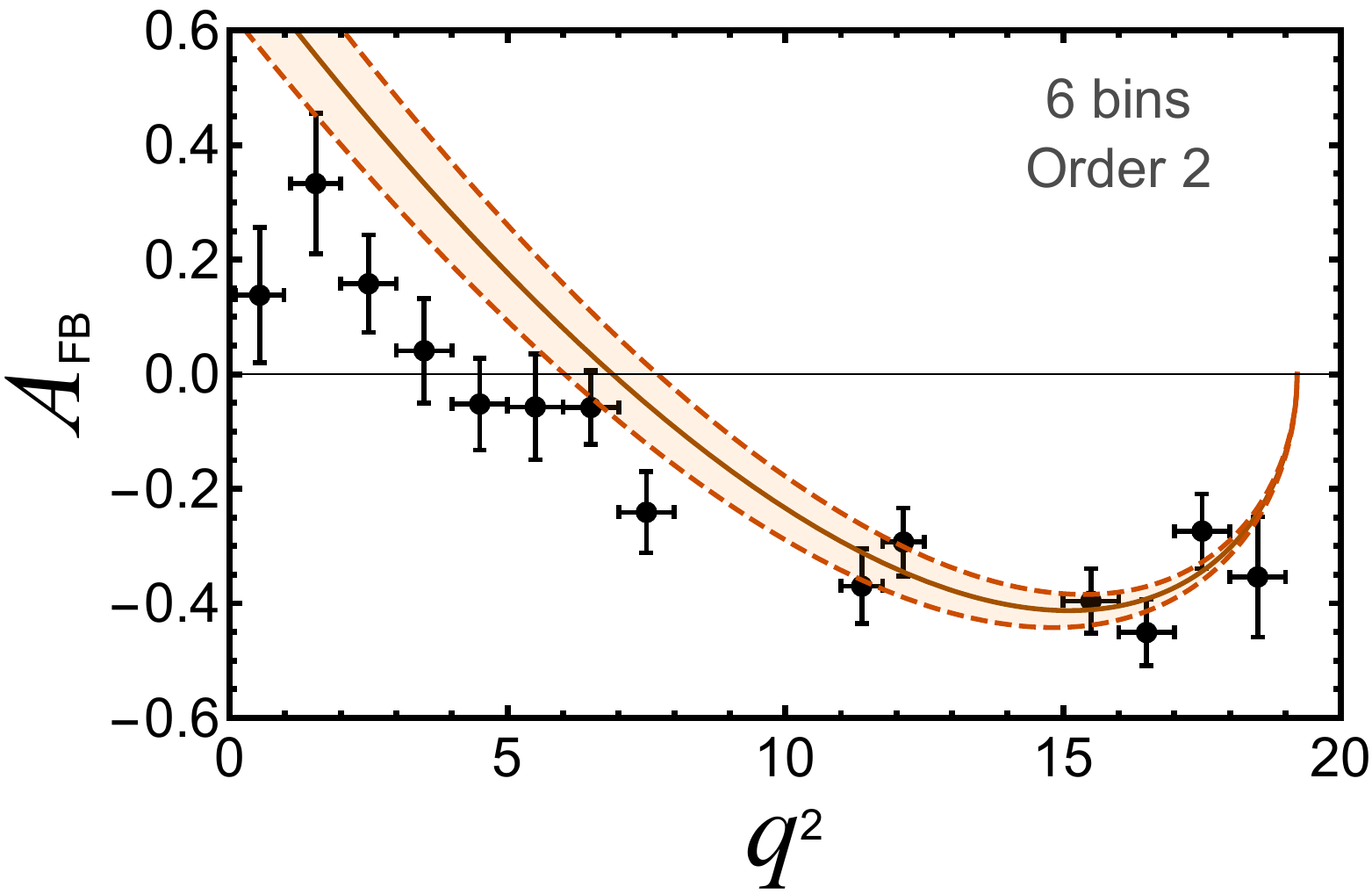}%
				\includegraphics*[width=1.5in]{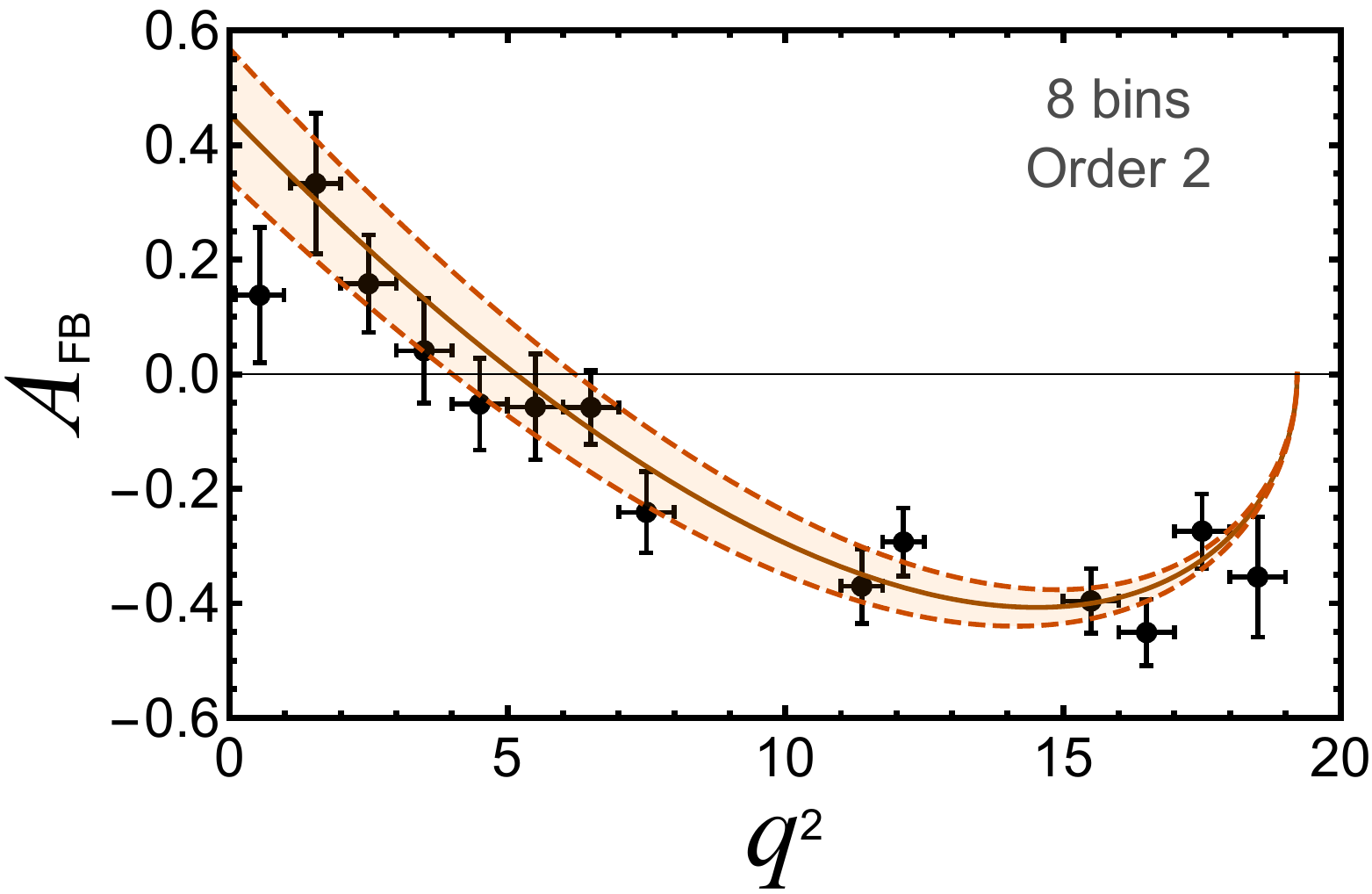}\\%
				\includegraphics*[width=1.5in]{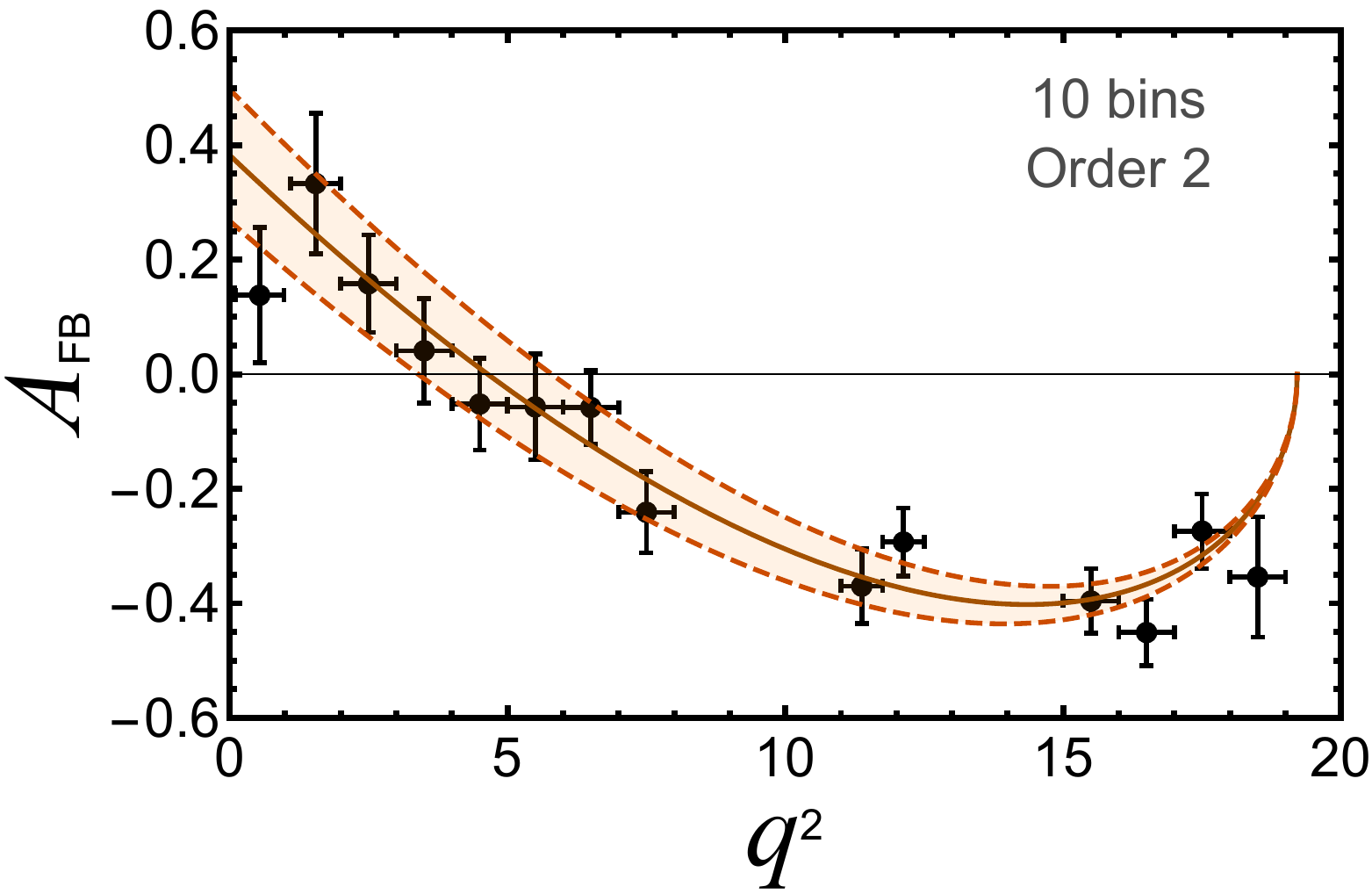}%
				\includegraphics*[width=1.5in]{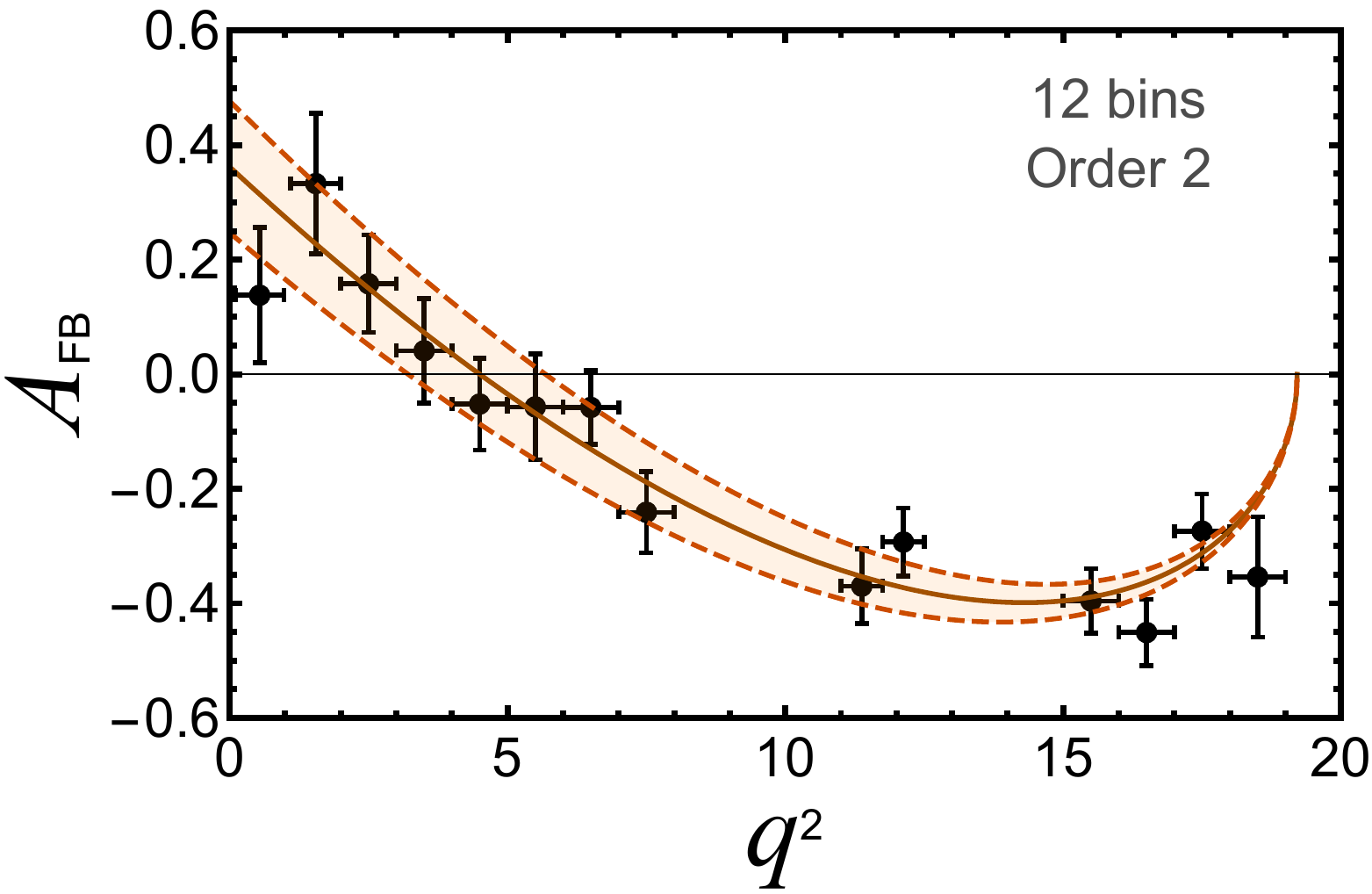}%
				\includegraphics*[width=1.5in]{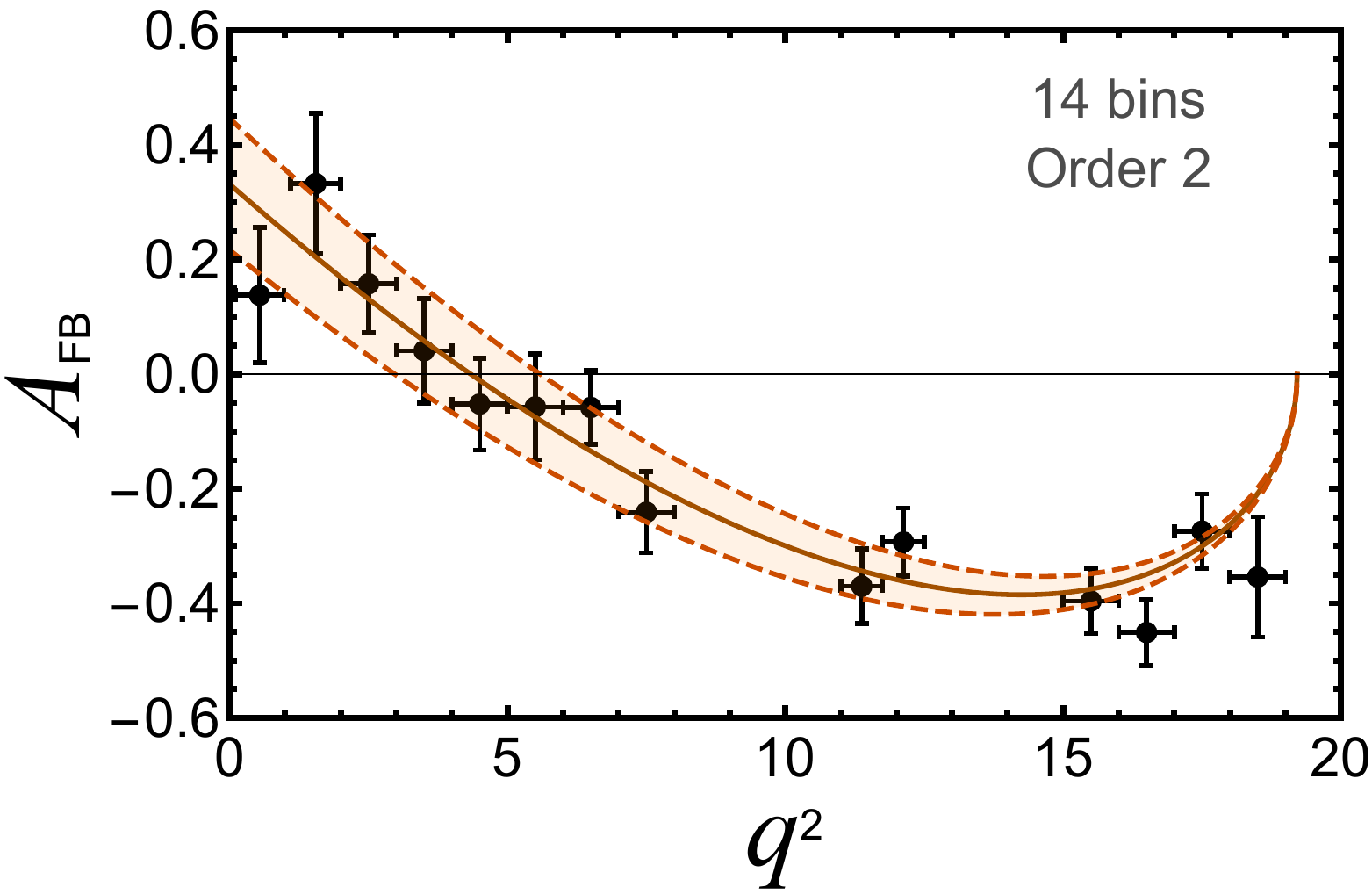}\\%
				\includegraphics*[width=1.5in]{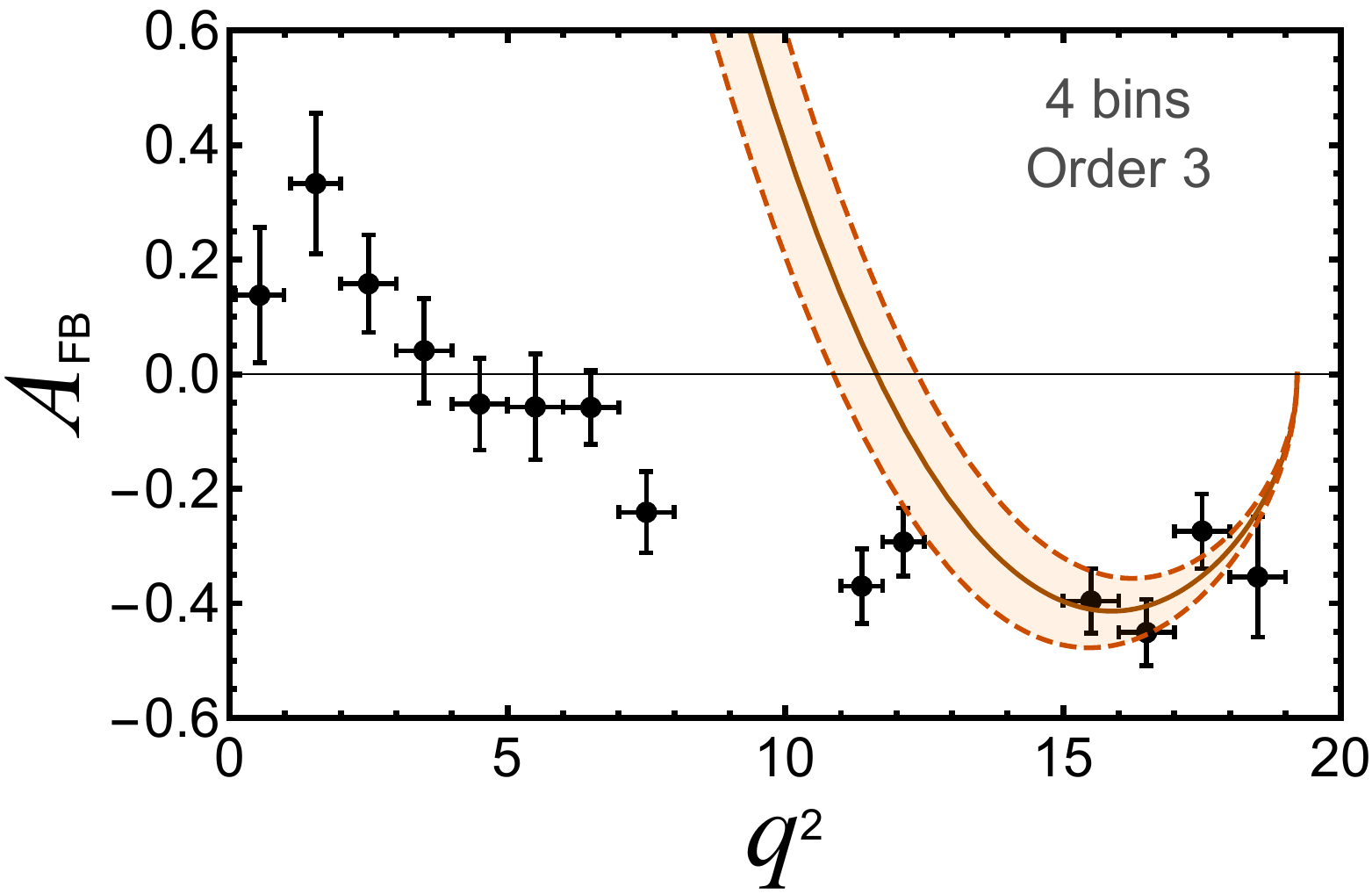}%
				\includegraphics*[width=1.5in]{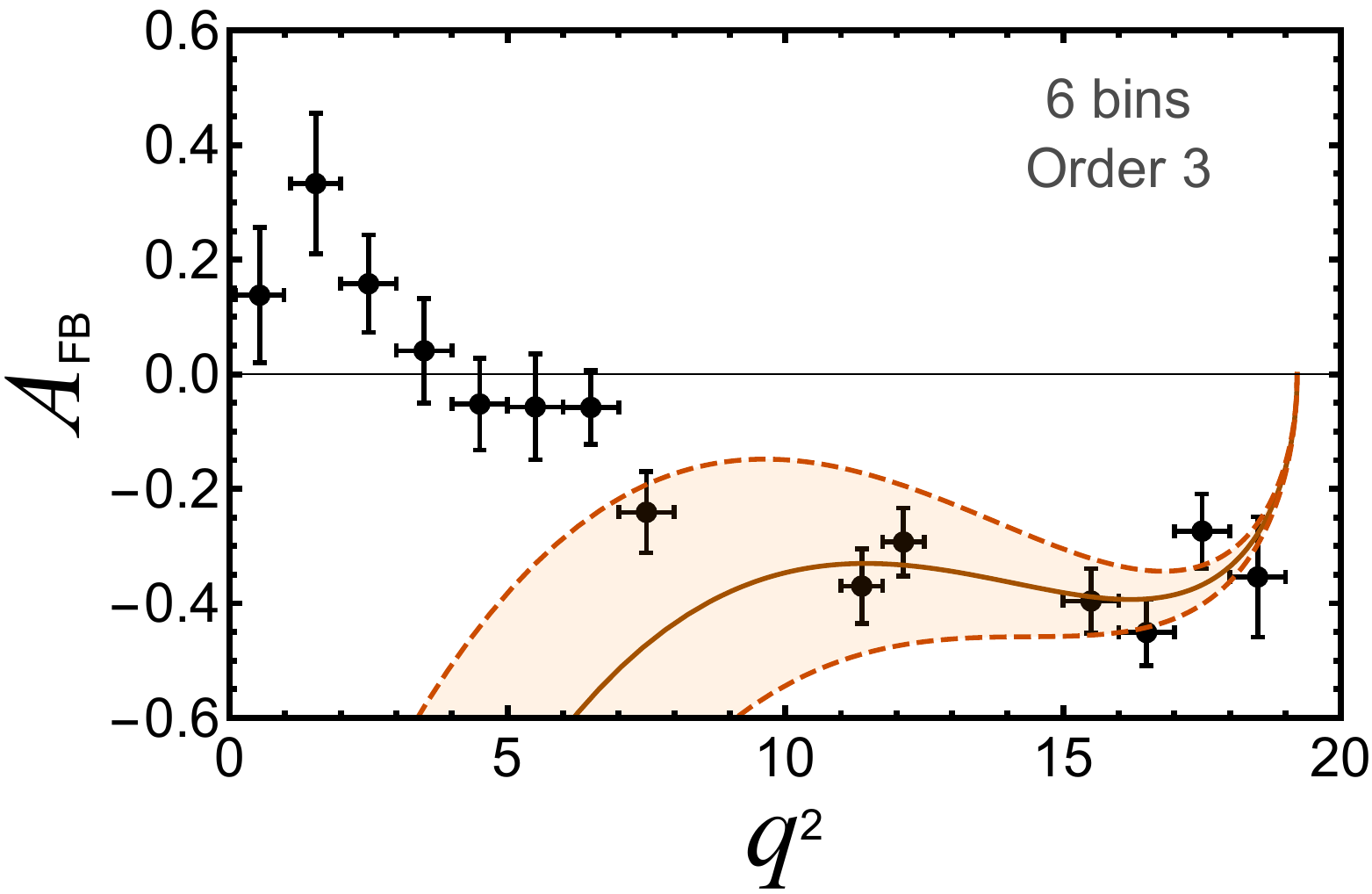}%
				\includegraphics*[width=1.5in]{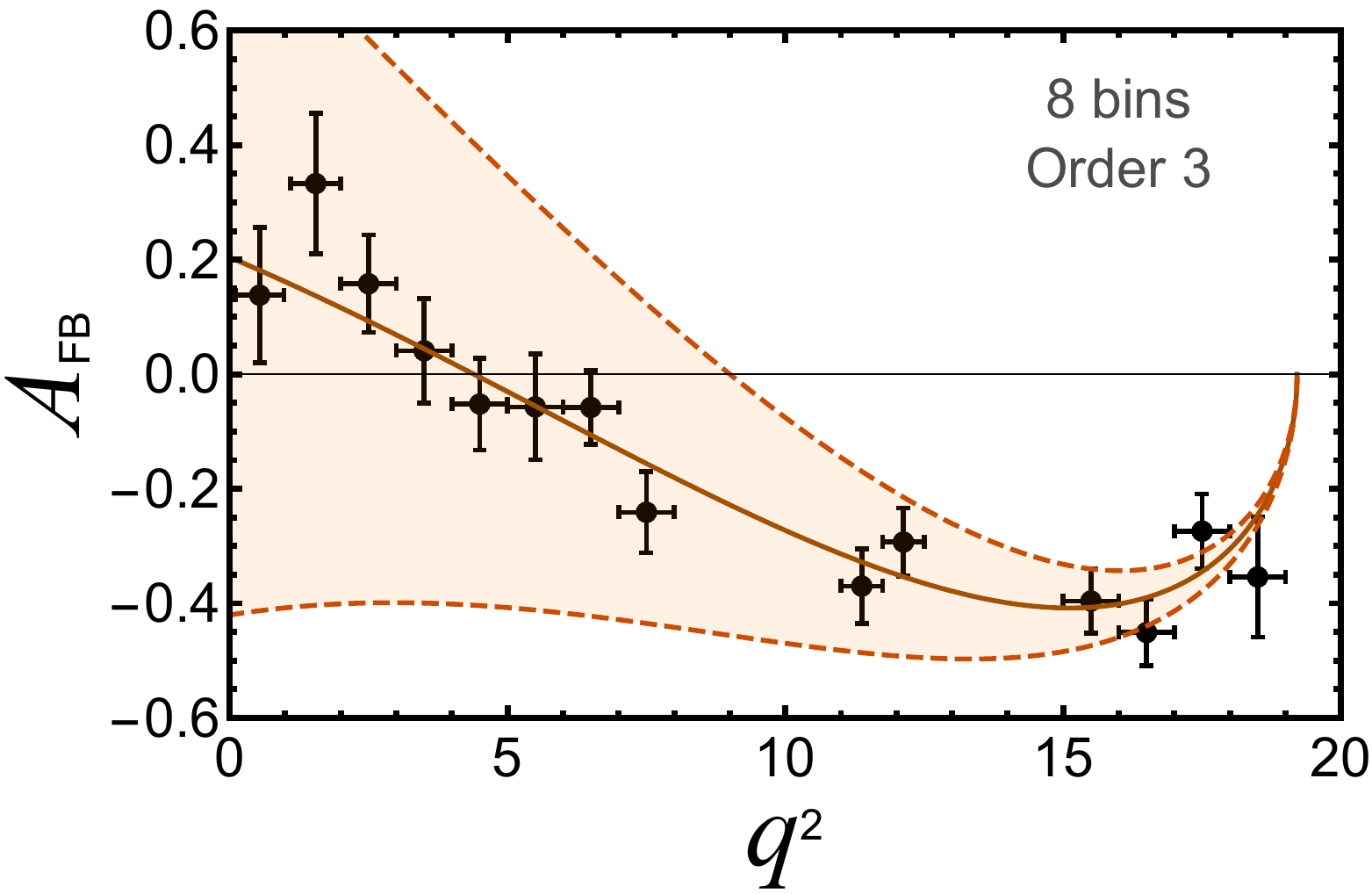}\\%
				\includegraphics*[width=1.5in]{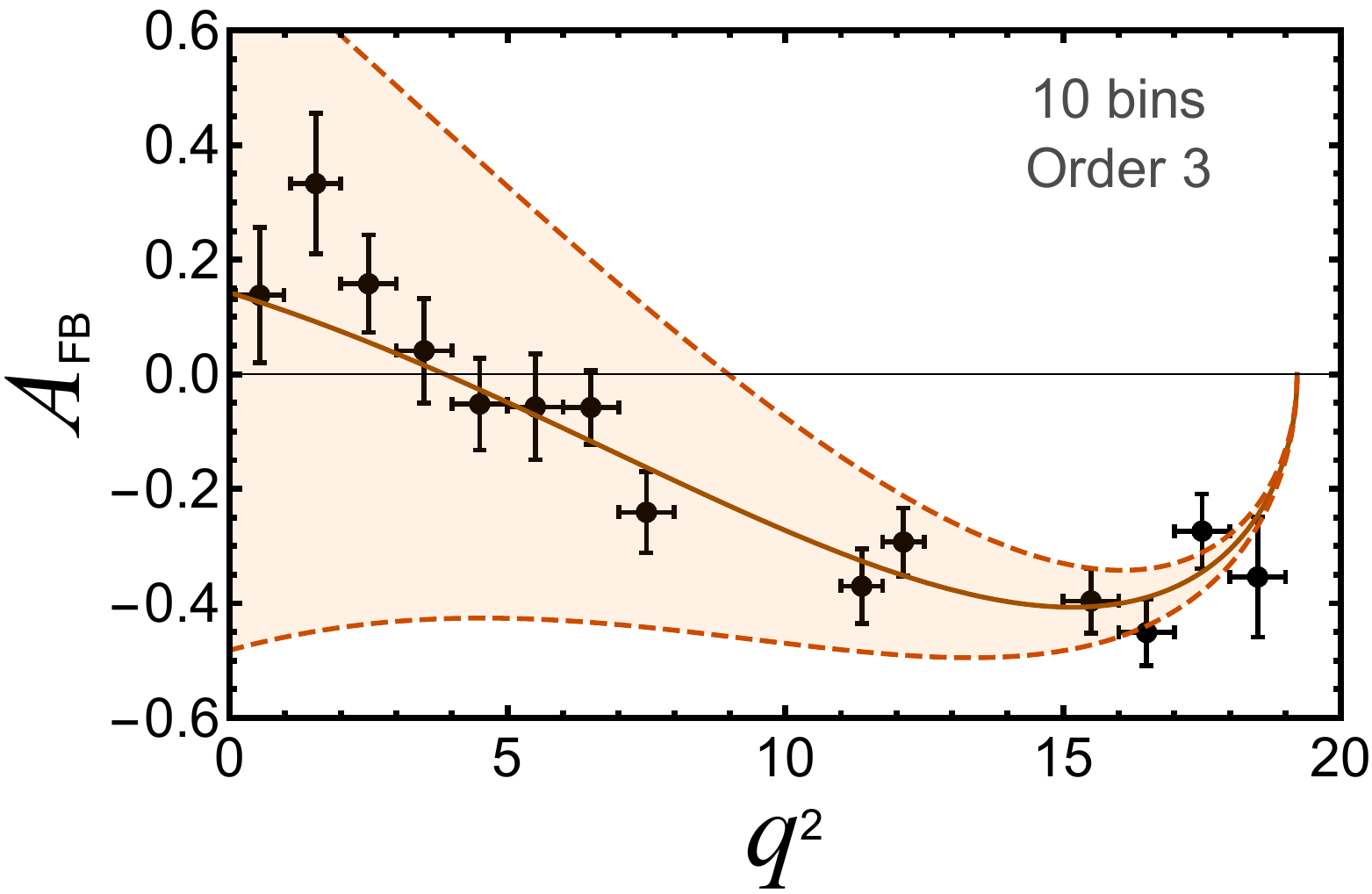}%
				\includegraphics*[width=1.5in]{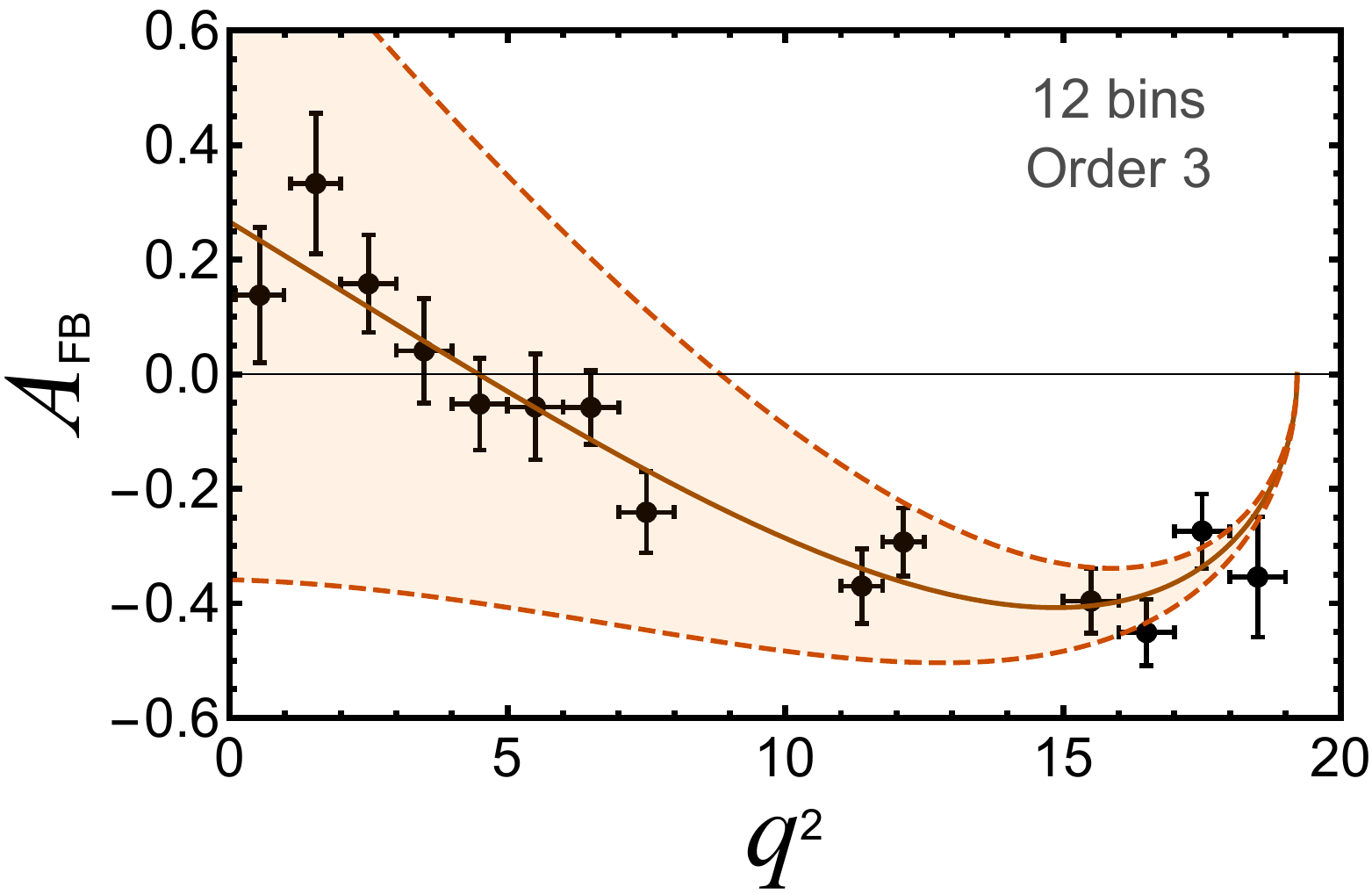}%
				\includegraphics*[width=1.5in]{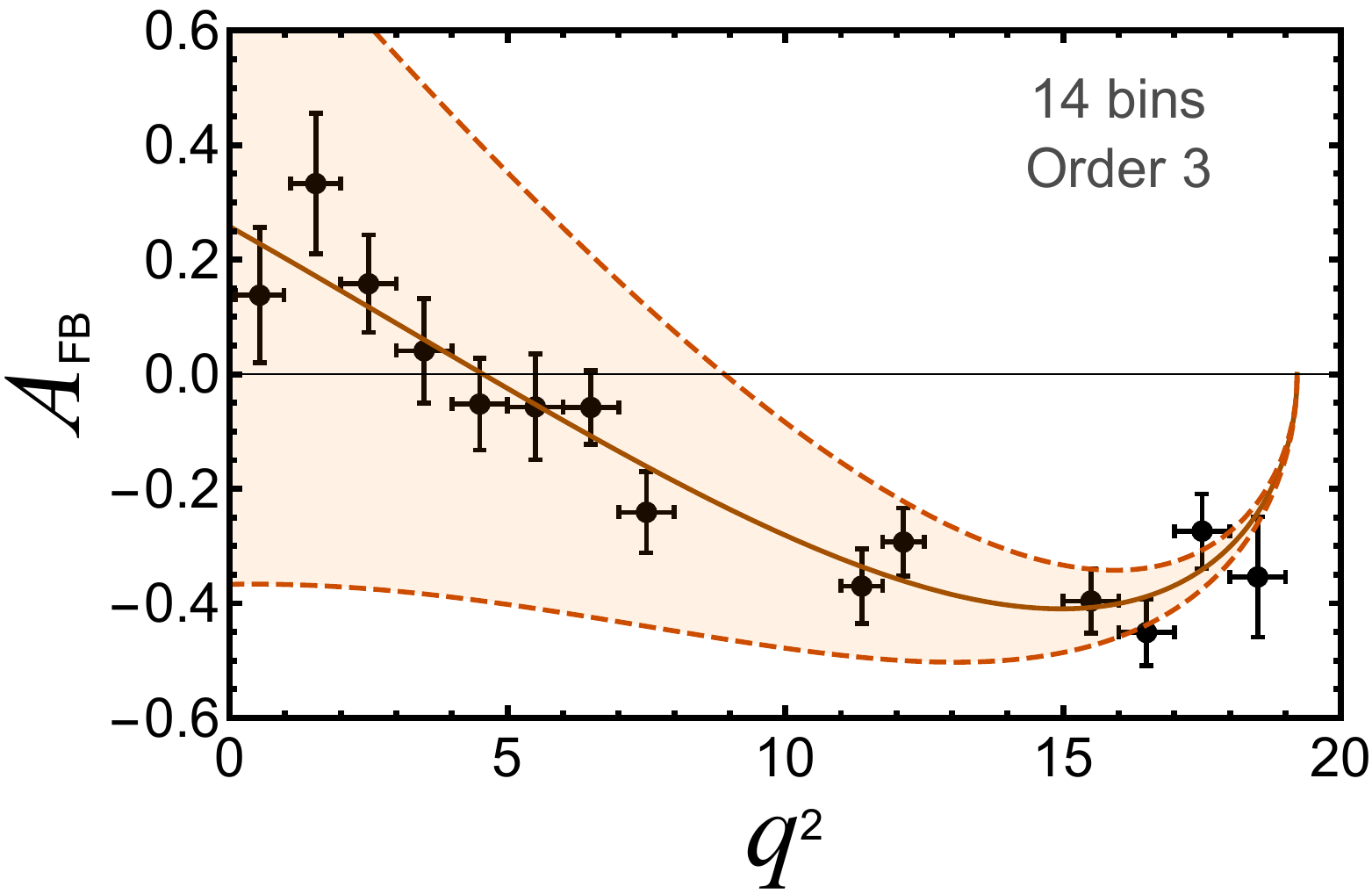}\\%
				\includegraphics*[width=1.5in]{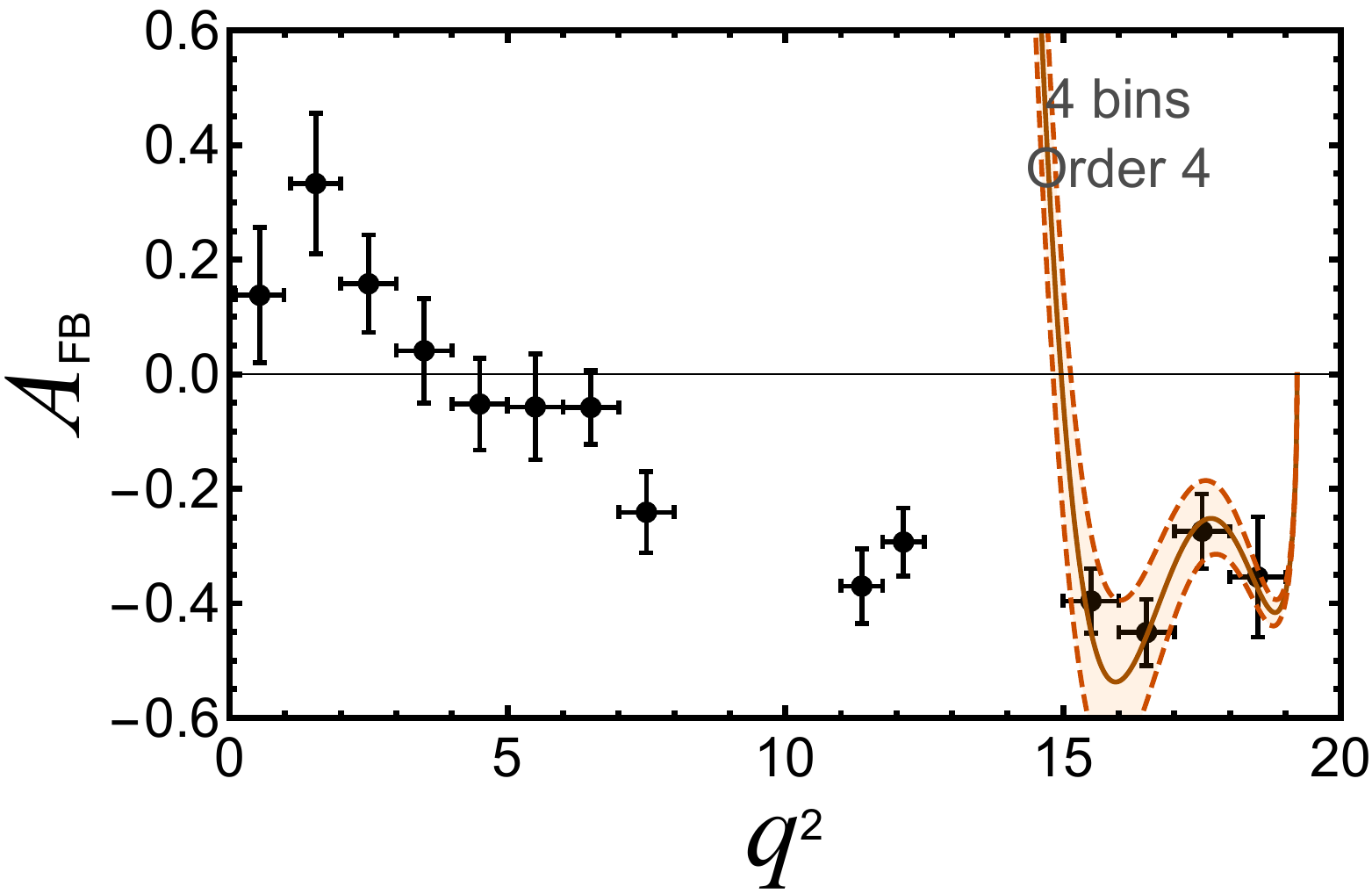}%
				\includegraphics*[width=1.5in]{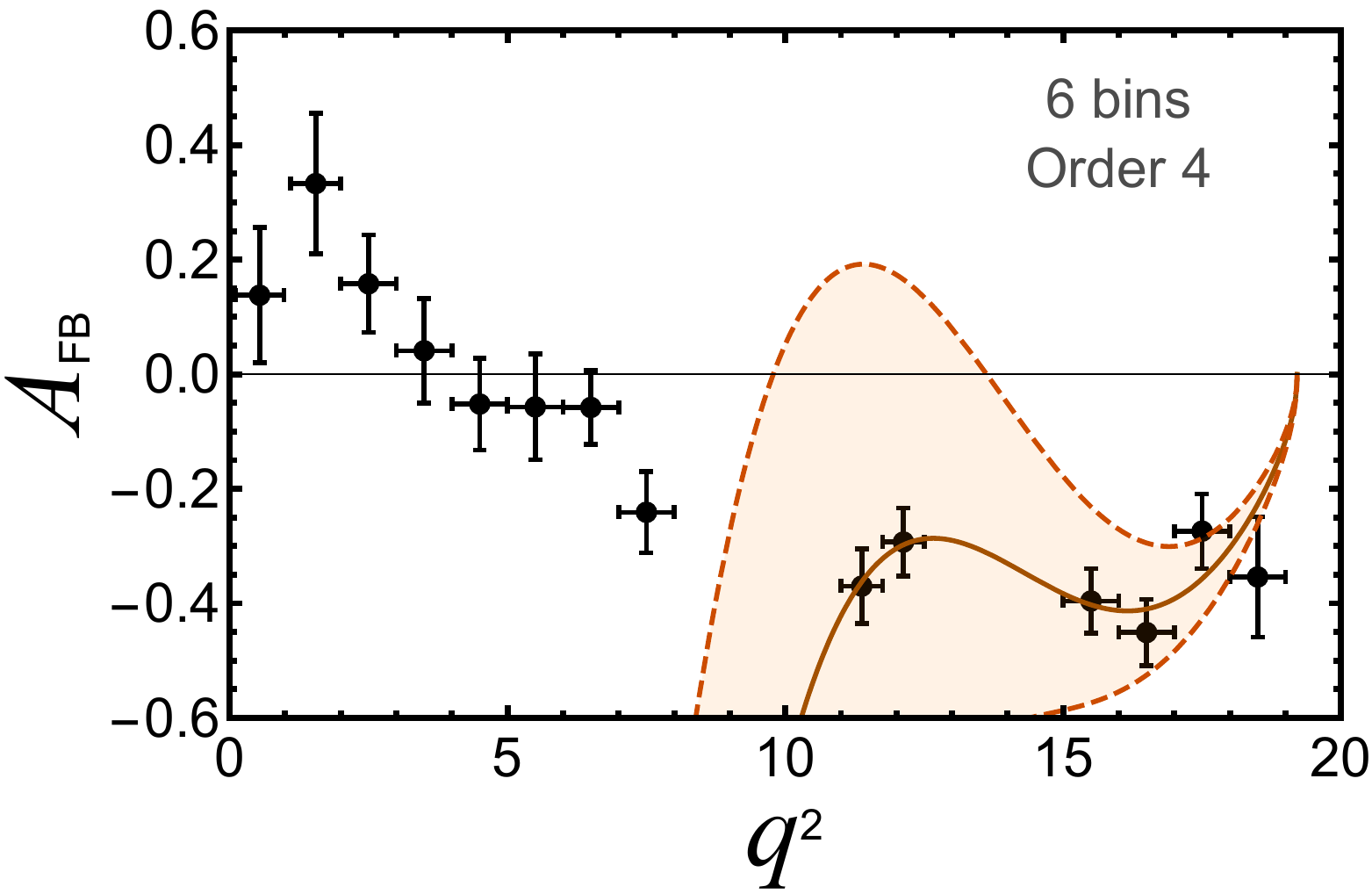}%
				\includegraphics*[width=1.5in]{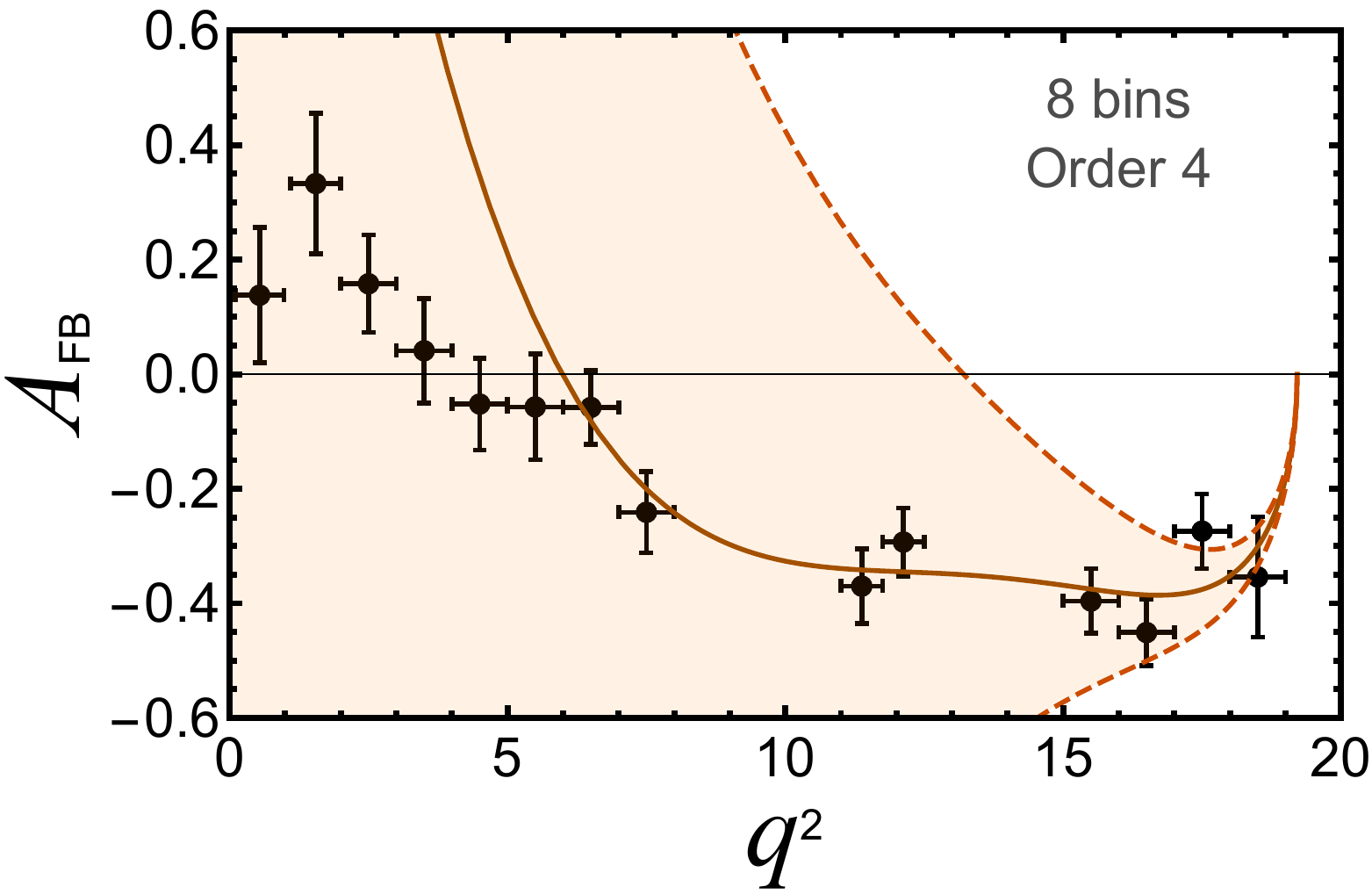}\\%
				\includegraphics*[width=1.5in]{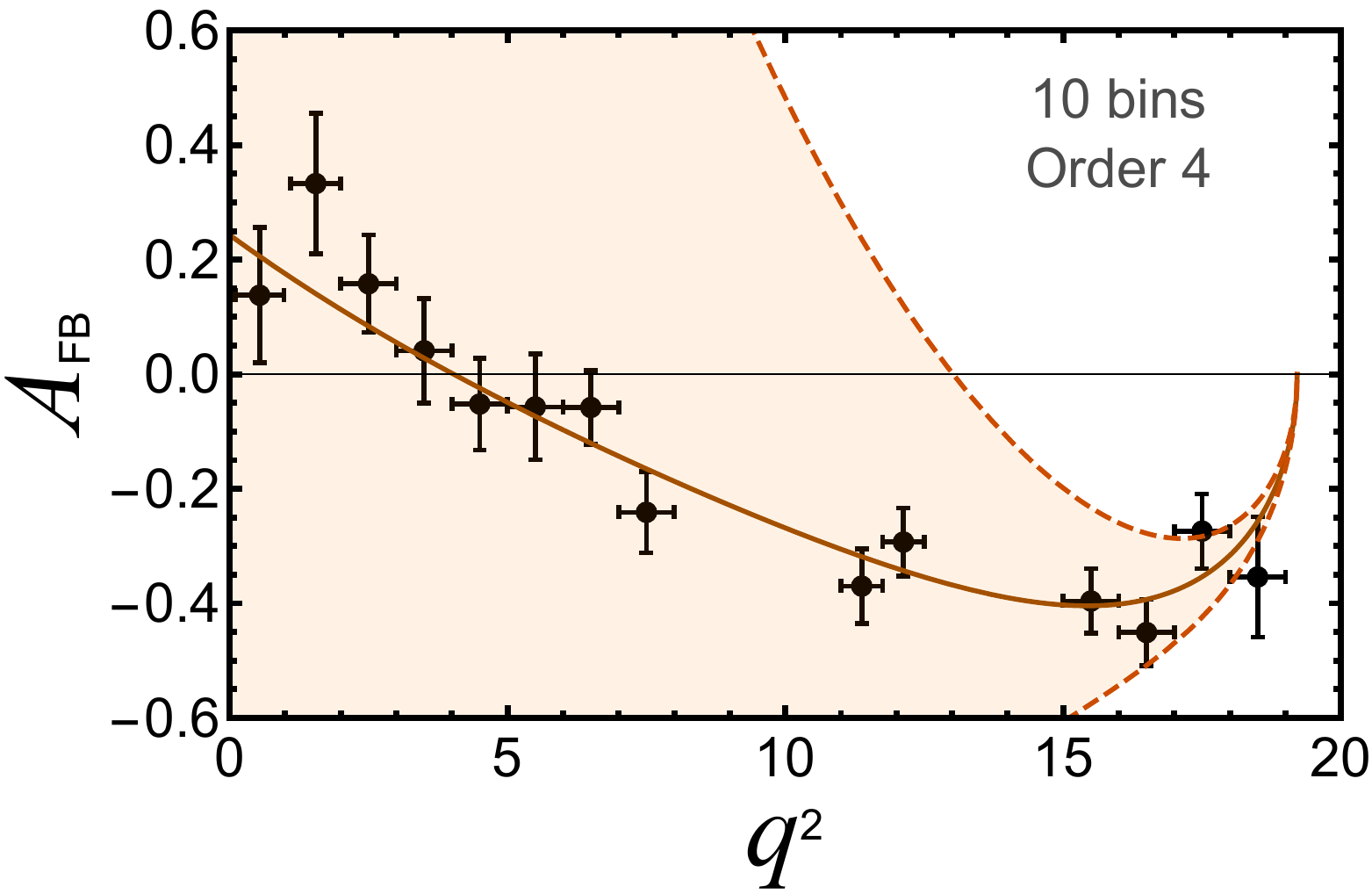}%
				\includegraphics*[width=1.5in]{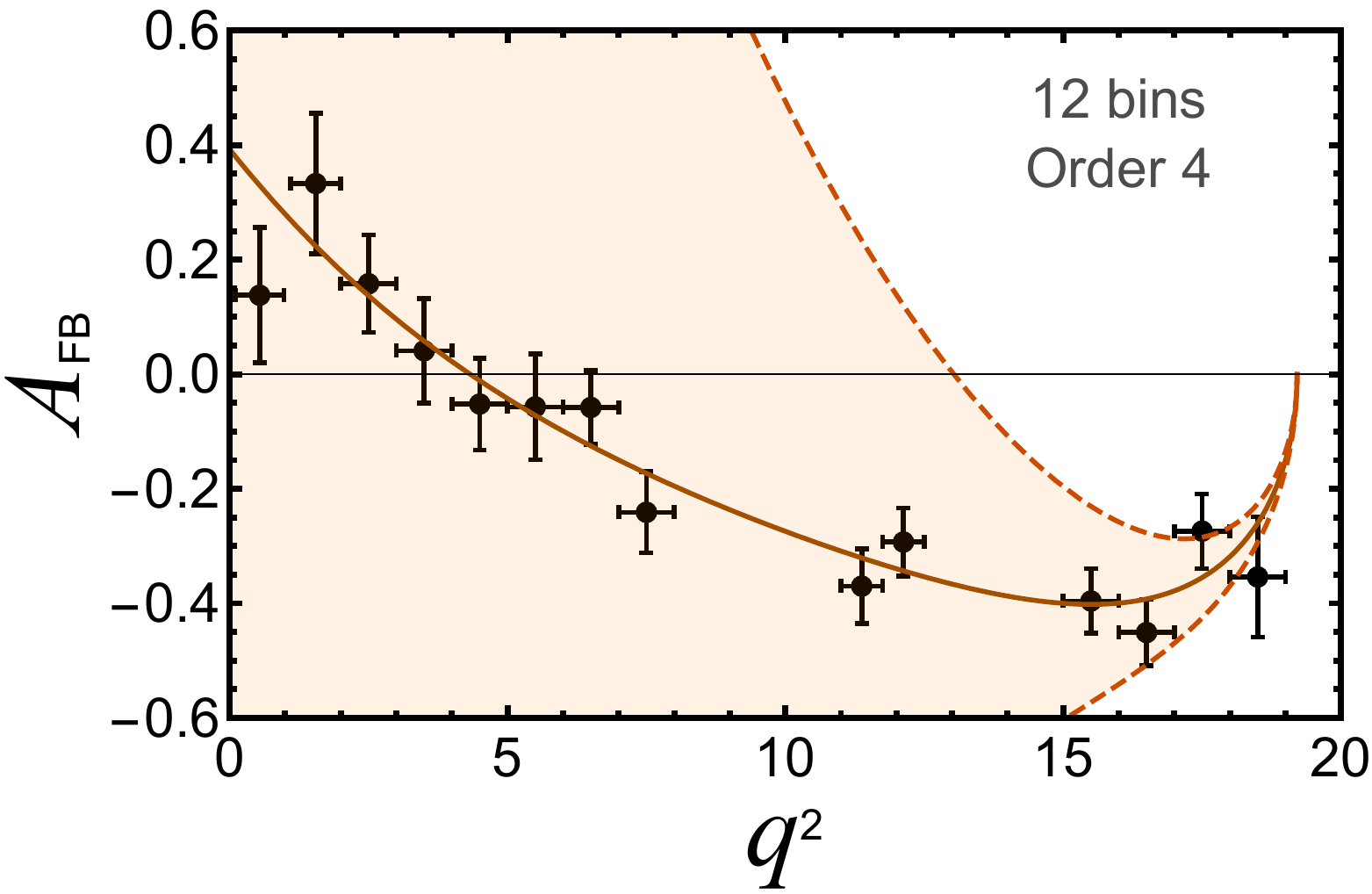}%
				\includegraphics*[width=1.5in]{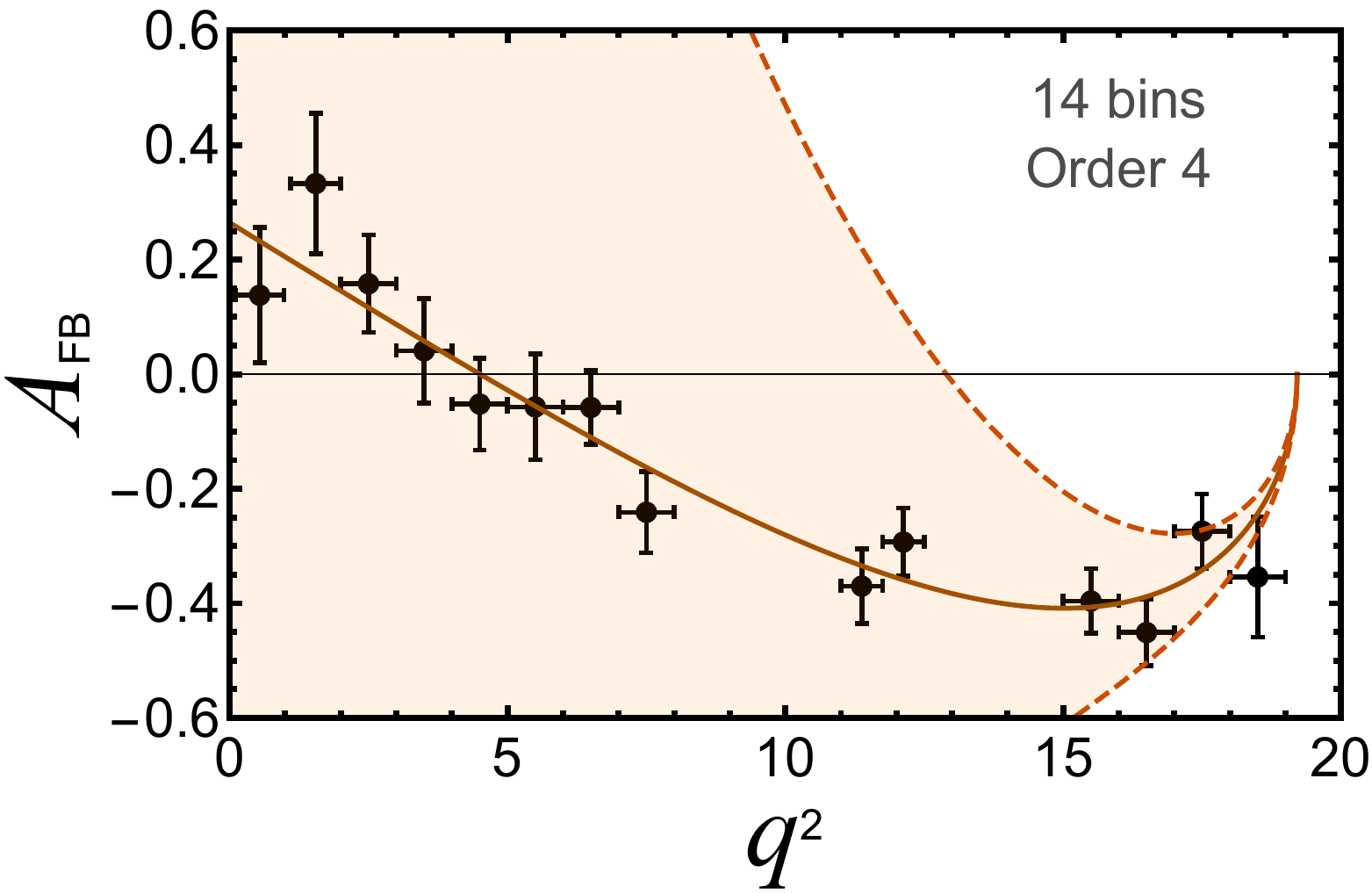}\\%
				\caption{Fits to $\AFB$ using various numbers of bins and polynomial parameterizations. The color code is the same as in Fig.~\ref{fig:1}} 
				\label{fig:0a}
			\end{center}
		\end{figure}
	\end{center}

	\begin{center}
		\begin{figure}[th]
			\begin{center}
				\includegraphics*[width=1.5in]{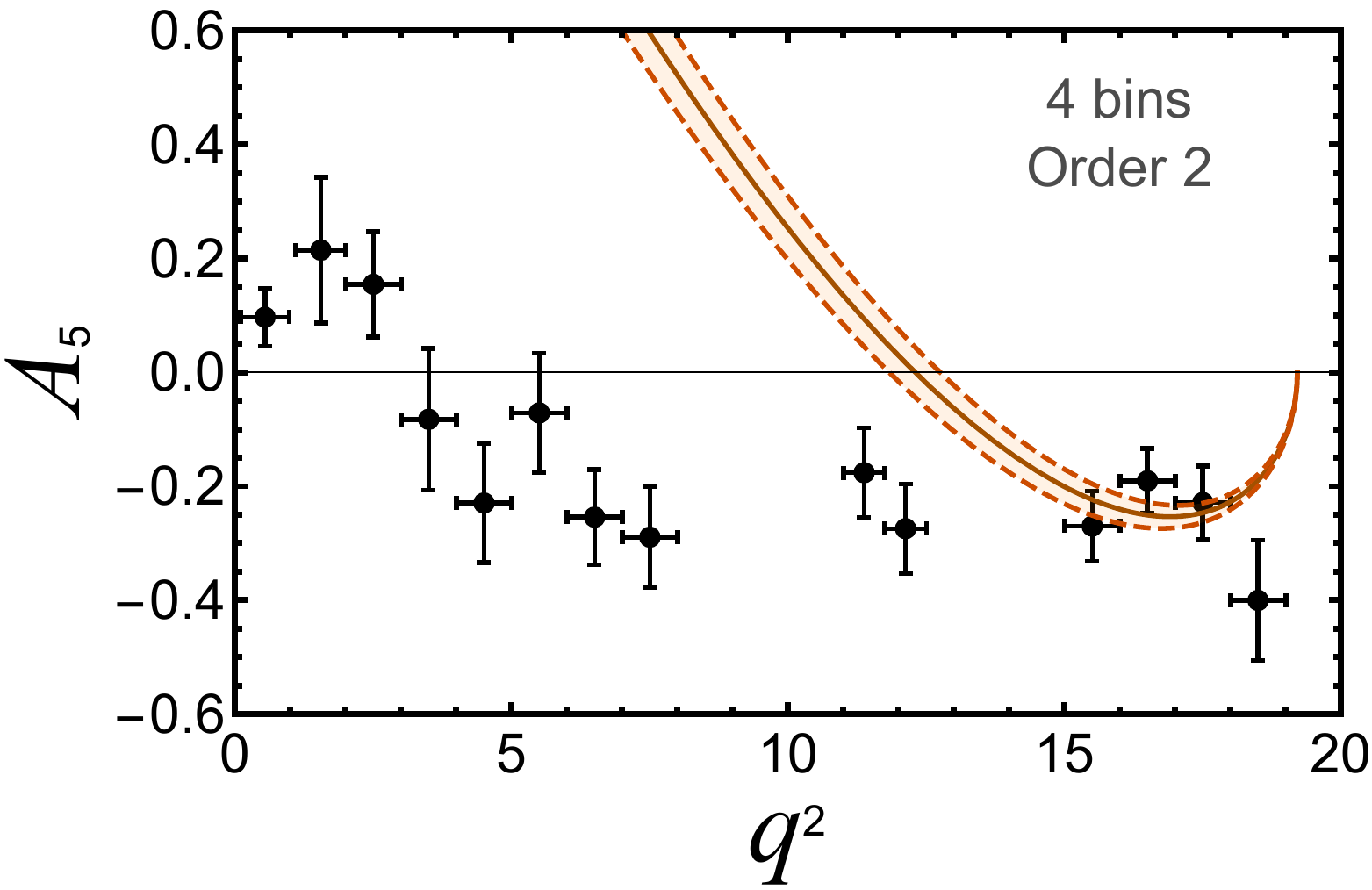}%
				\includegraphics*[width=1.5in]{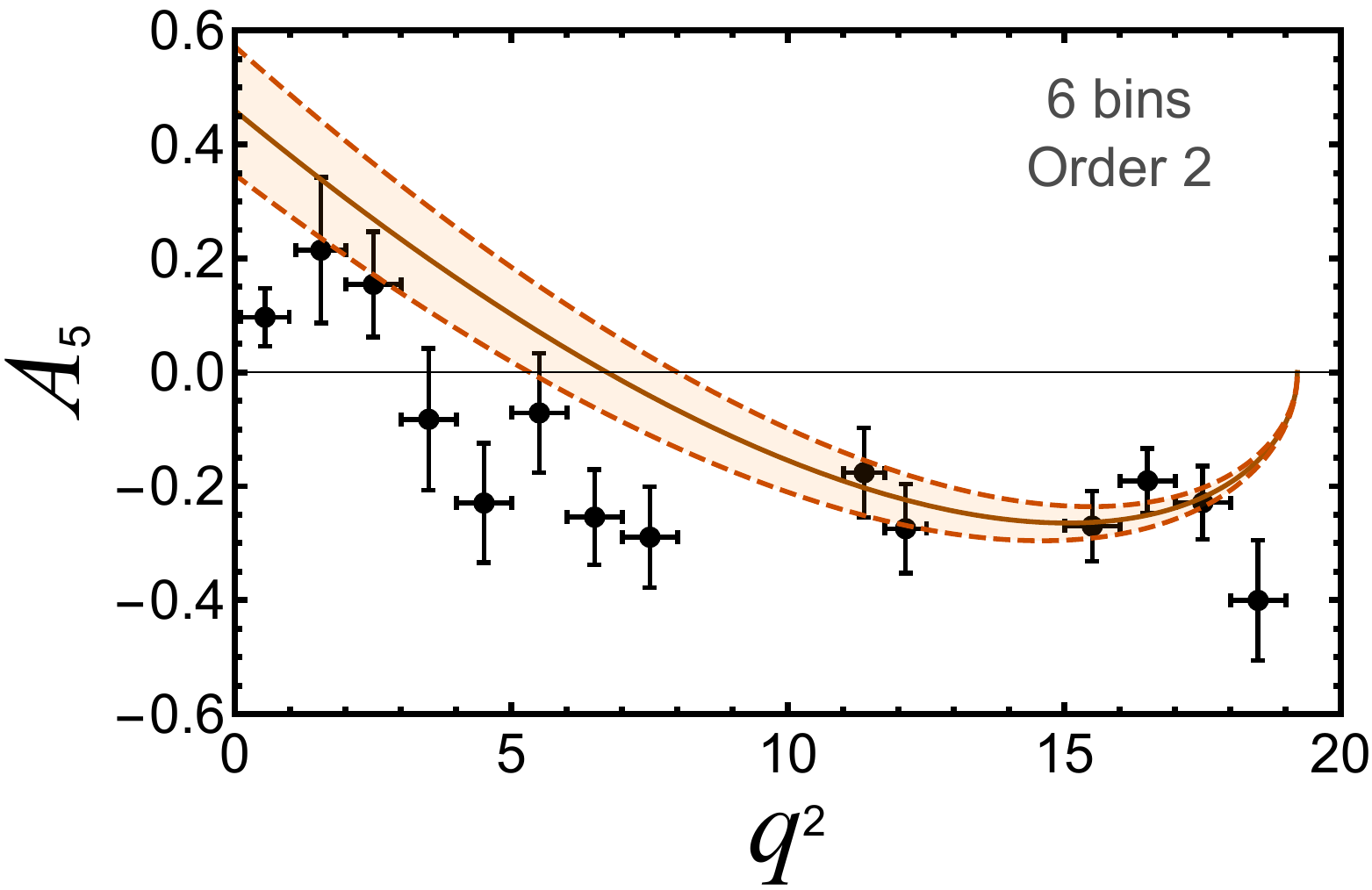}%
				\includegraphics*[width=1.5in]{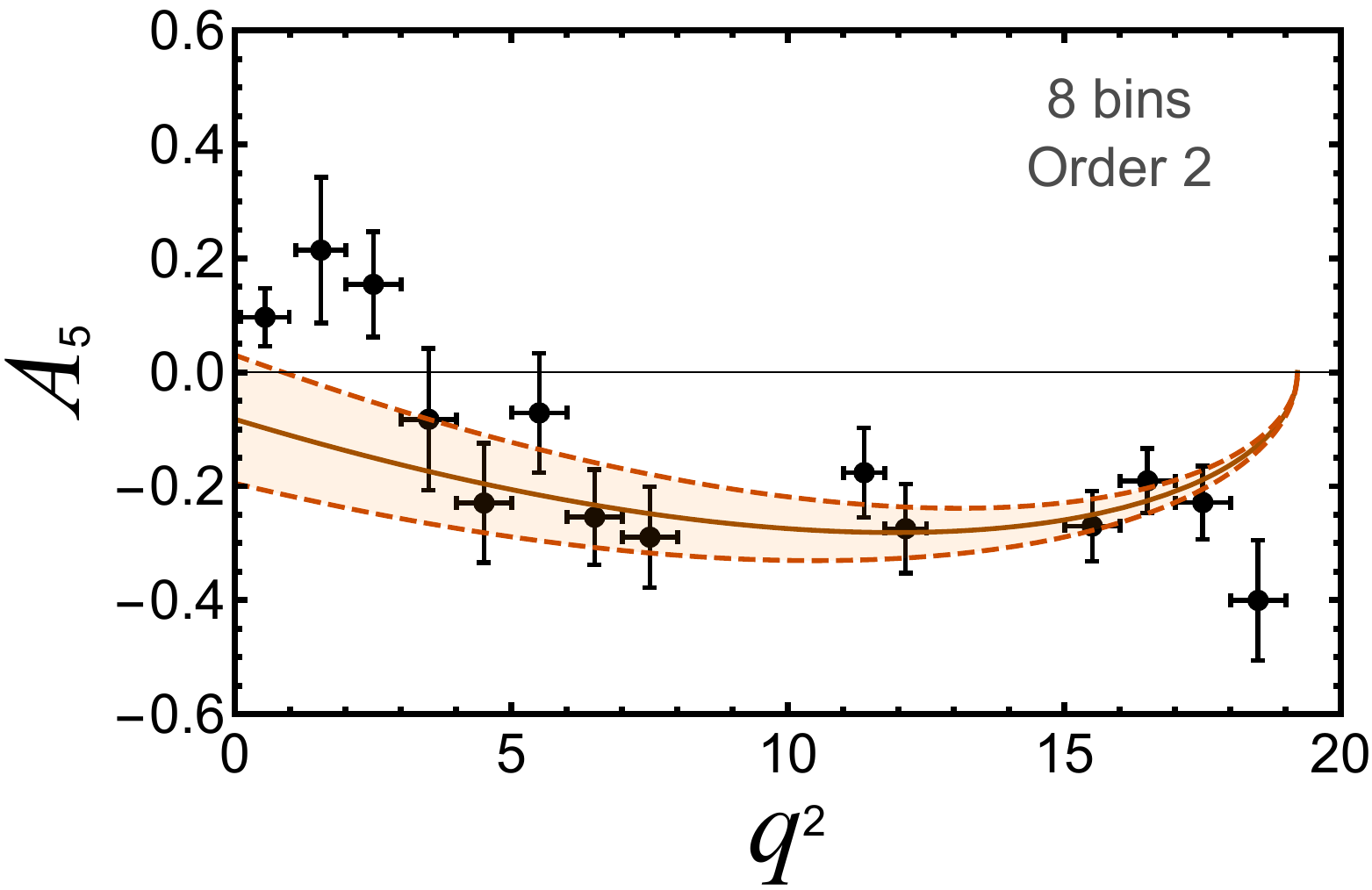}\\%
				\includegraphics*[width=1.5in]{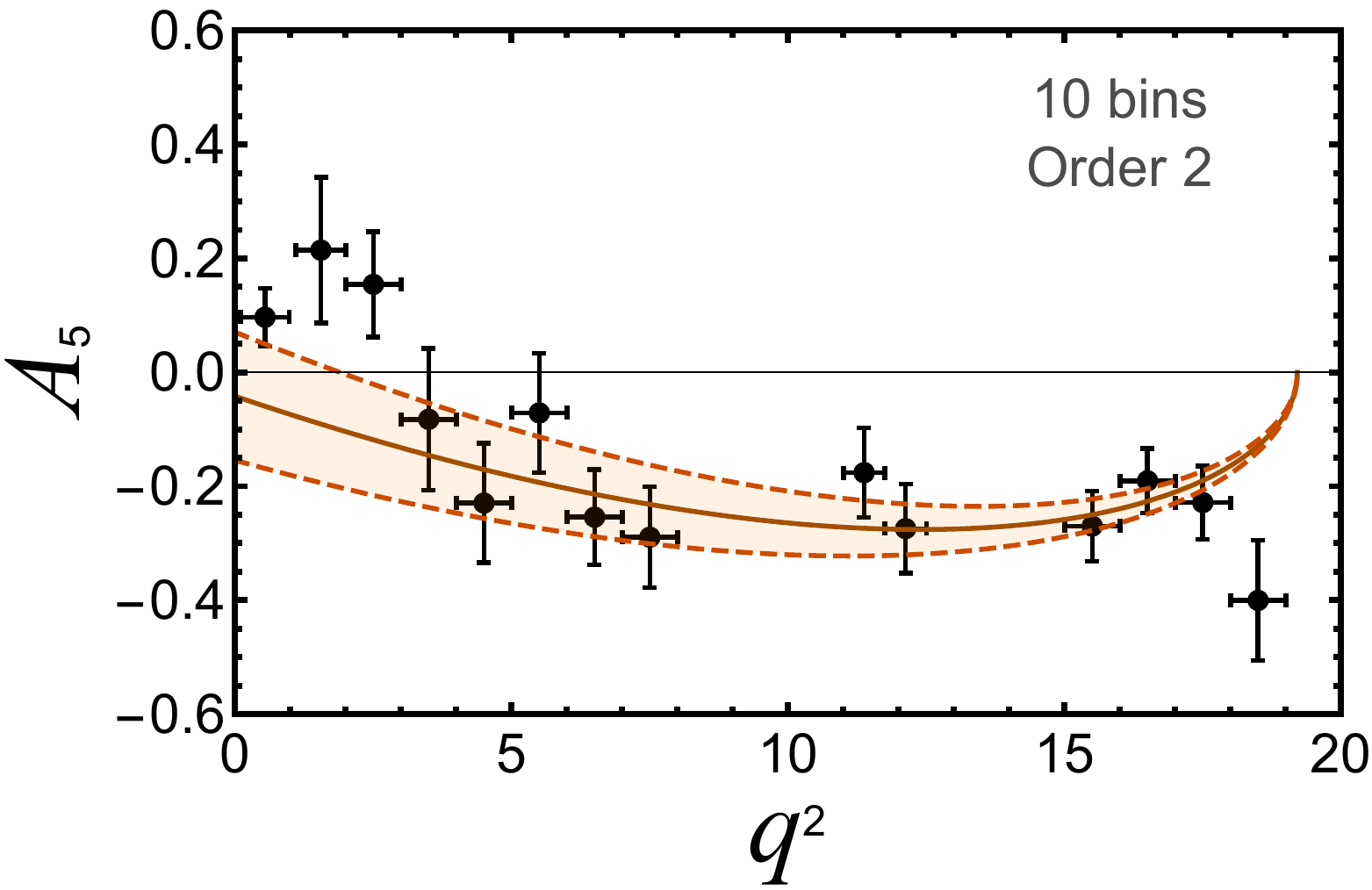}%
				\includegraphics*[width=1.5in]{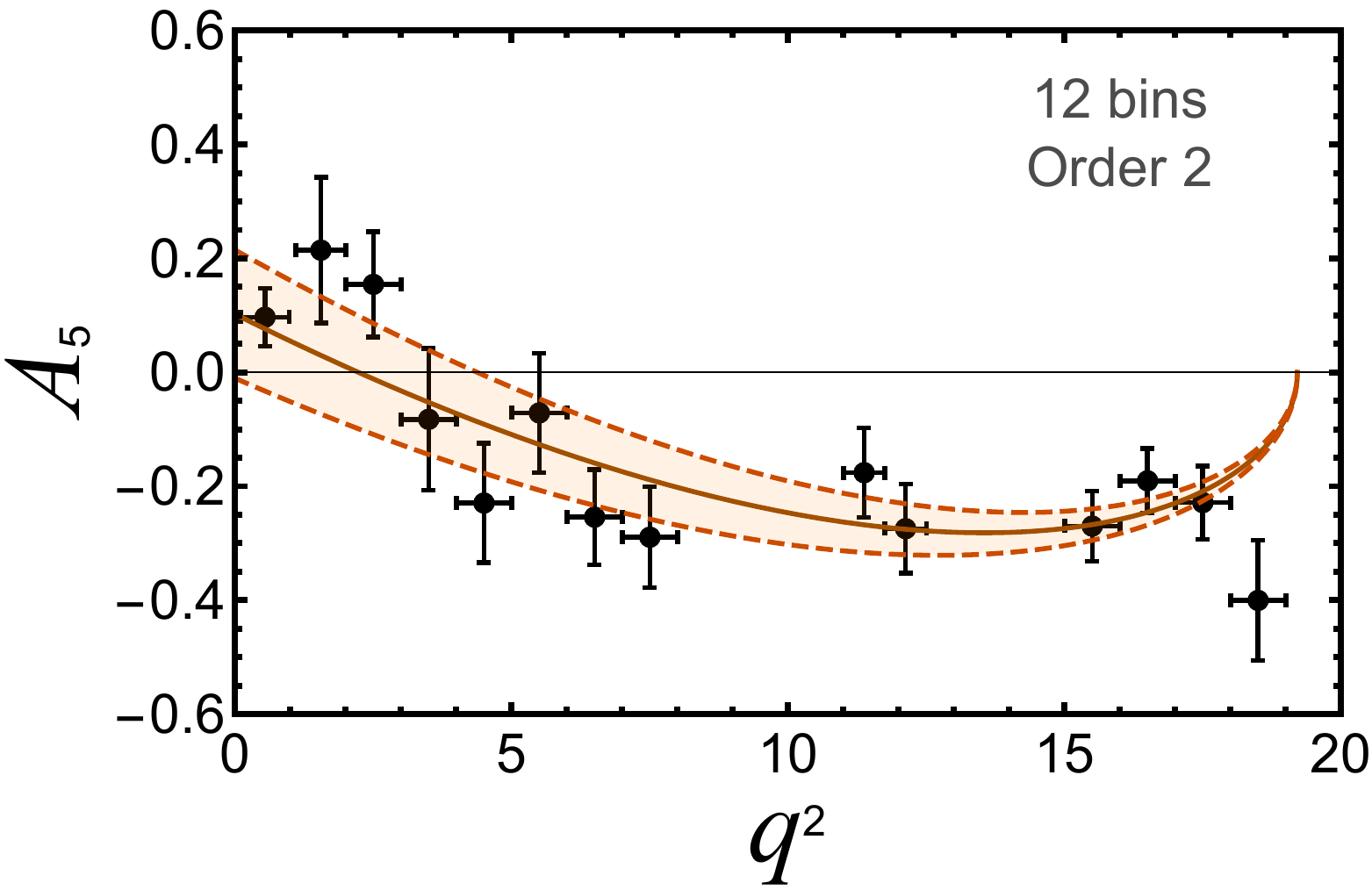}%
				\includegraphics*[width=1.5in]{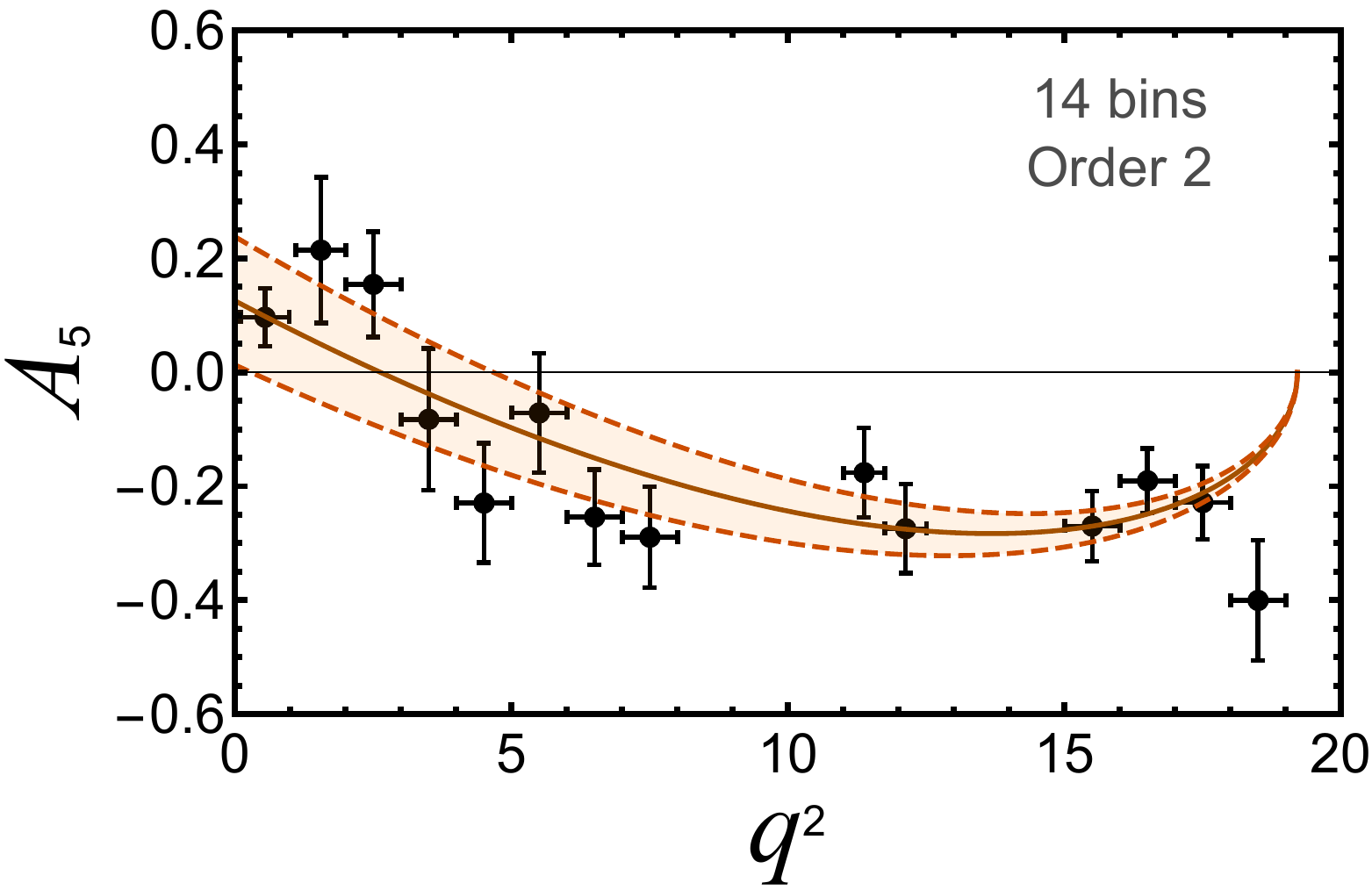}\\%
				\includegraphics*[width=1.5in]{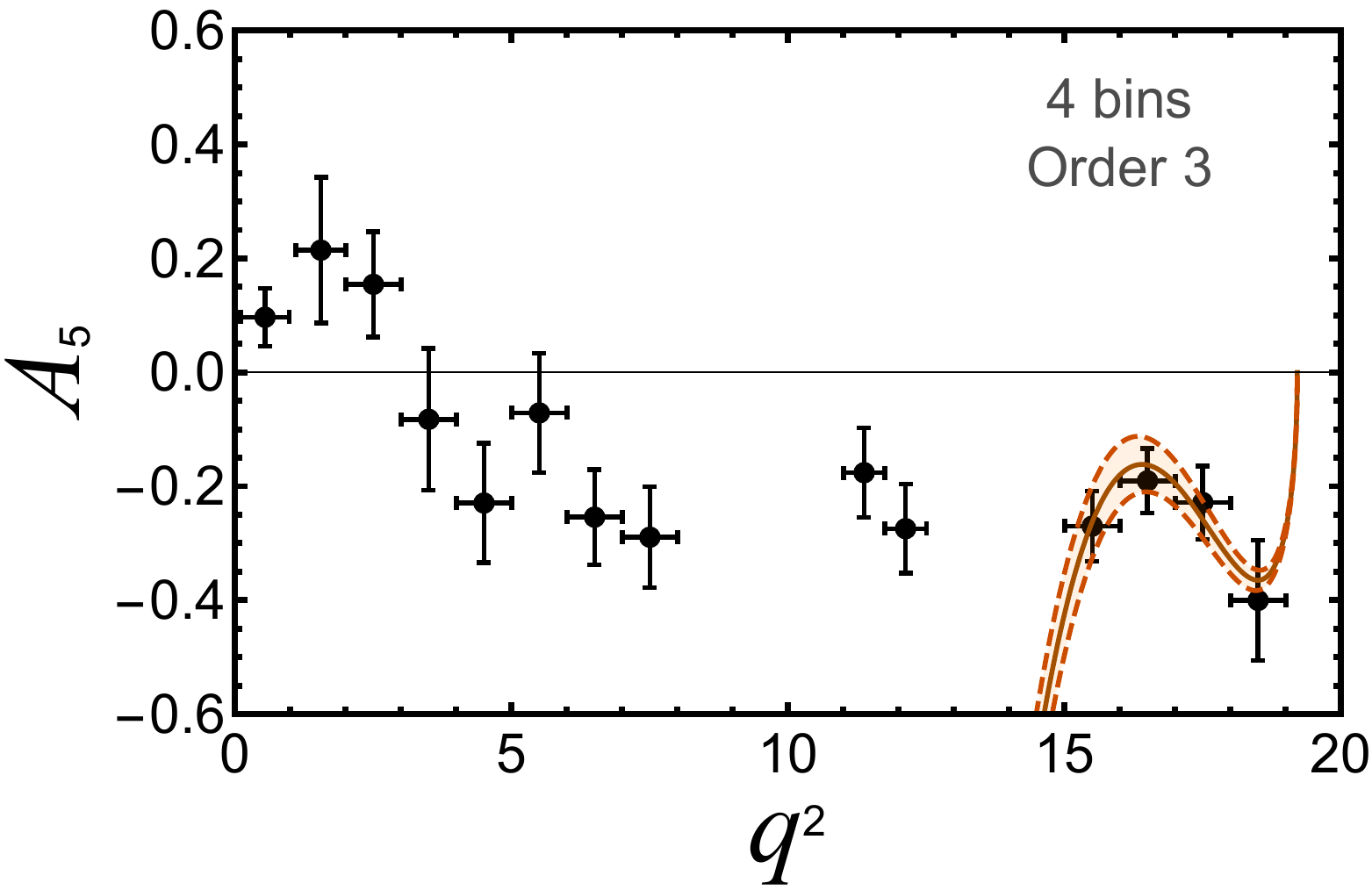}%
				\includegraphics*[width=1.5in]{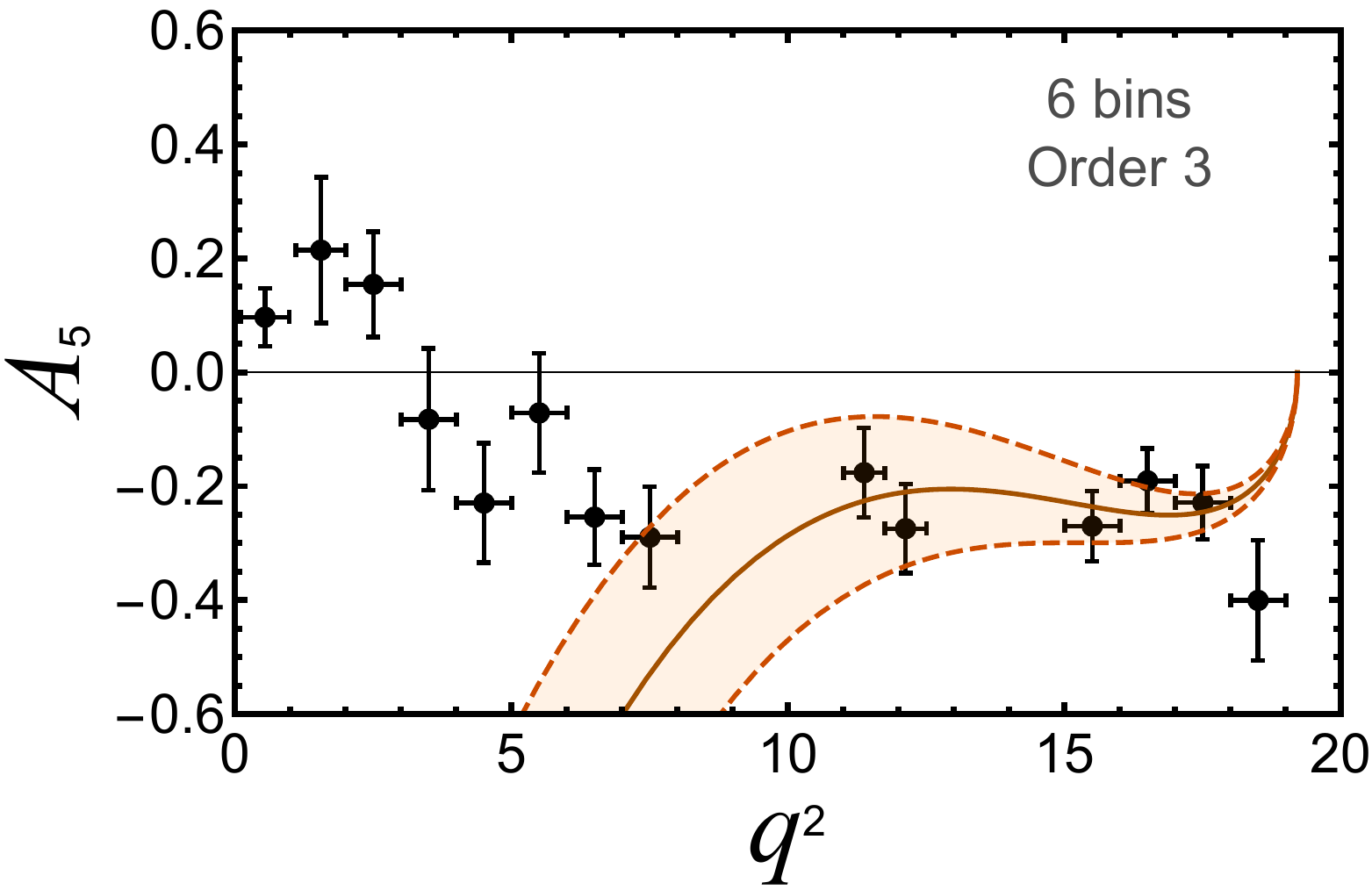}%
				\includegraphics*[width=1.5in]{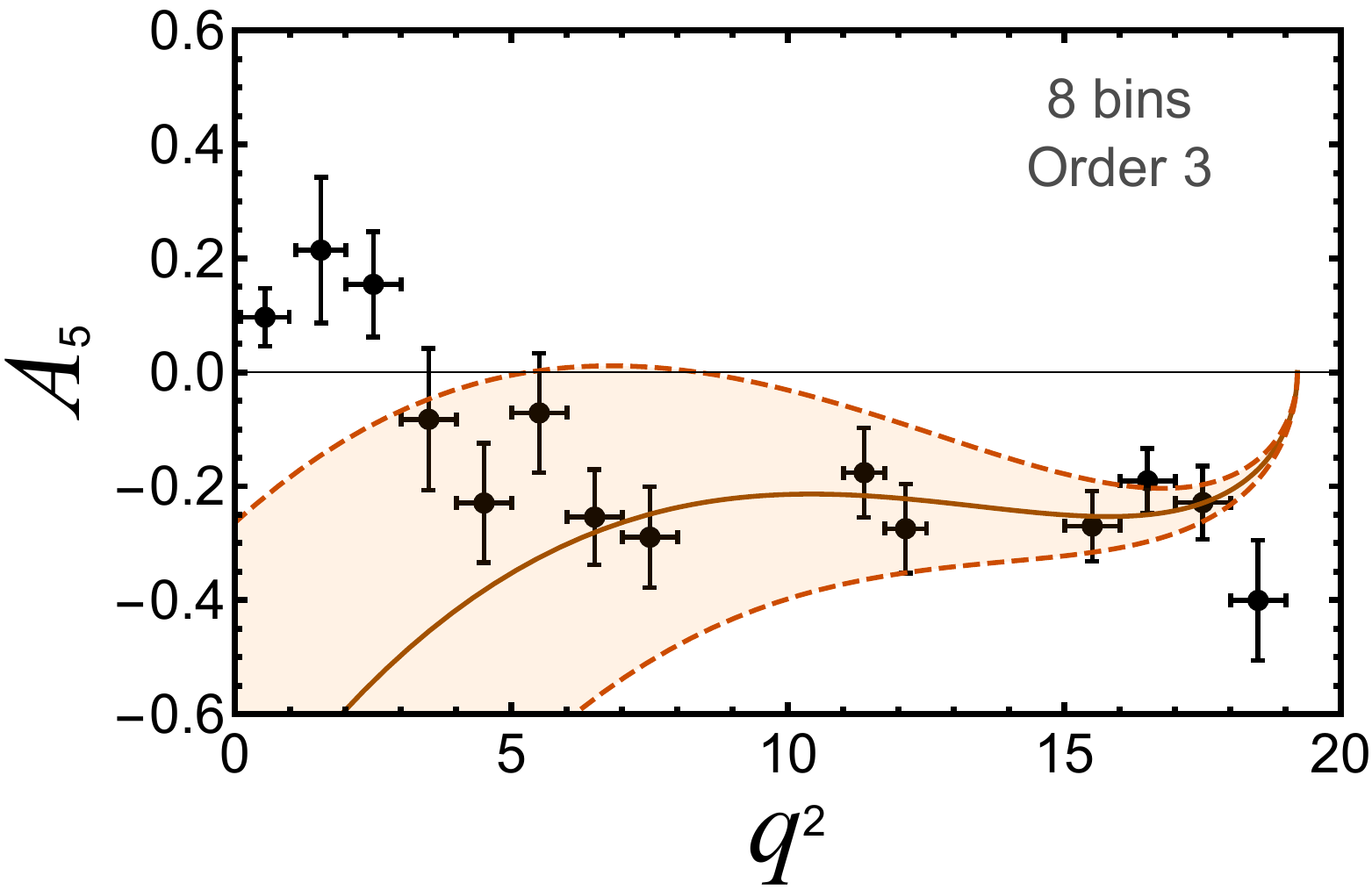}\\%
				\includegraphics*[width=1.5in]{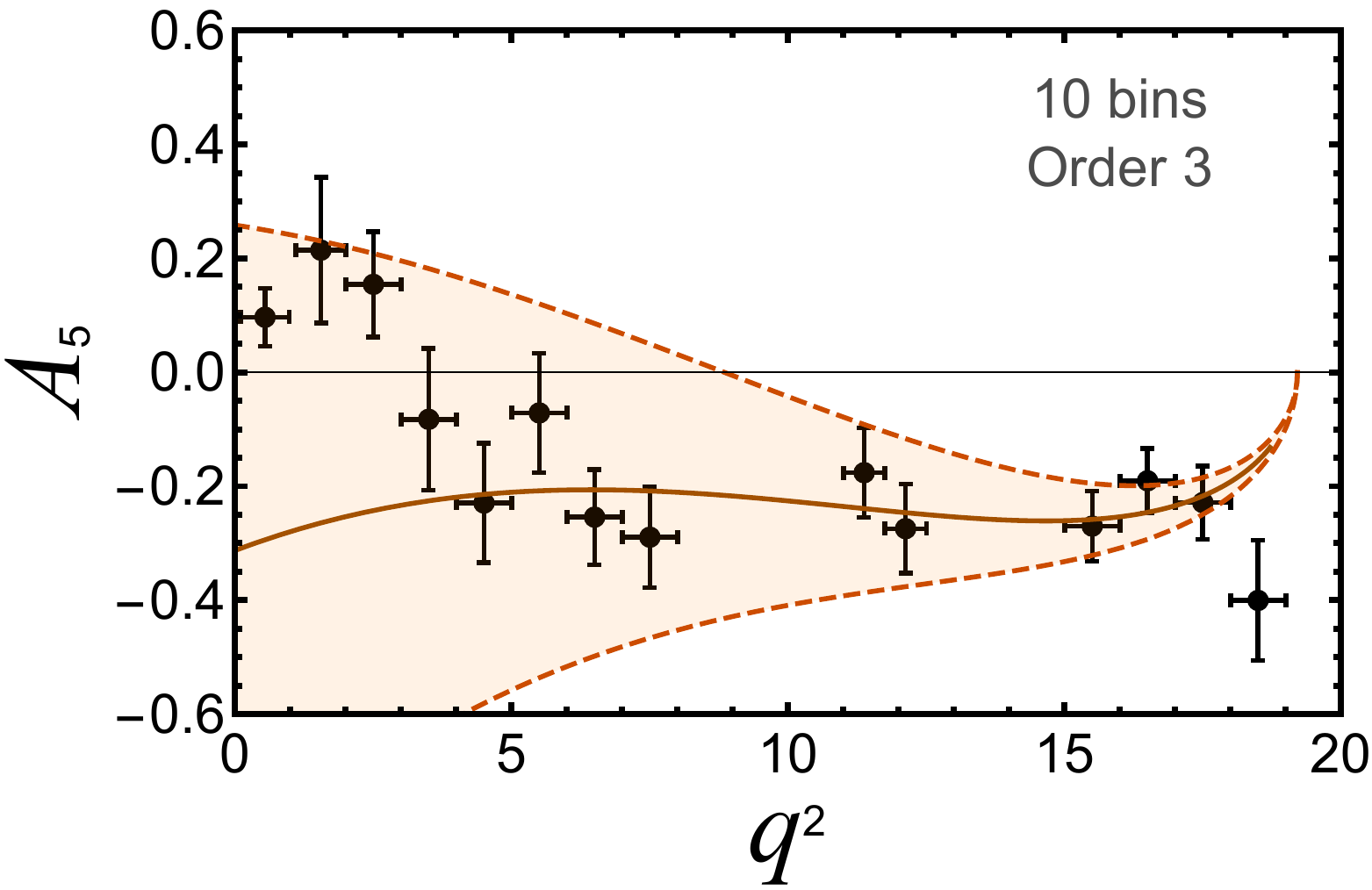}%
				\includegraphics*[width=1.5in]{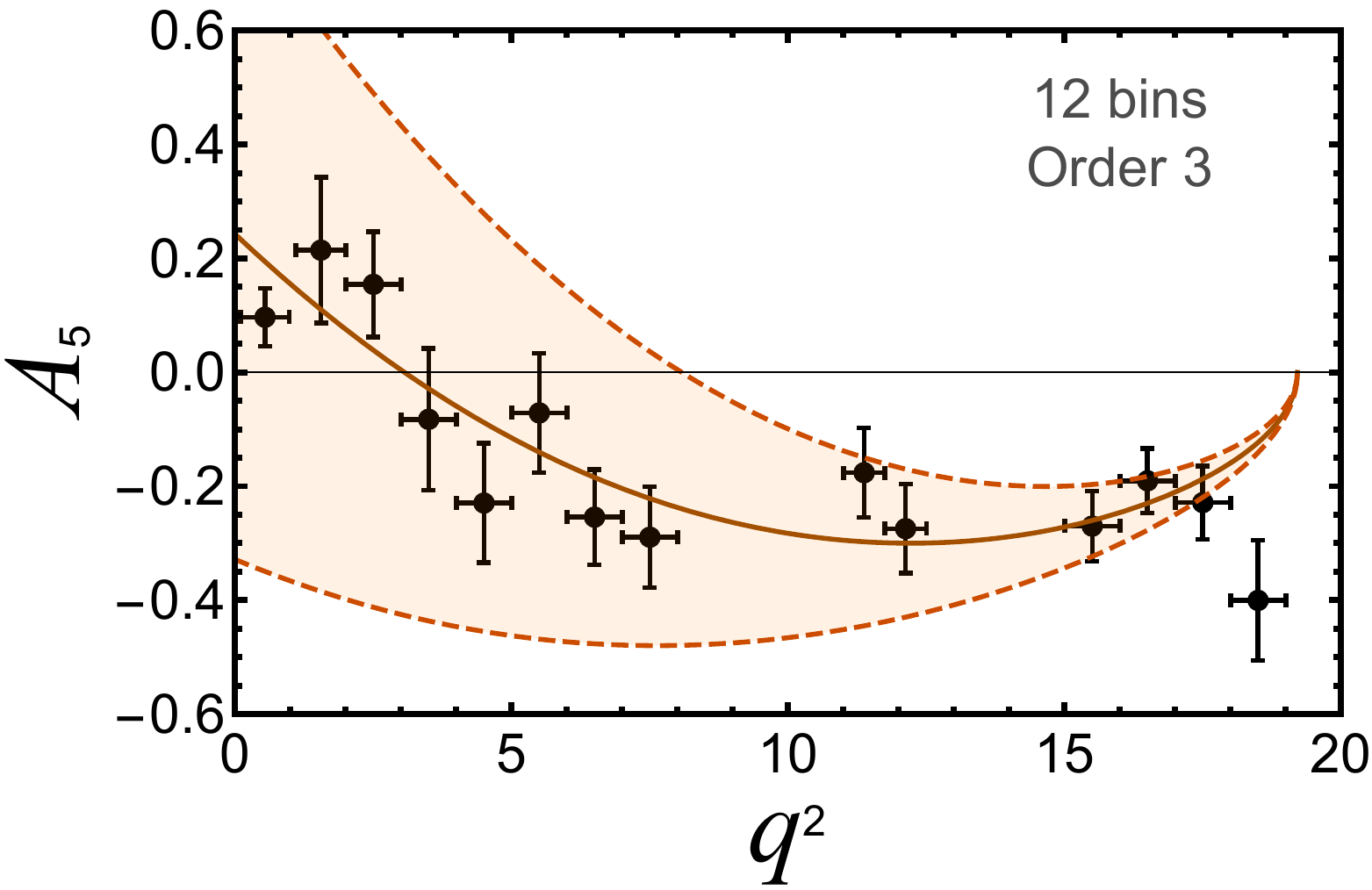}%
				\includegraphics*[width=1.5in]{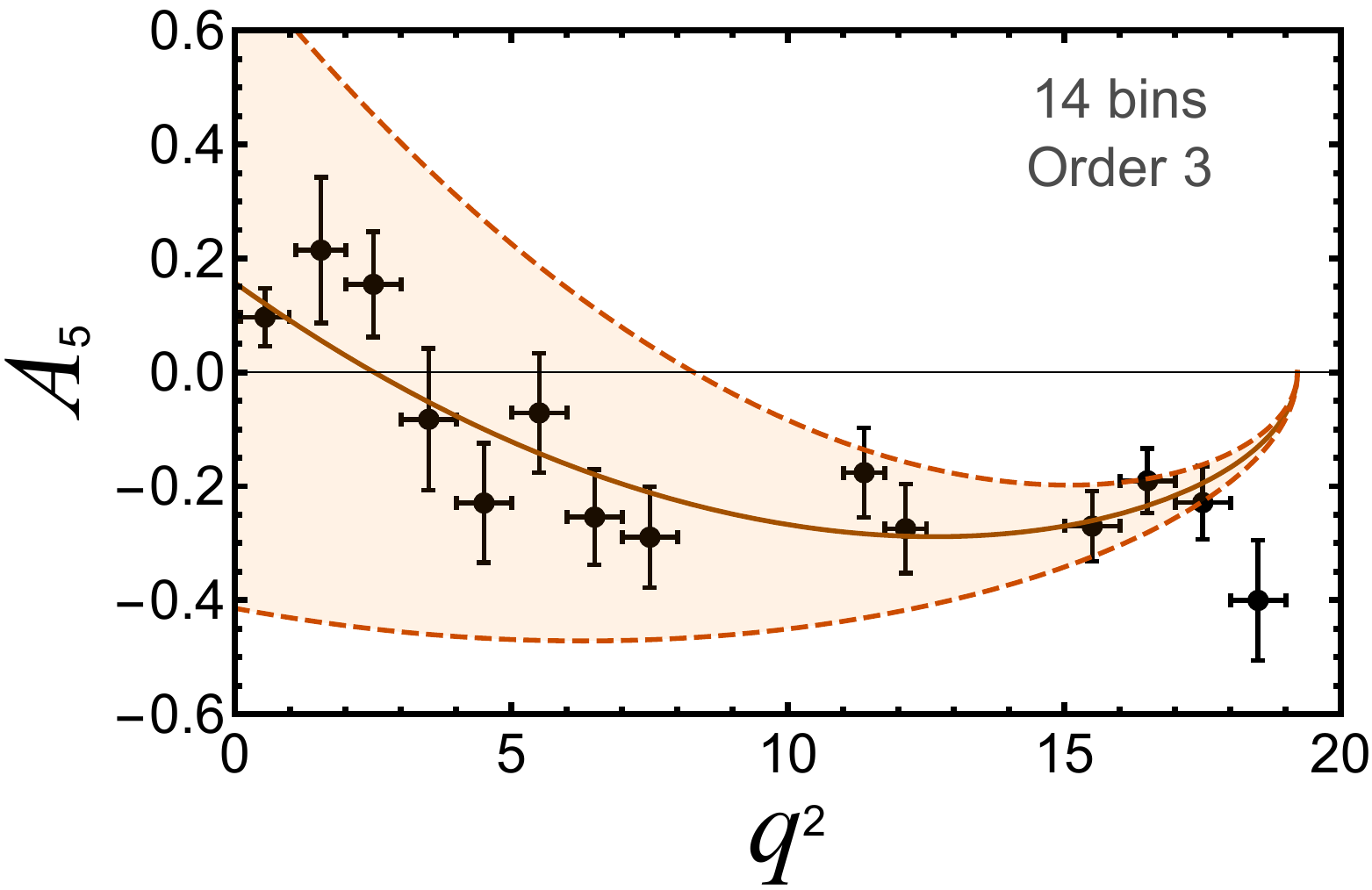}\\%
				\includegraphics*[width=1.5in]{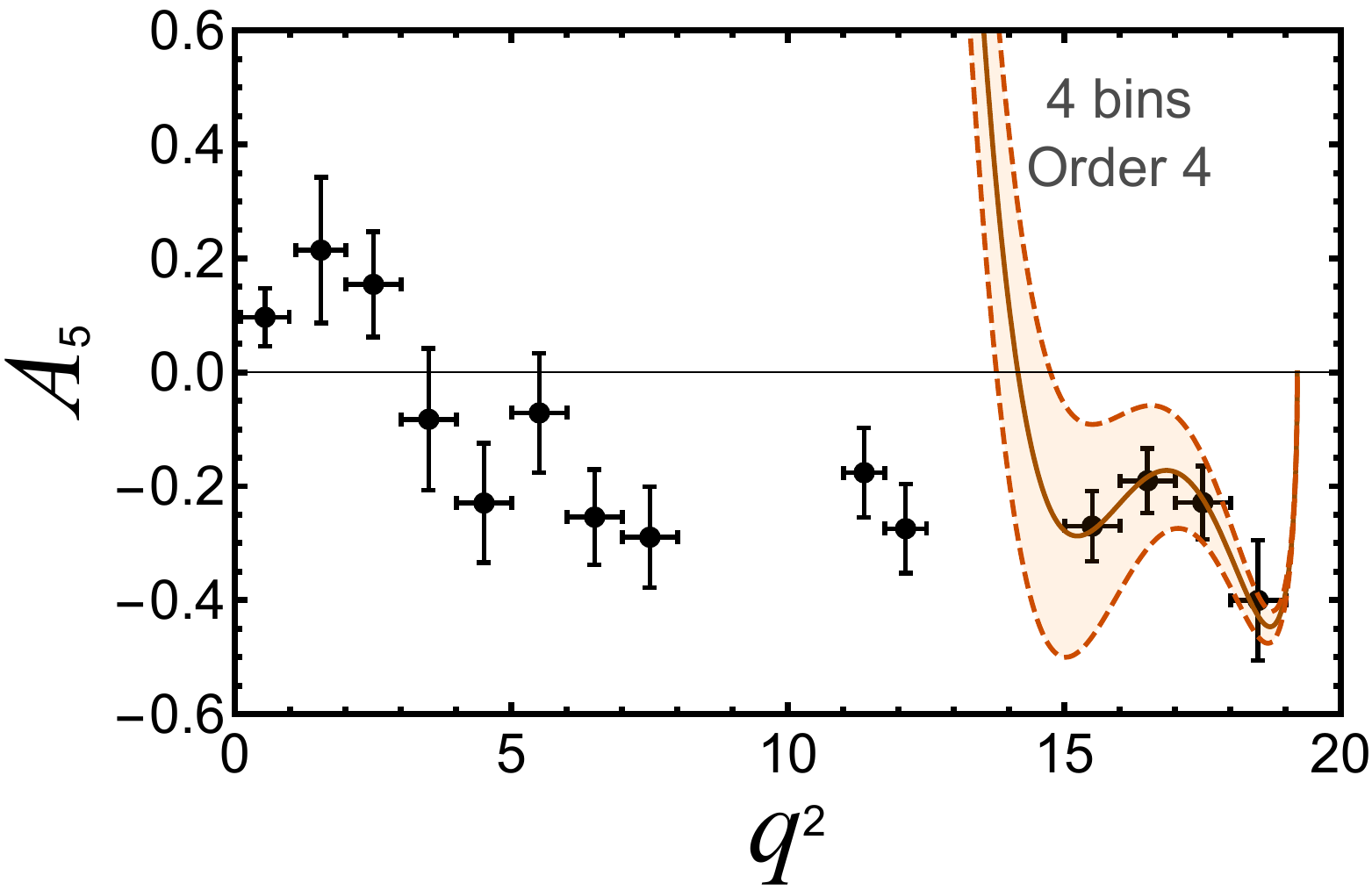}%
				\includegraphics*[width=1.5in]{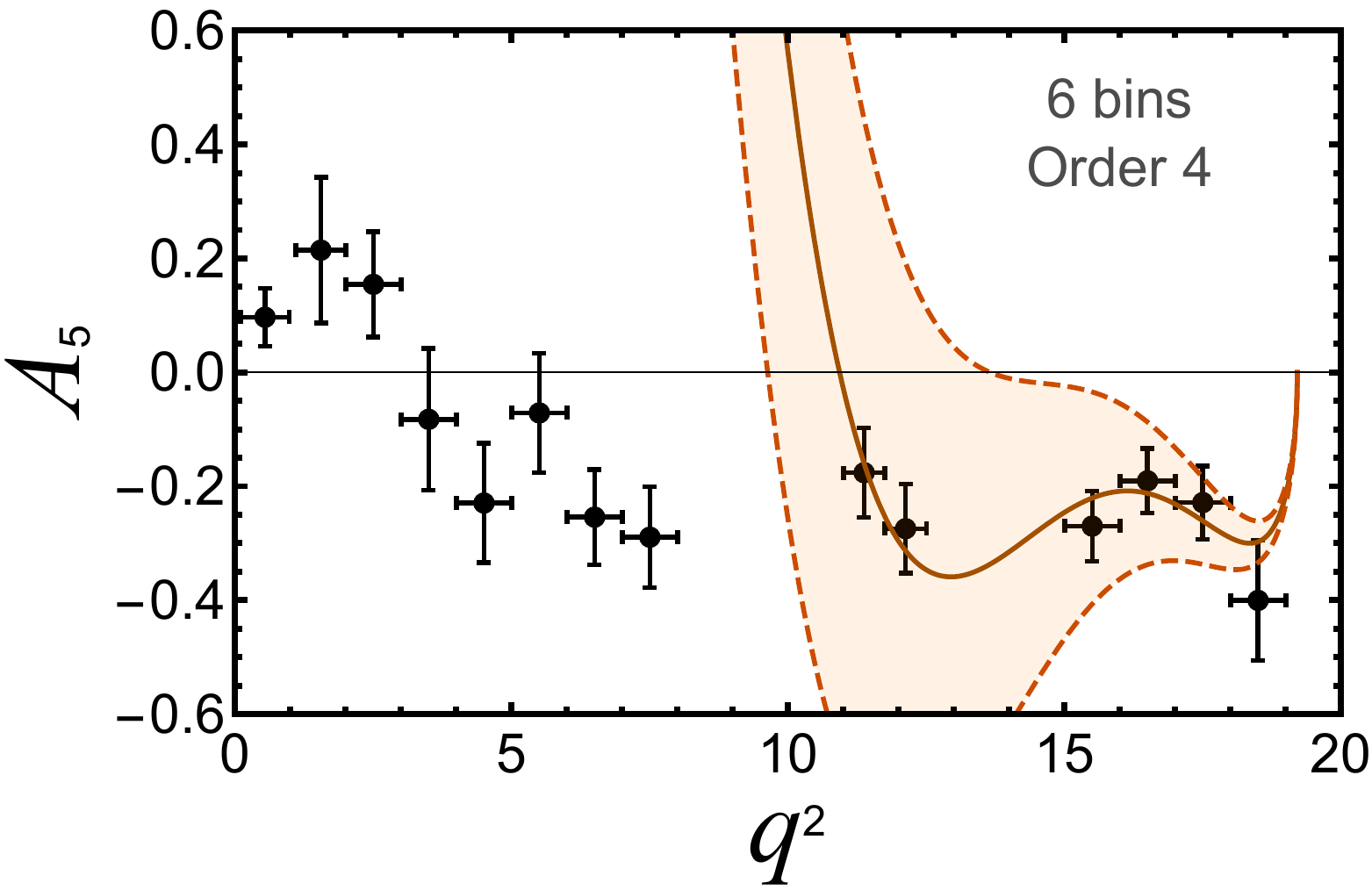}%
				\includegraphics*[width=1.5in]{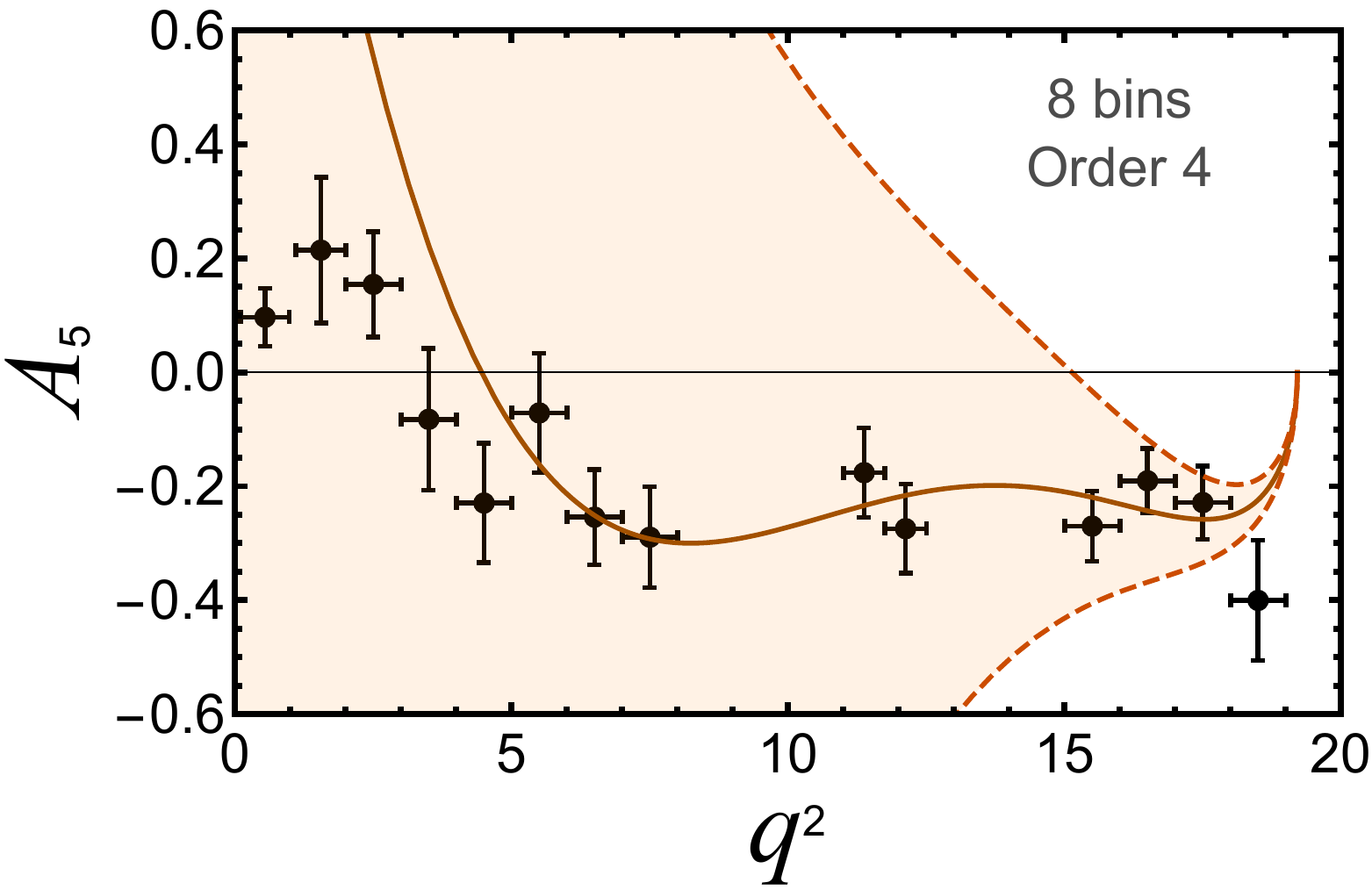}\\%
				\includegraphics*[width=1.5in]{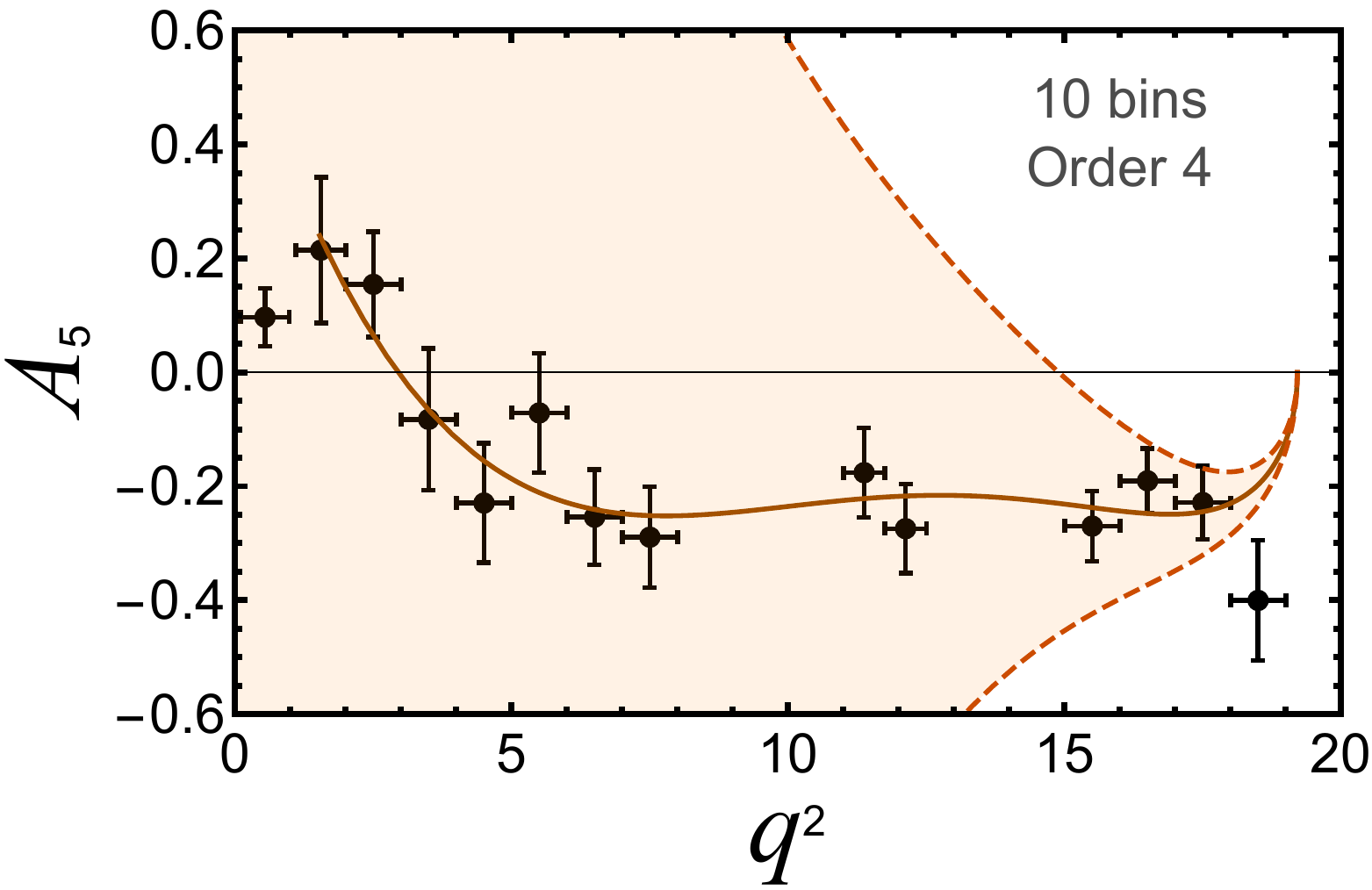}%
				\includegraphics*[width=1.5in]{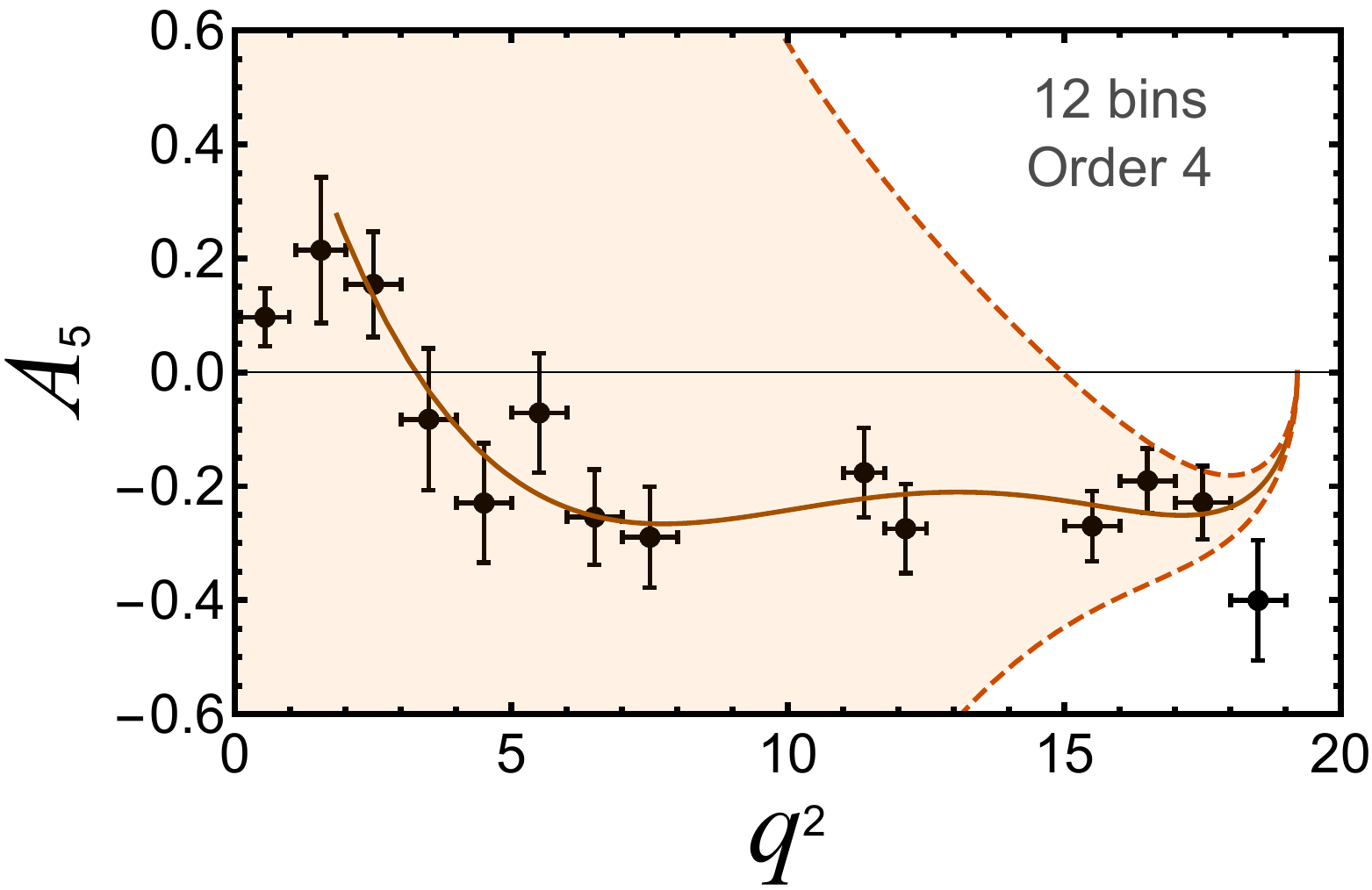}%
				\includegraphics*[width=1.5in]{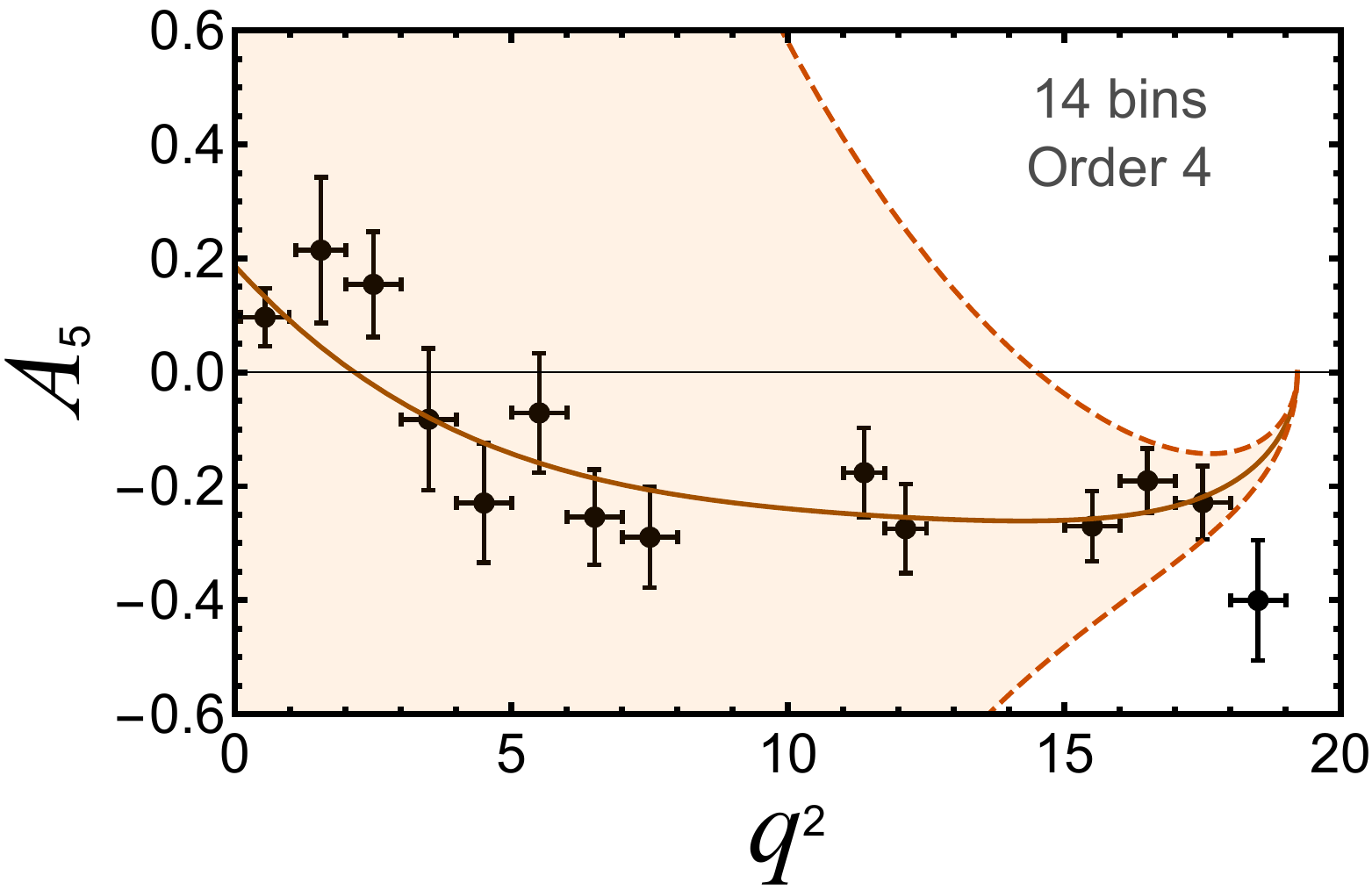}\\%
				\caption{Fits to $A_5$ various numbers of bins and polynomial parameterizations.  The color code is the same as in Fig.~\ref{fig:1}} 
				\label{fig:0b}
			\end{center}
		\end{figure}
	\end{center}

	\begin{center}
		\begin{figure}[th]
			\begin{center}
				\includegraphics*[width=1.5in]{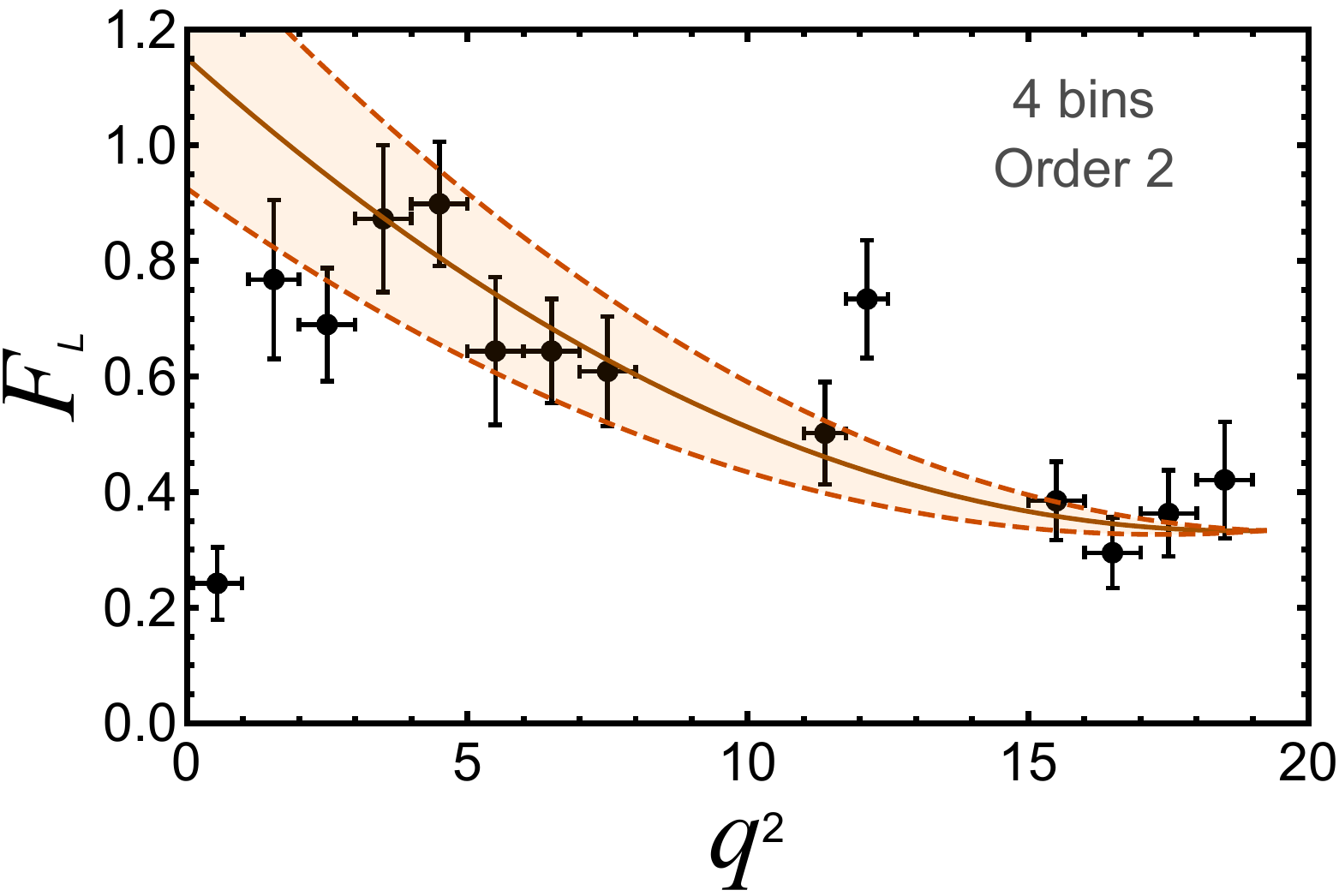}%
				\includegraphics*[width=1.5in]{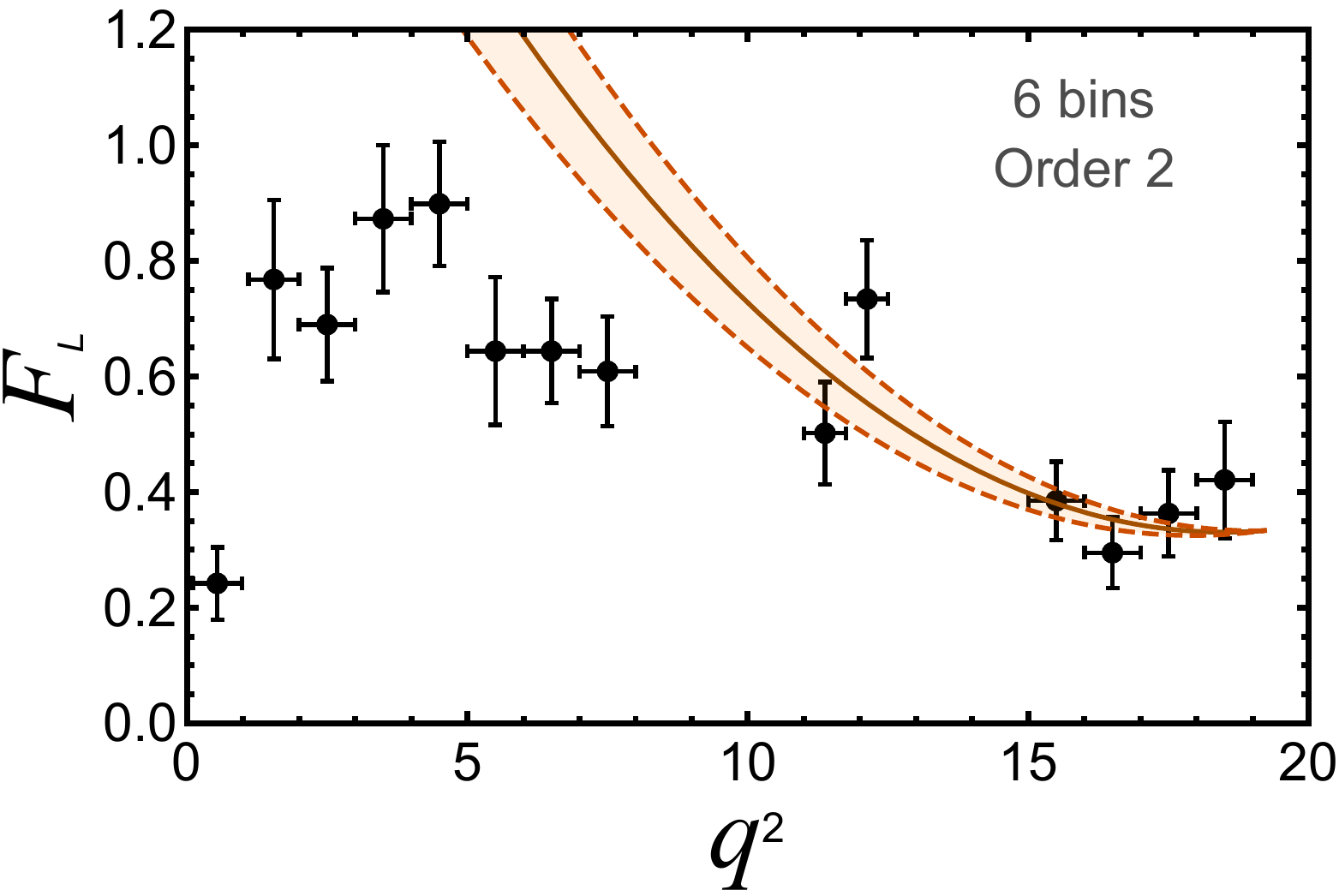}%
				\includegraphics*[width=1.5in]{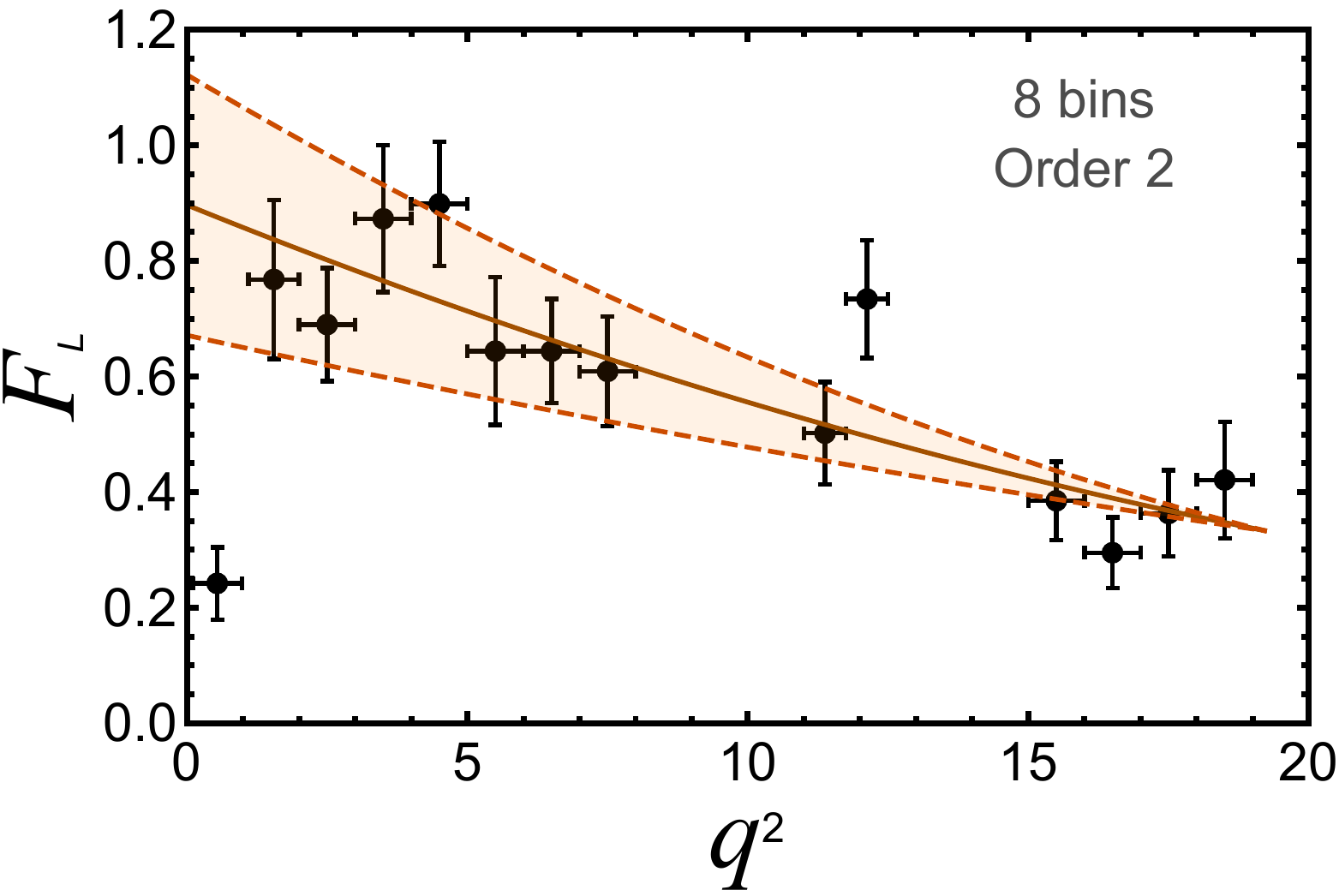}\\%
				\includegraphics*[width=1.5in]{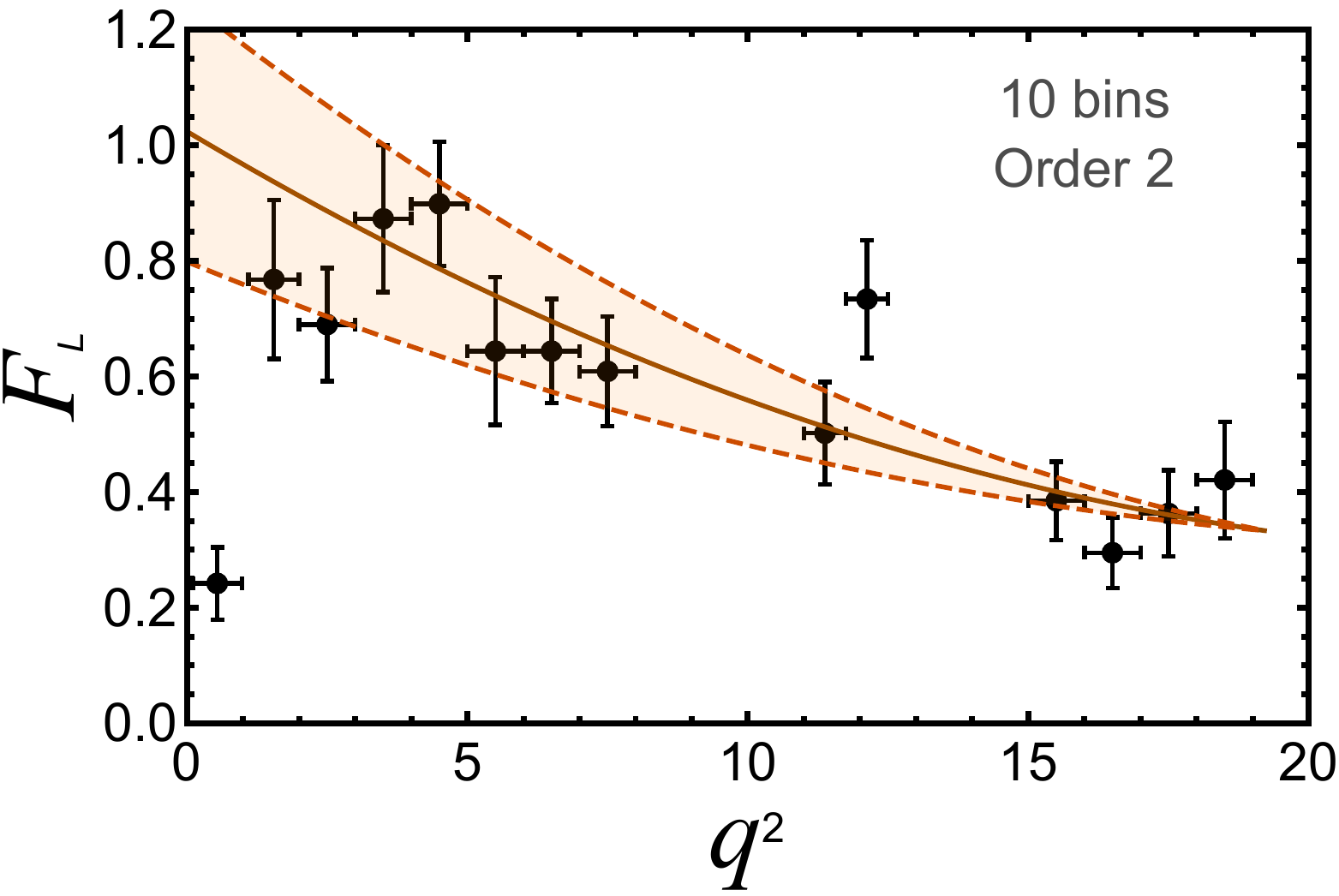}%
				\includegraphics*[width=1.5in]{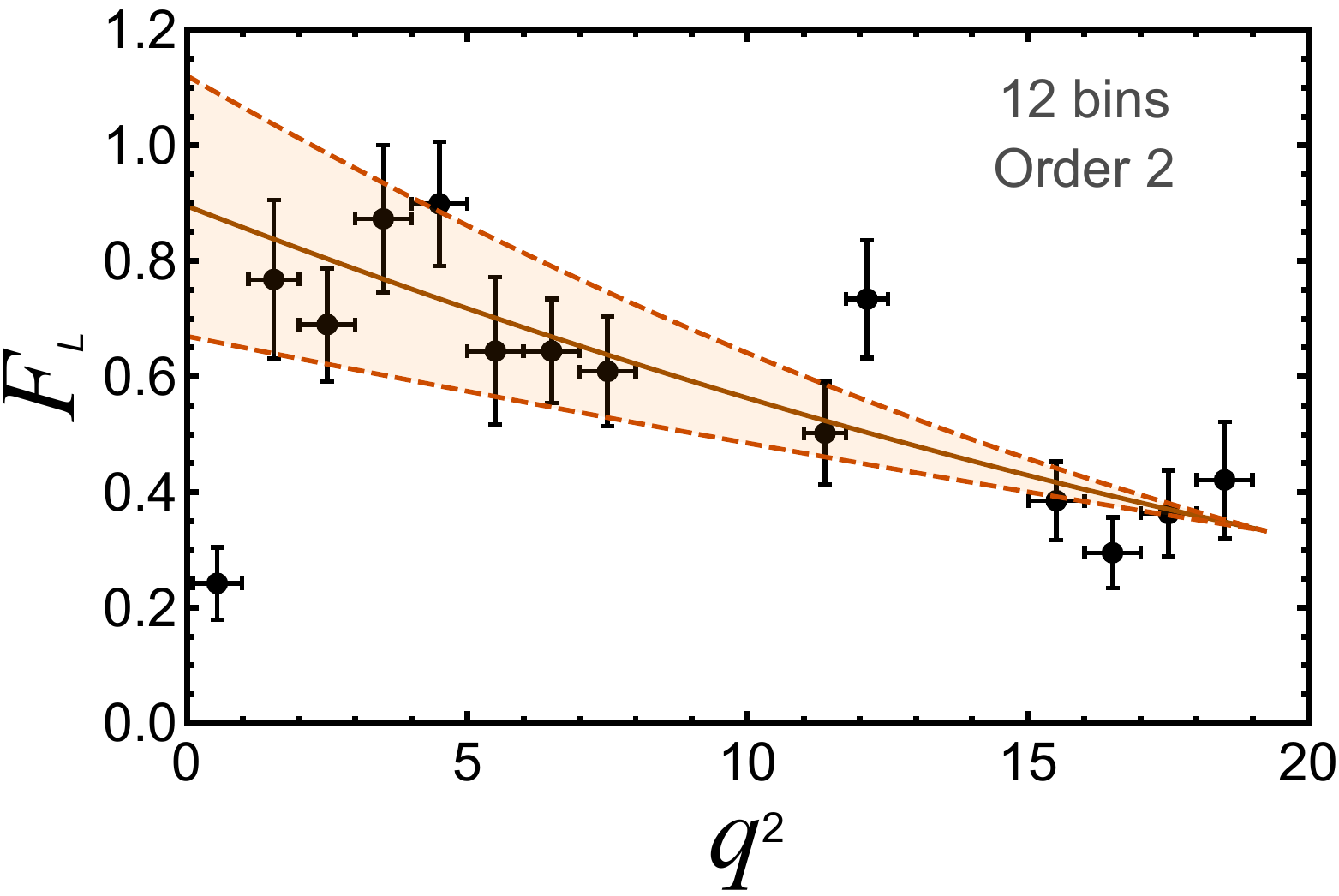}%
				\includegraphics*[width=1.5in]{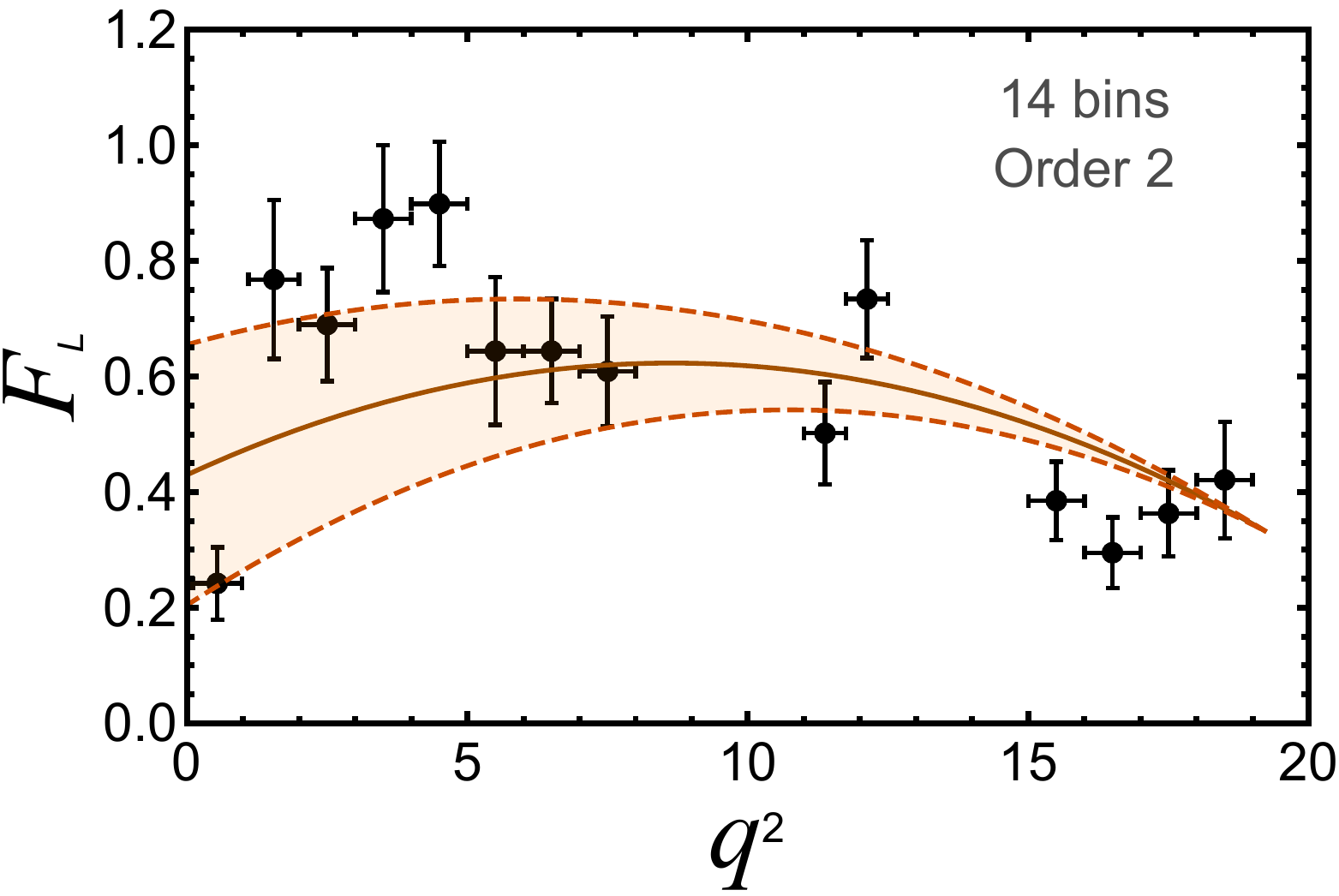}\\%
				\includegraphics*[width=1.5in]{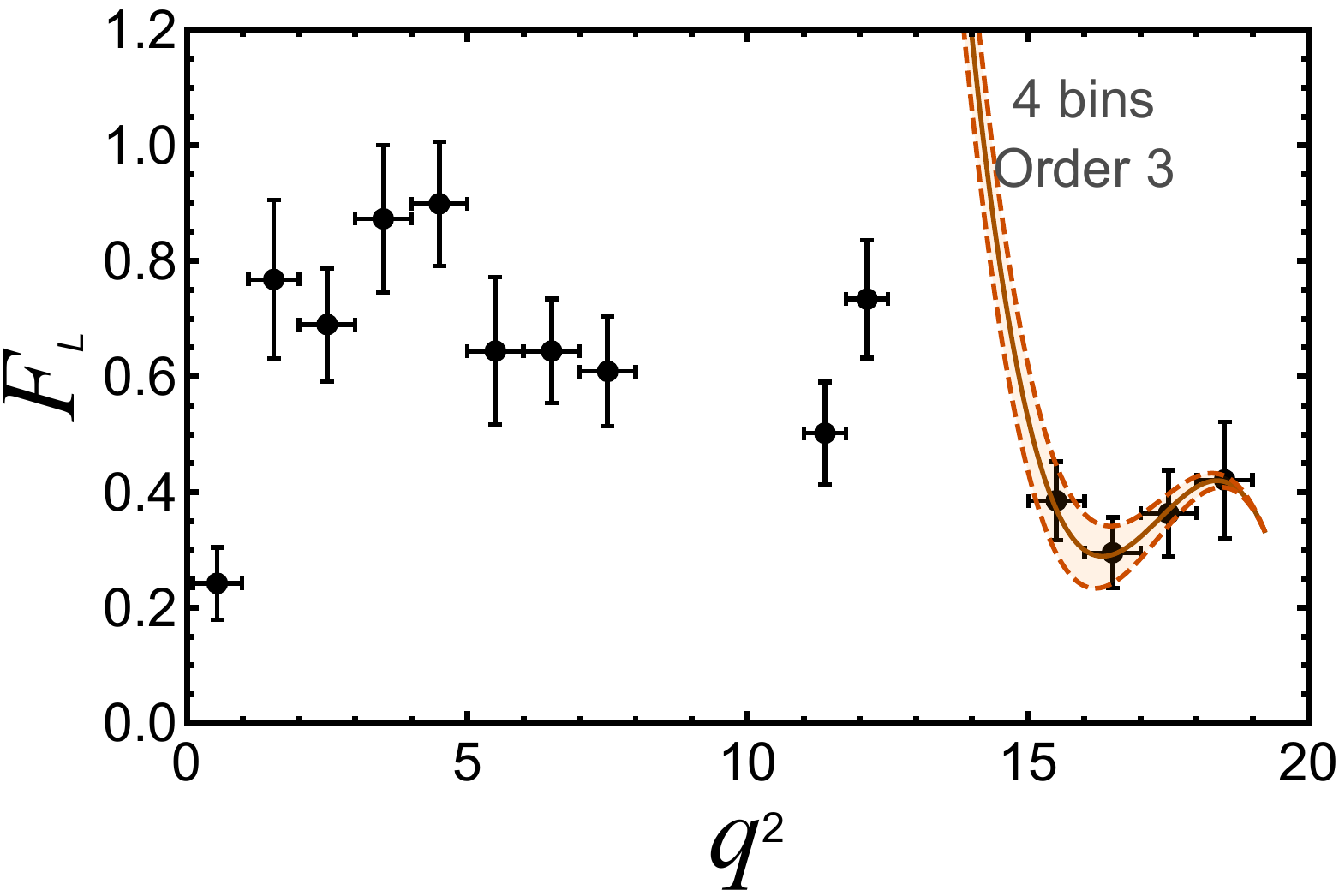}%
				\includegraphics*[width=1.5in]{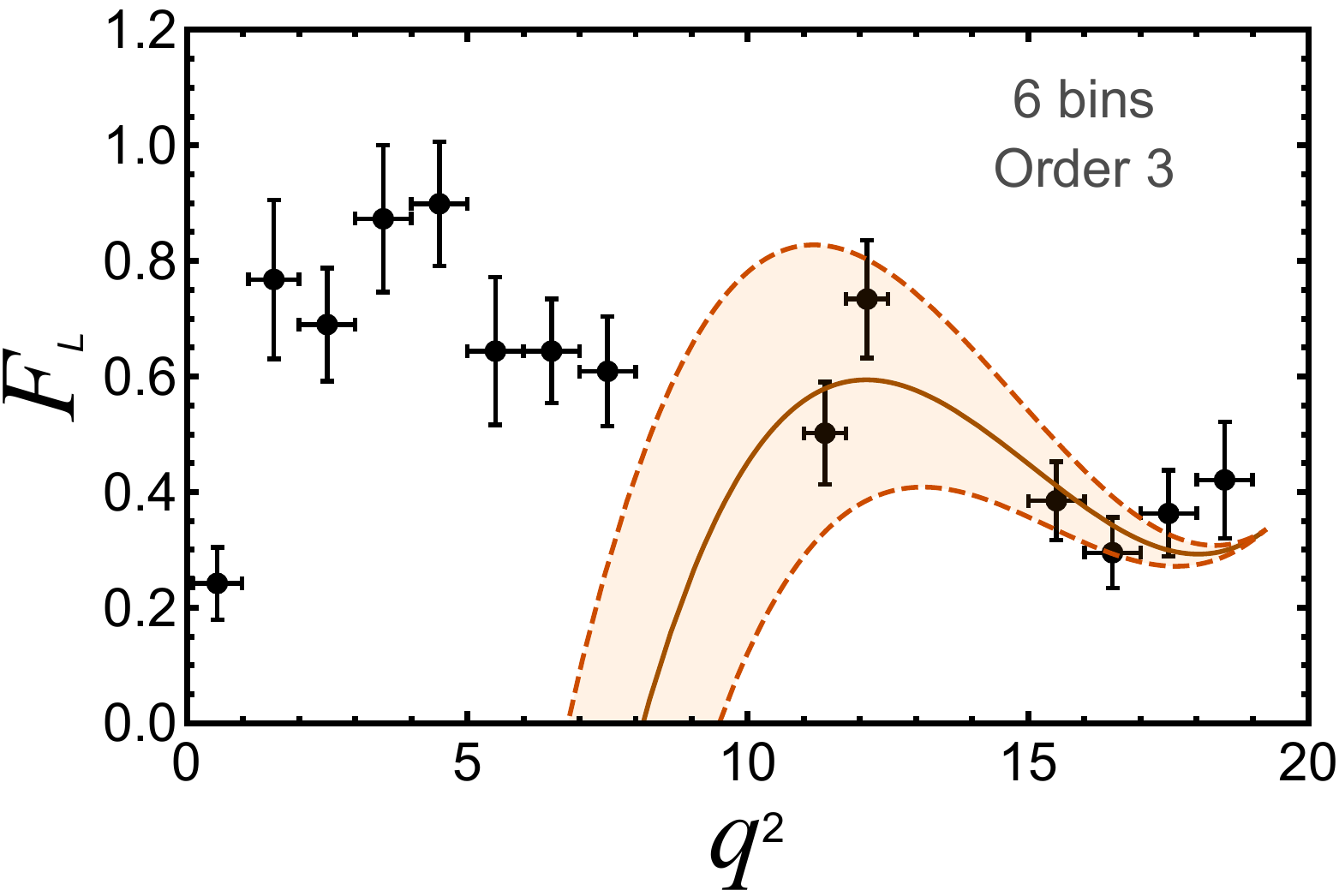}%
				\includegraphics*[width=1.5in]{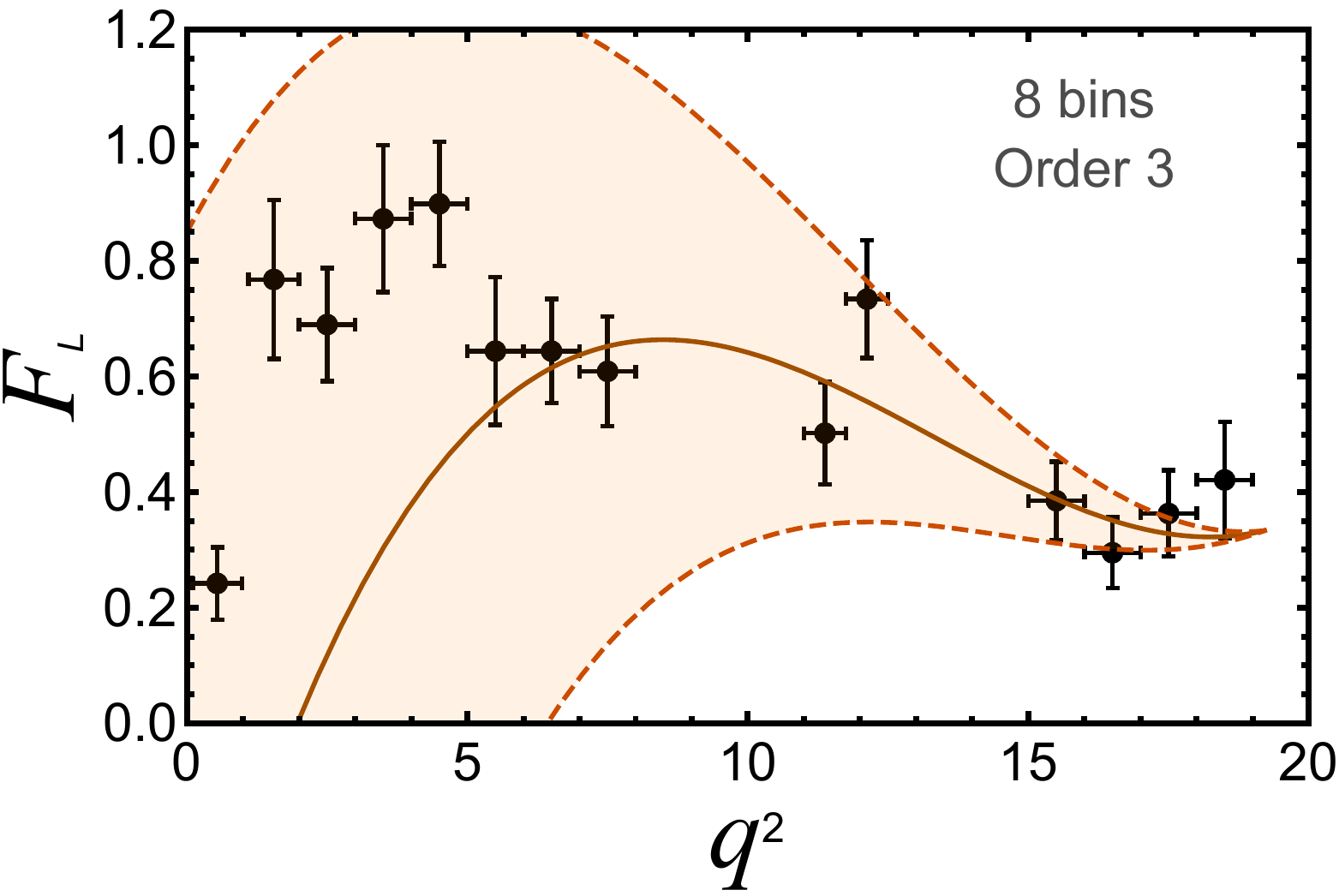}\\%
				\includegraphics*[width=1.5in]{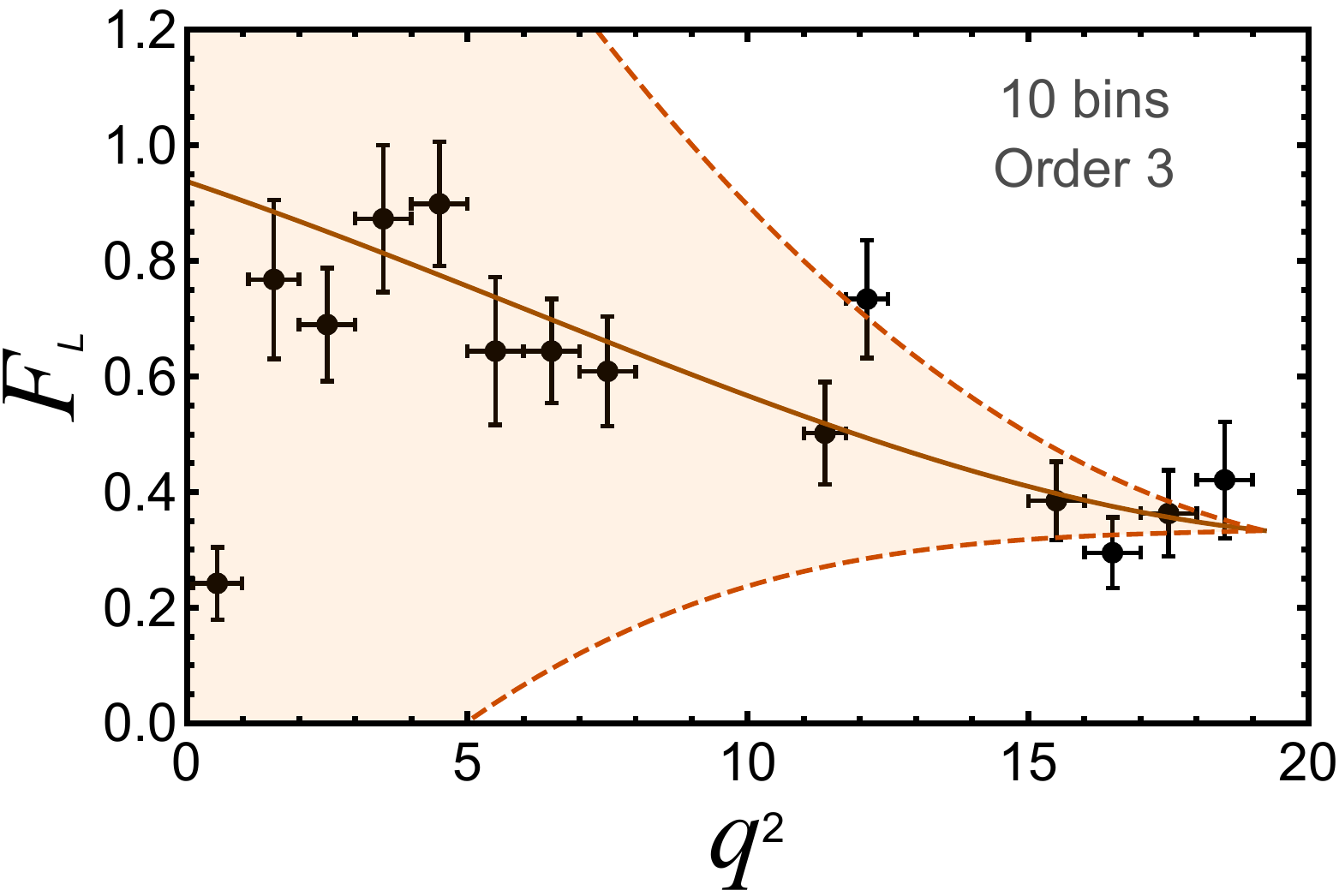}%
				\includegraphics*[width=1.5in]{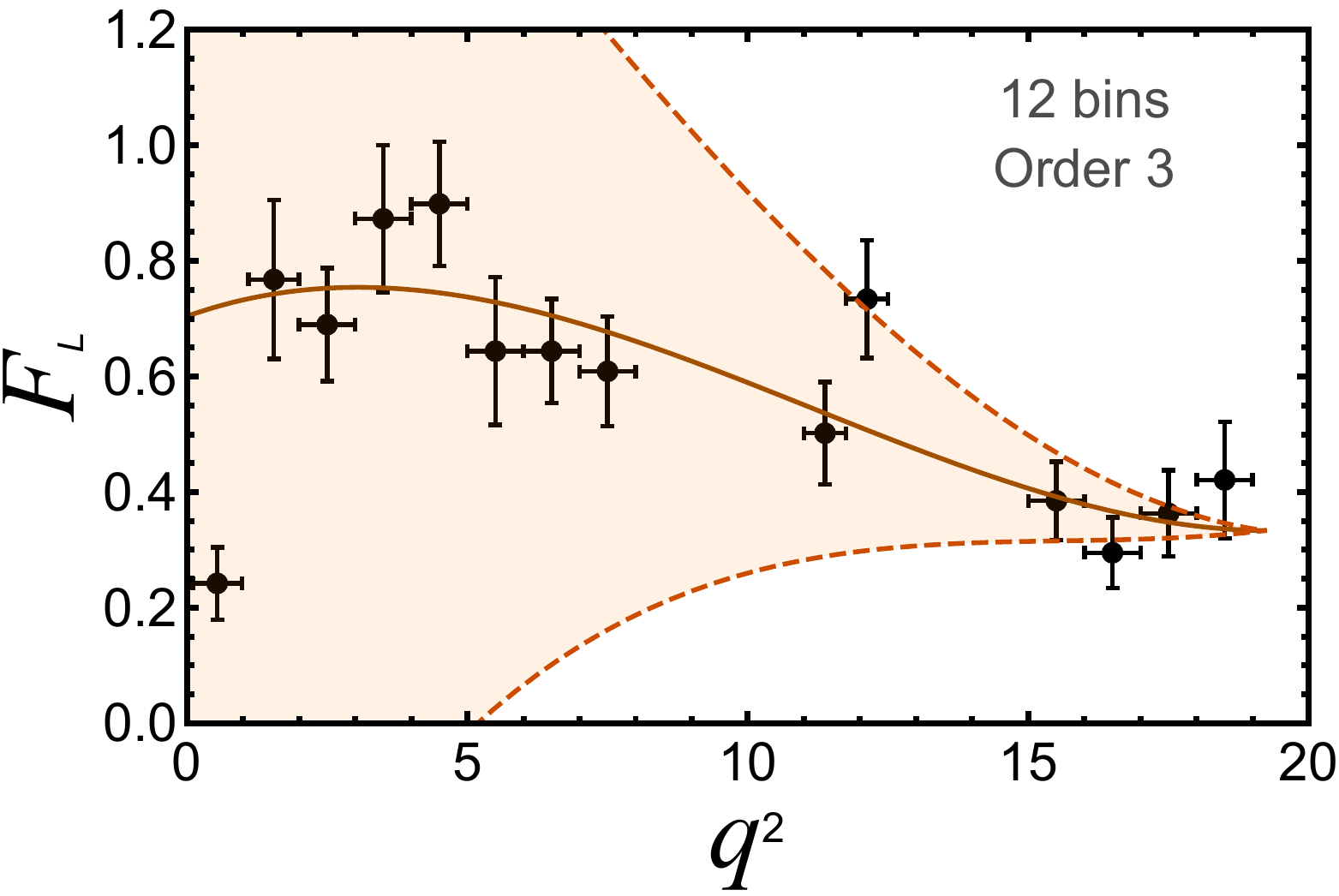}%
				\includegraphics*[width=1.5in]{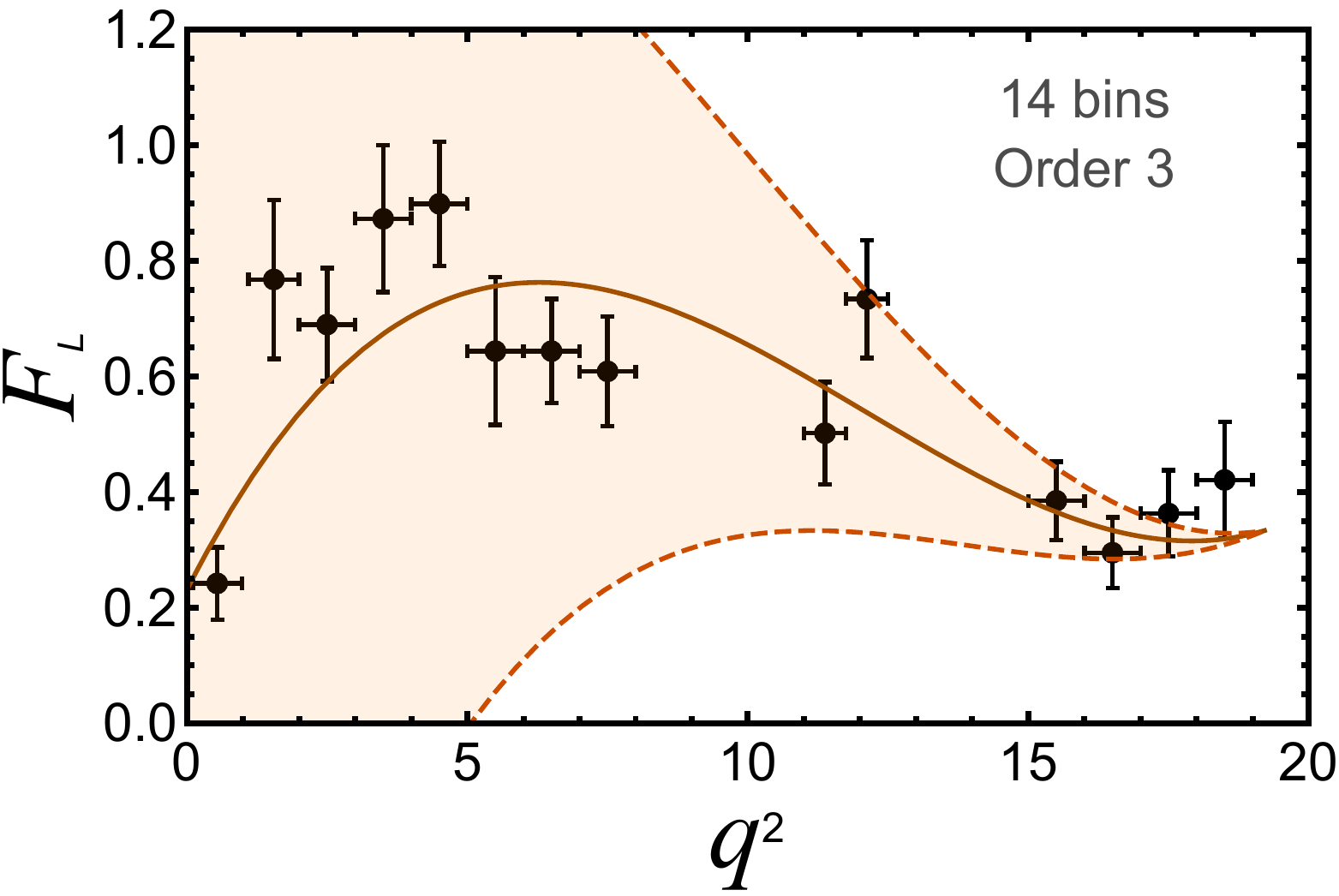}\\%
				\includegraphics*[width=1.5in]{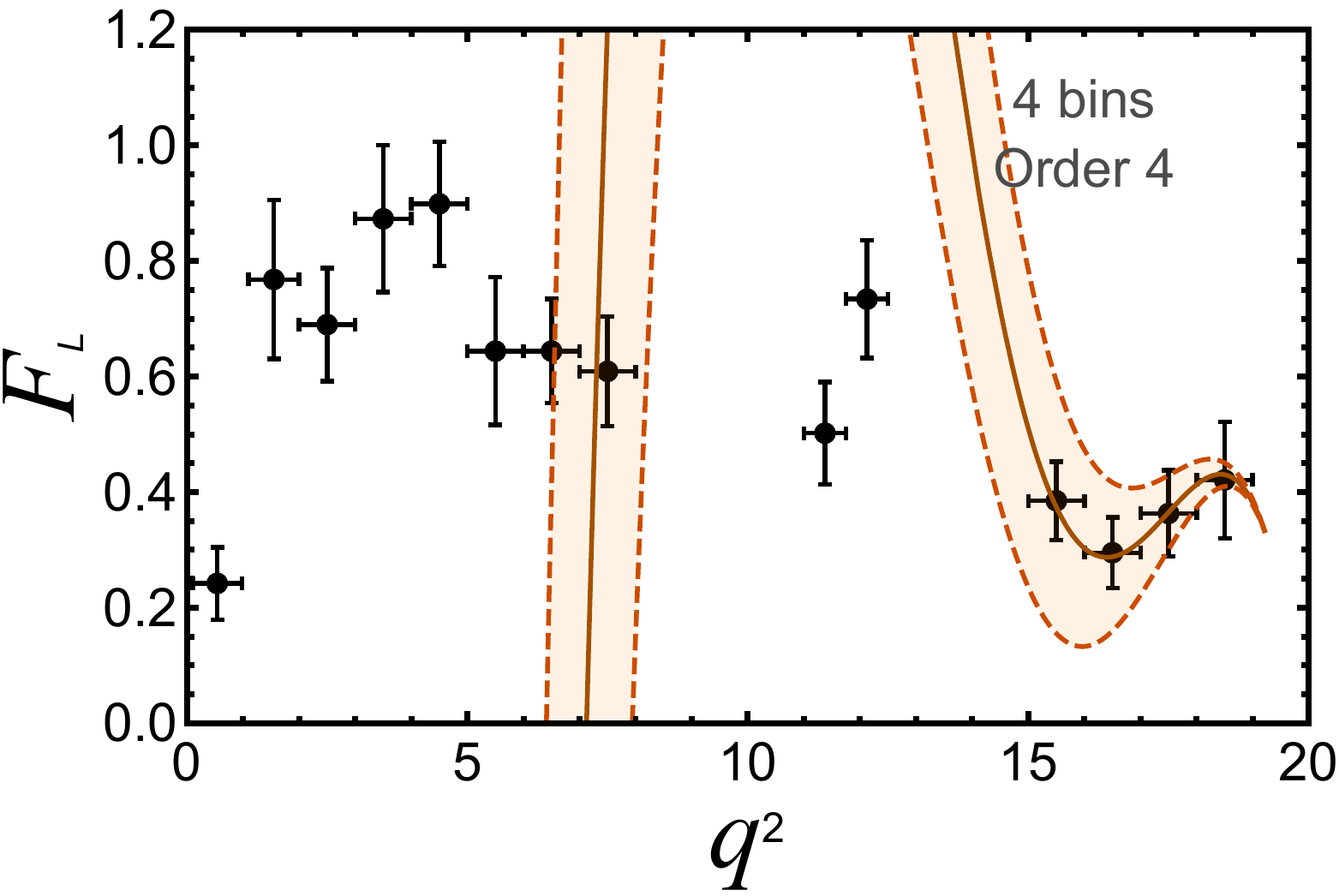}%
				\includegraphics*[width=1.5in]{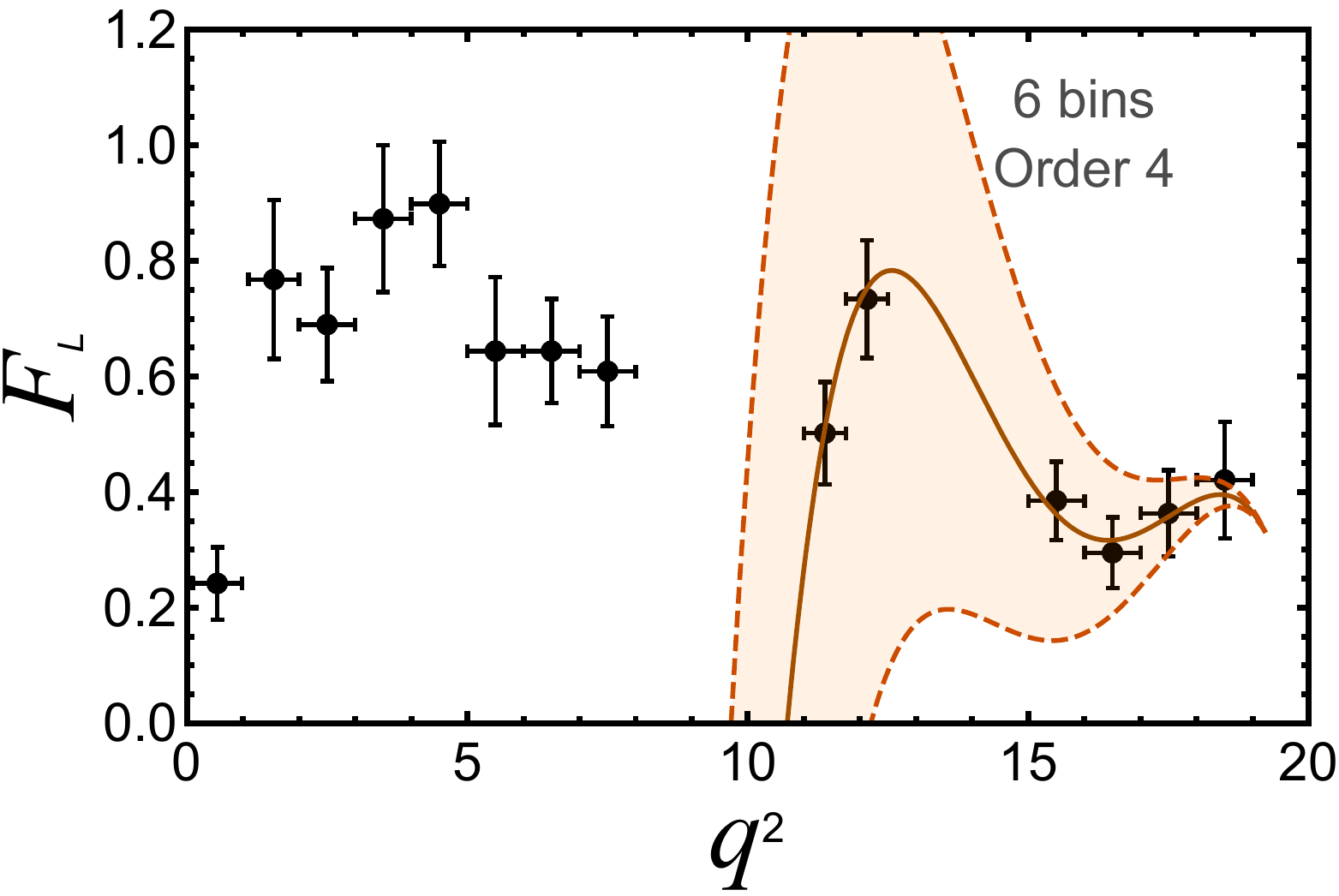}%
				\includegraphics*[width=1.5in]{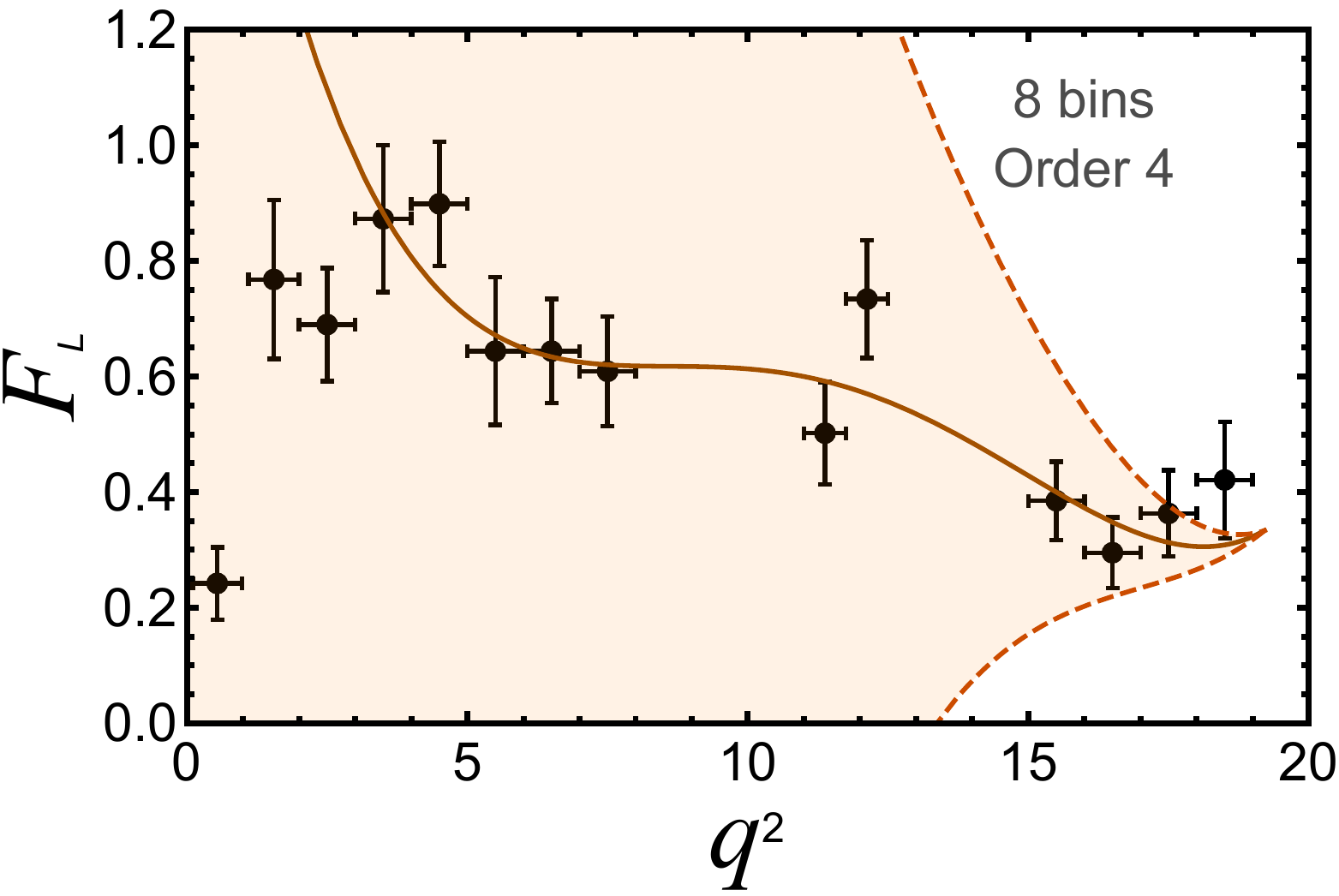}\\%
				\includegraphics*[width=1.5in]{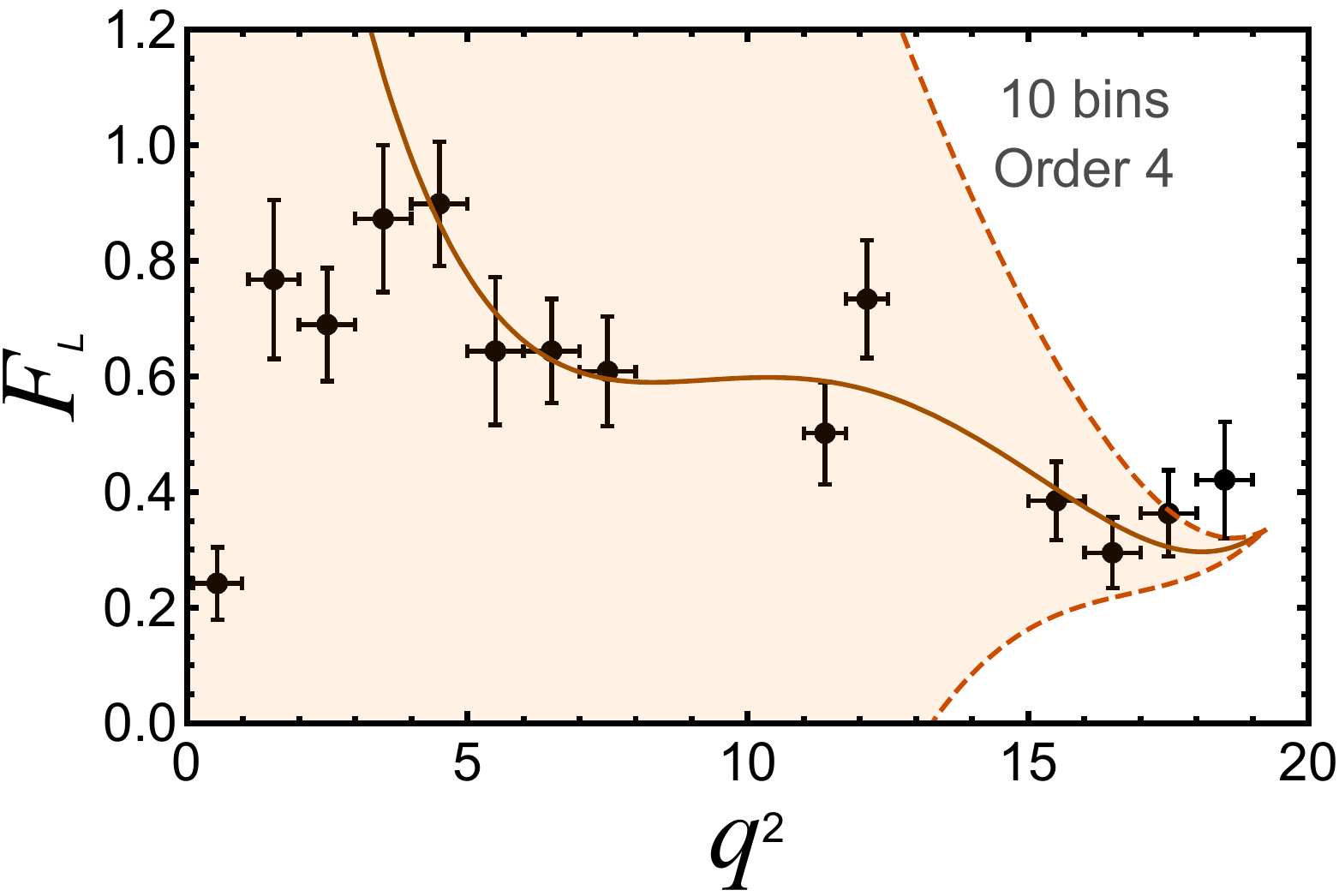}%
				\includegraphics*[width=1.5in]{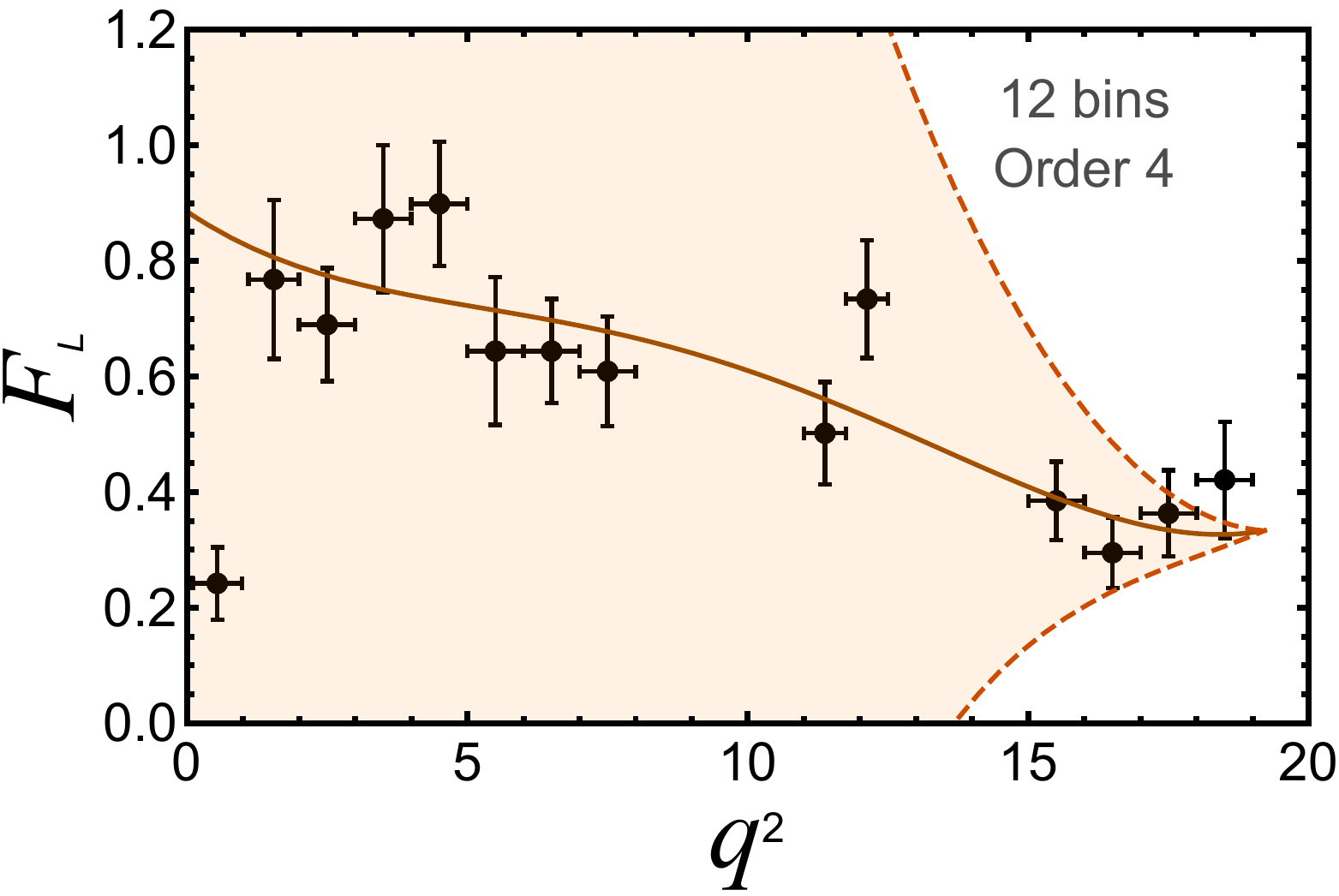}%
				\includegraphics*[width=1.5in]{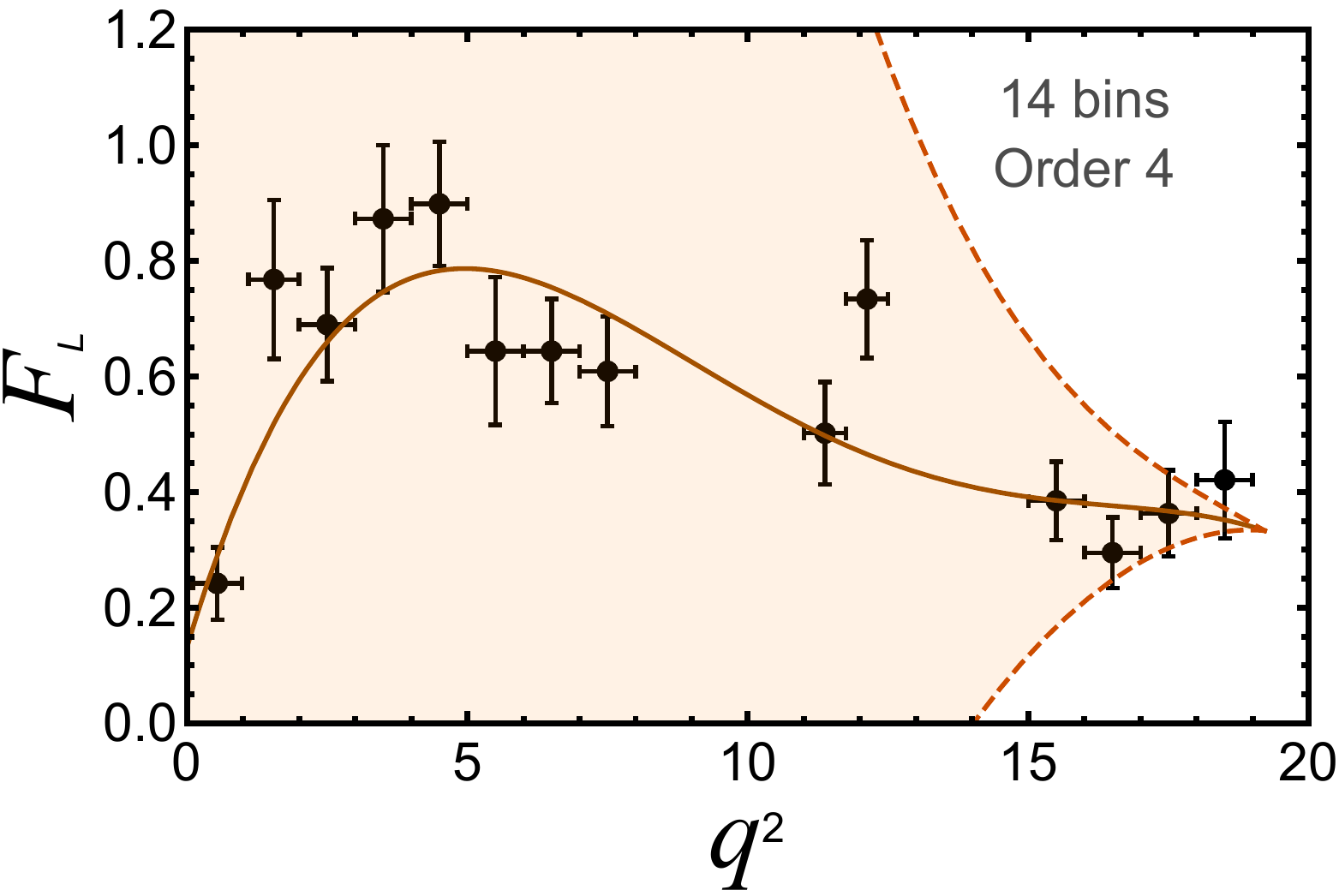}\\%
				\caption{Fits to $F_L$ various numbers of bins and polynomial parameterizations.  The color code is the same as in Fig.~\ref{fig:1}} 
				\label{fig:0c}
			\end{center}
		\end{figure}
	\end{center}

	\begin{center}
		\begin{figure}[!h]
			\begin{center}
				\includegraphics*[width=1.5in]{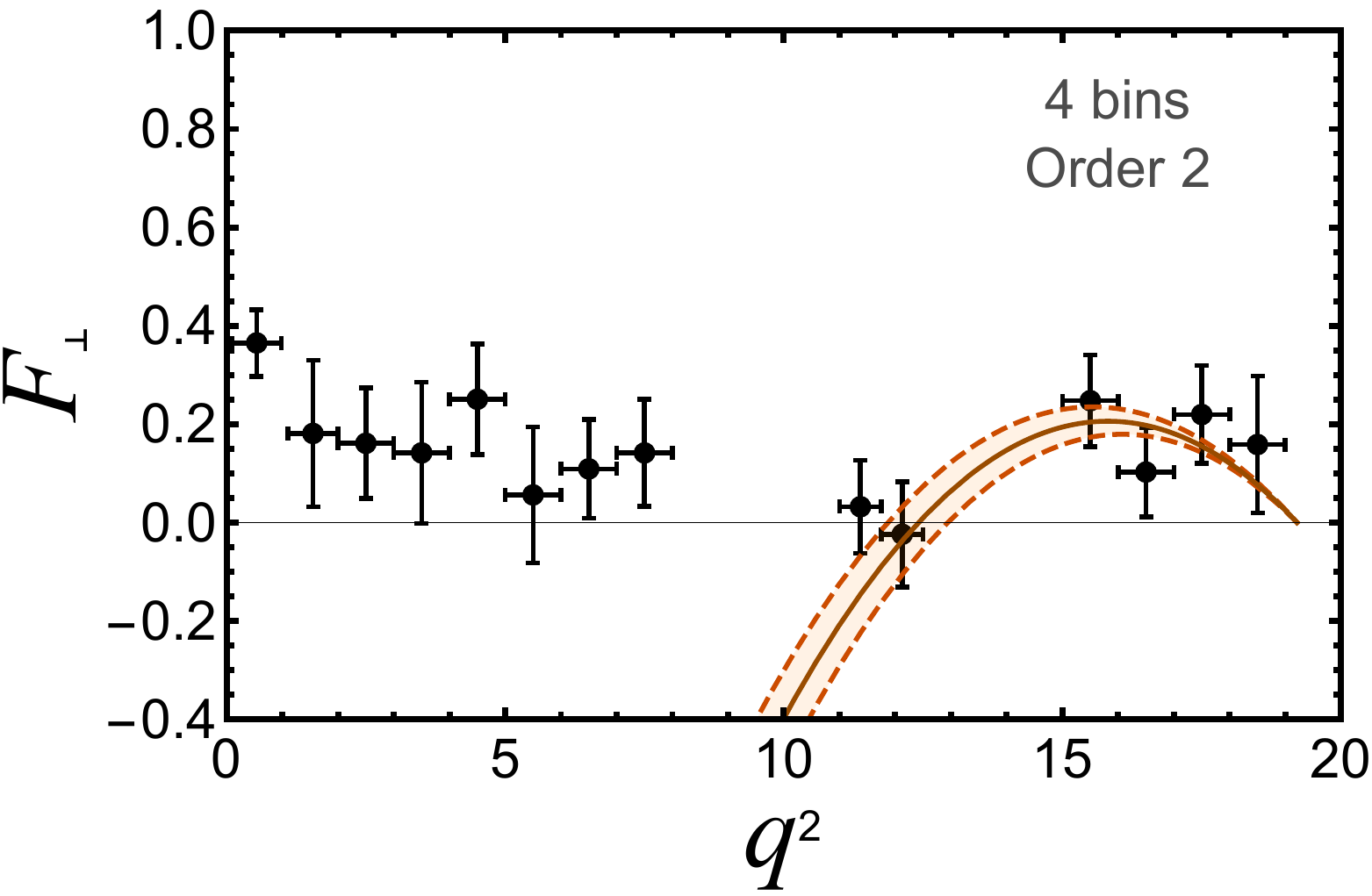}%
				\includegraphics*[width=1.5in]{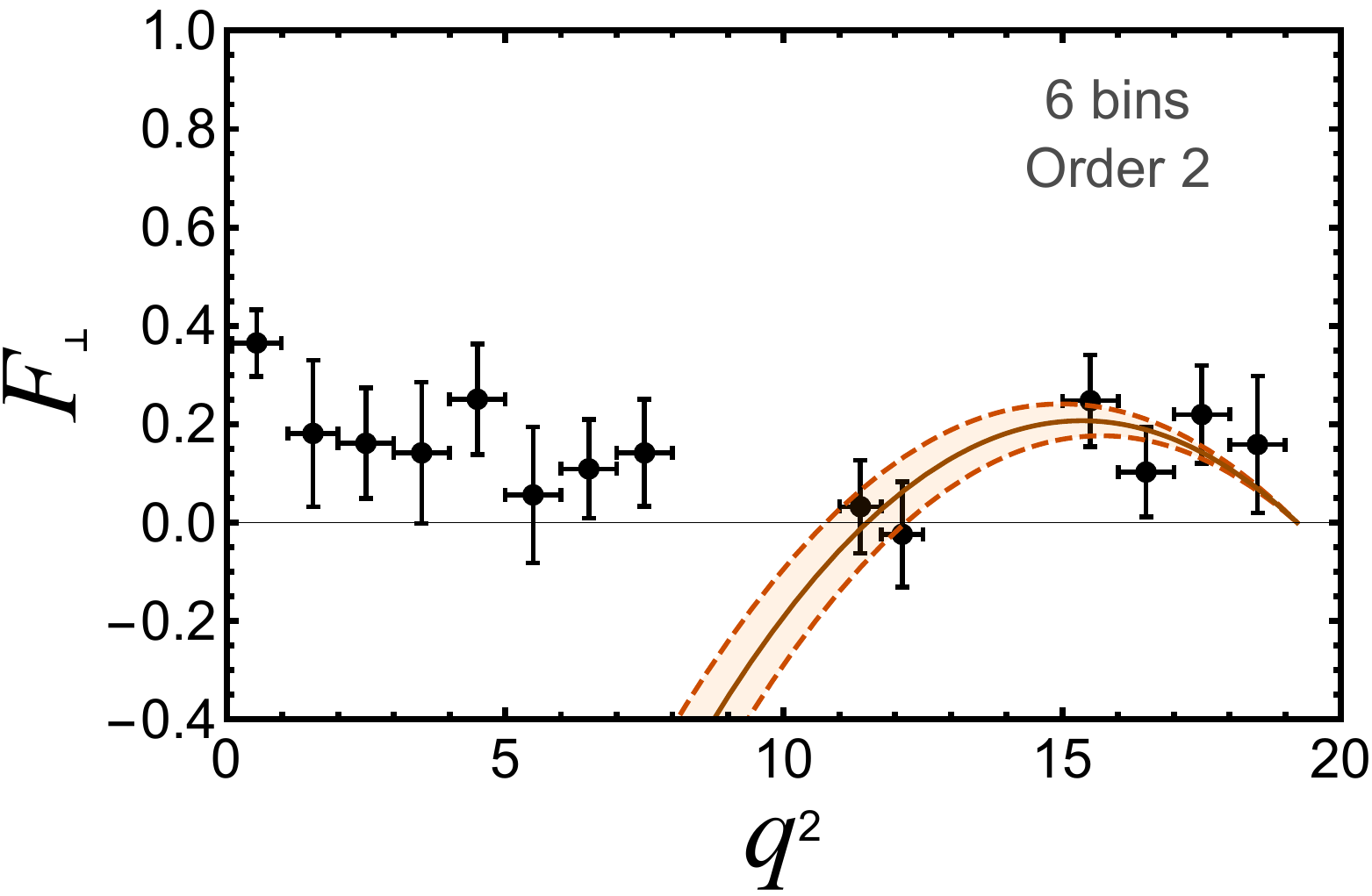}%
				\includegraphics*[width=1.5in]{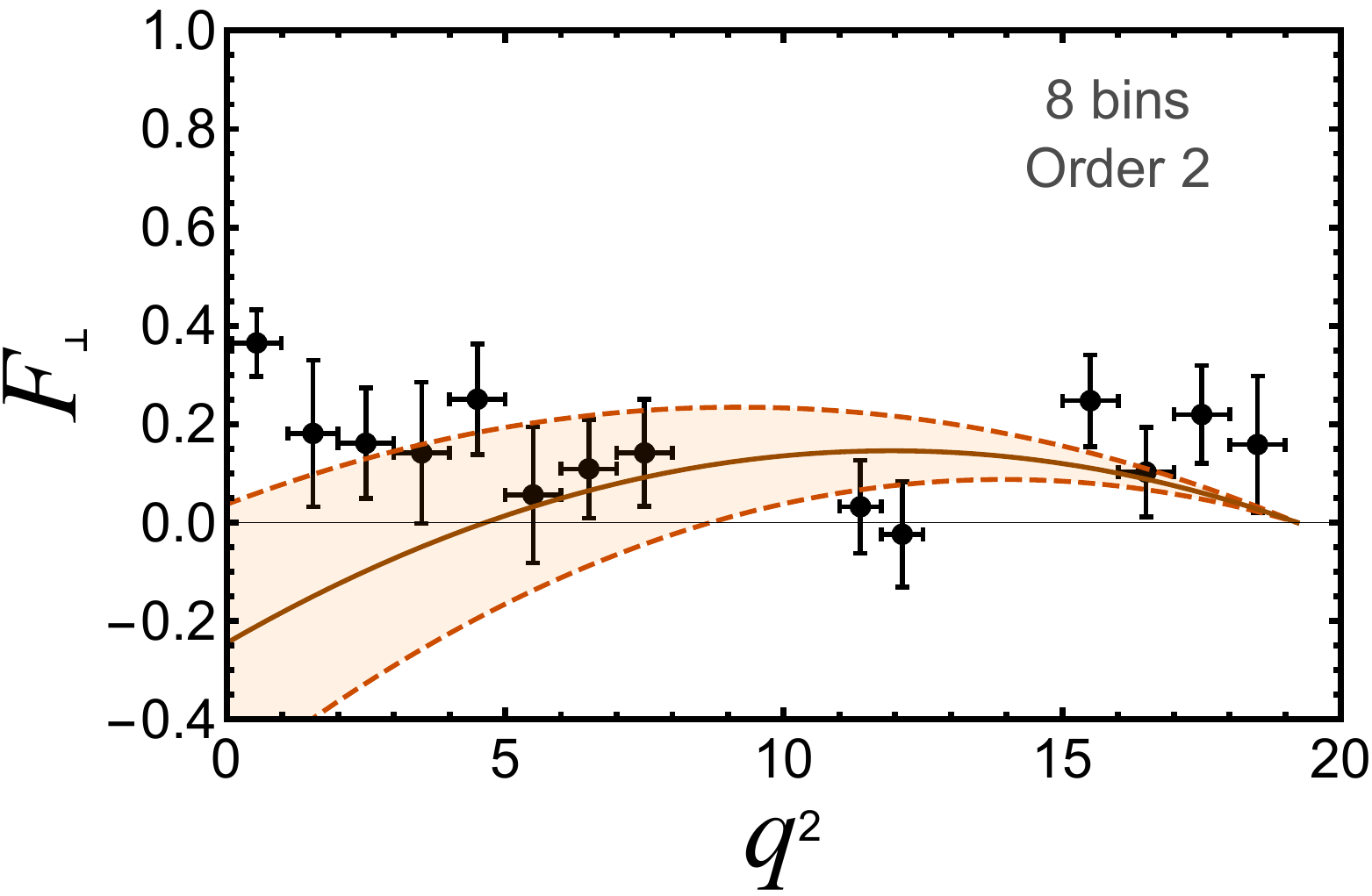}\\%
				\includegraphics*[width=1.5in]{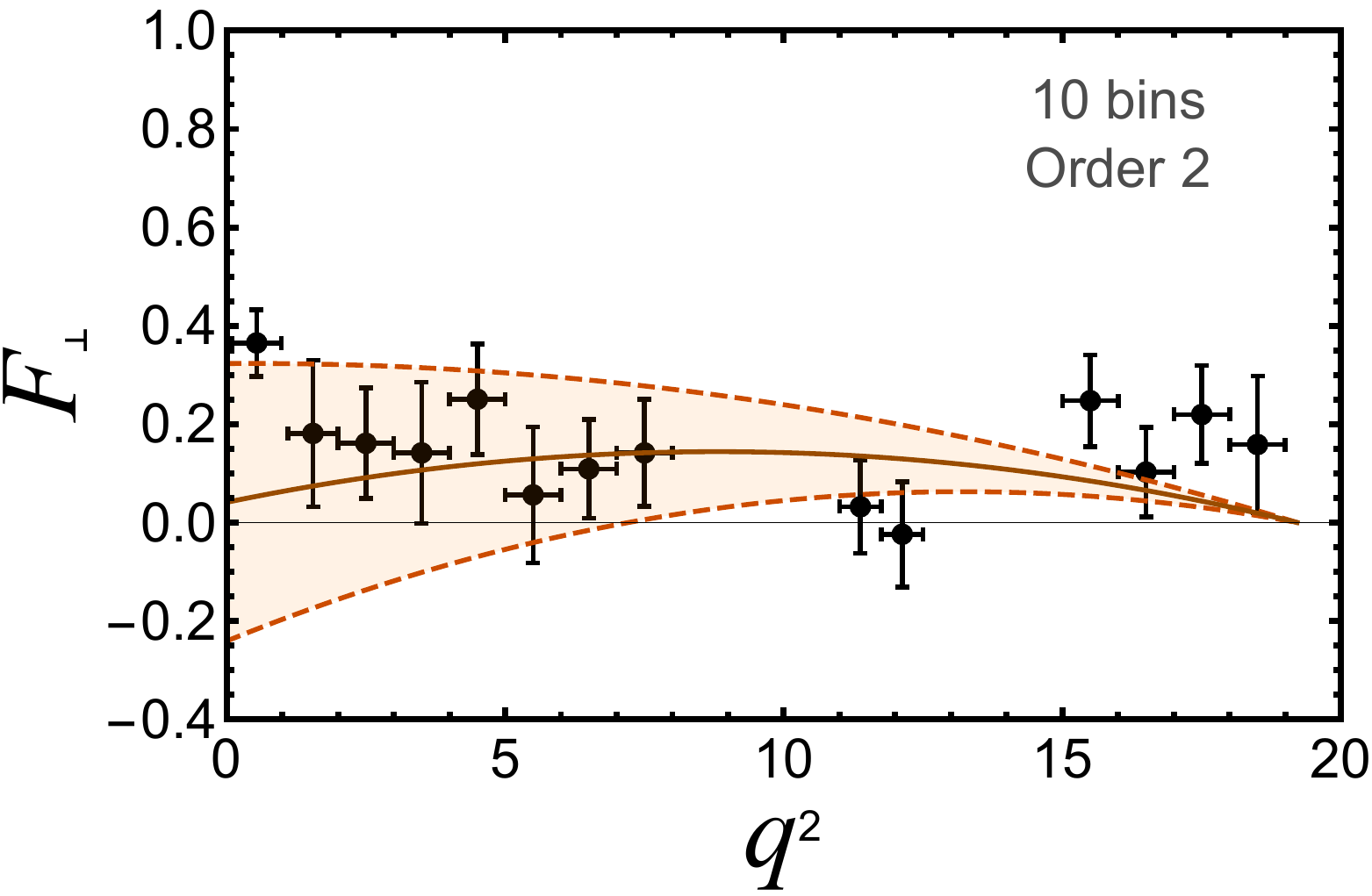}%
				\includegraphics*[width=1.5in]{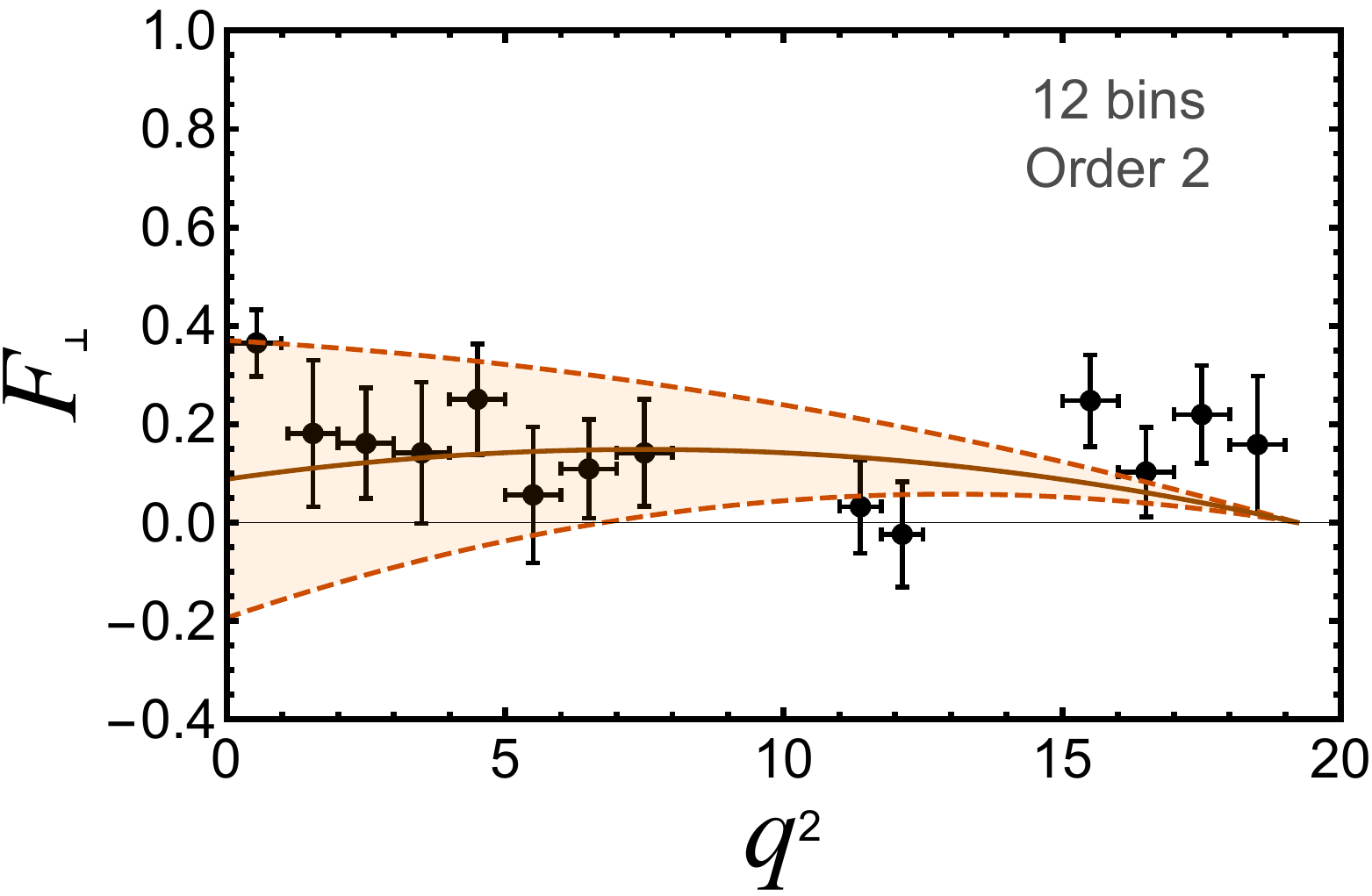}%
				\includegraphics*[width=1.5in]{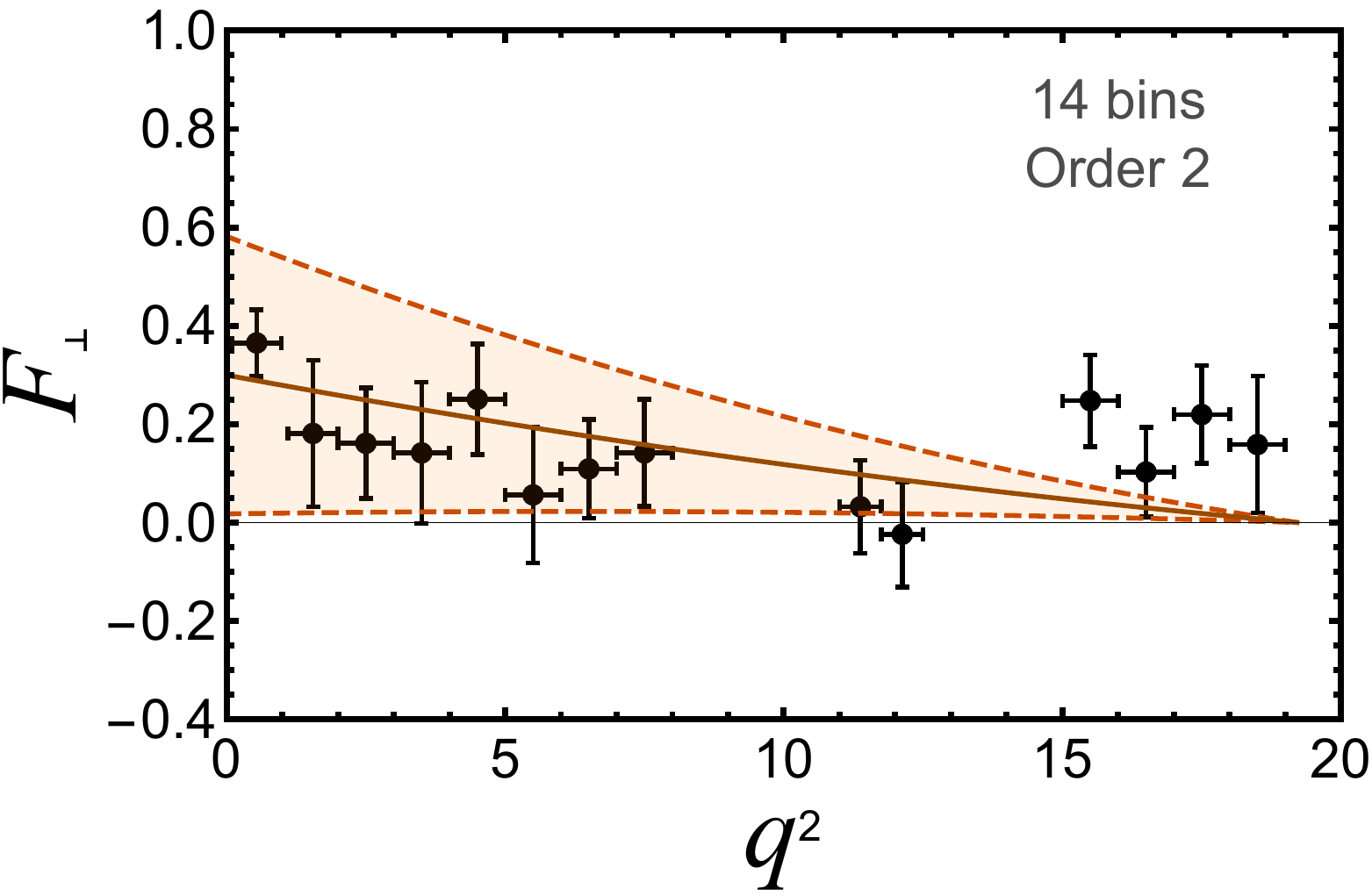}\\%
				\includegraphics*[width=1.5in]{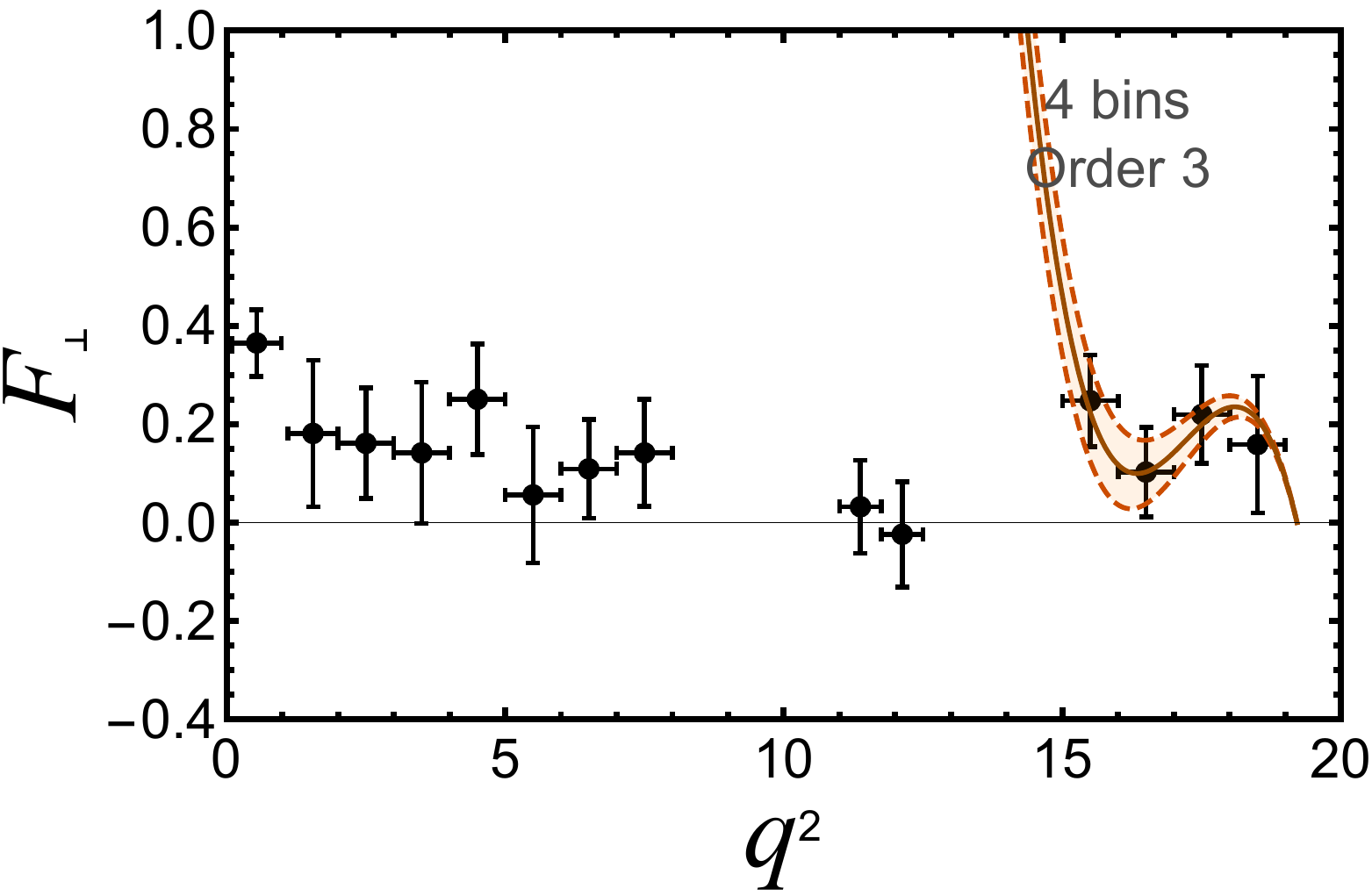}%
				\includegraphics*[width=1.5in]{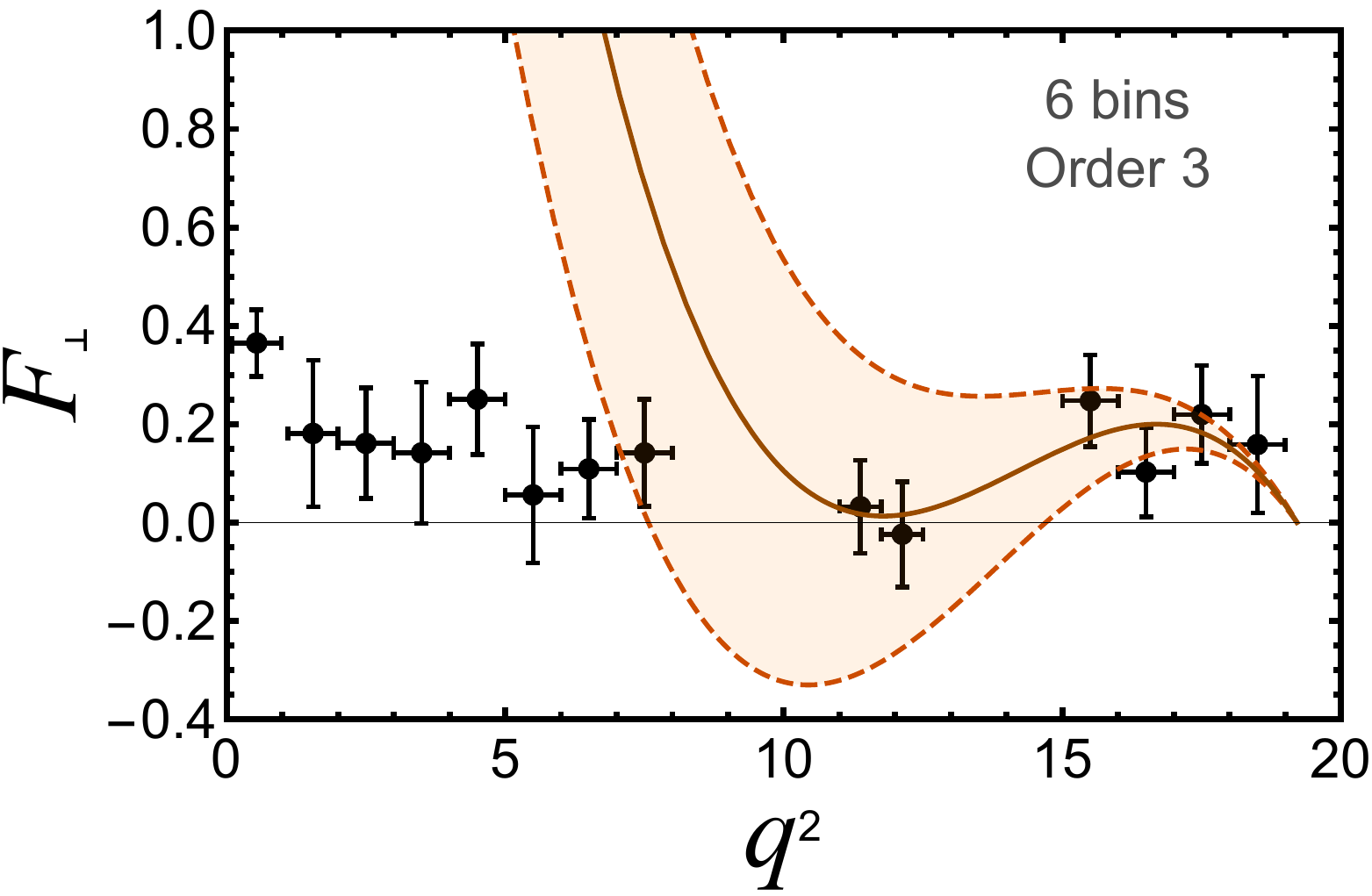}%
				\includegraphics*[width=1.5in]{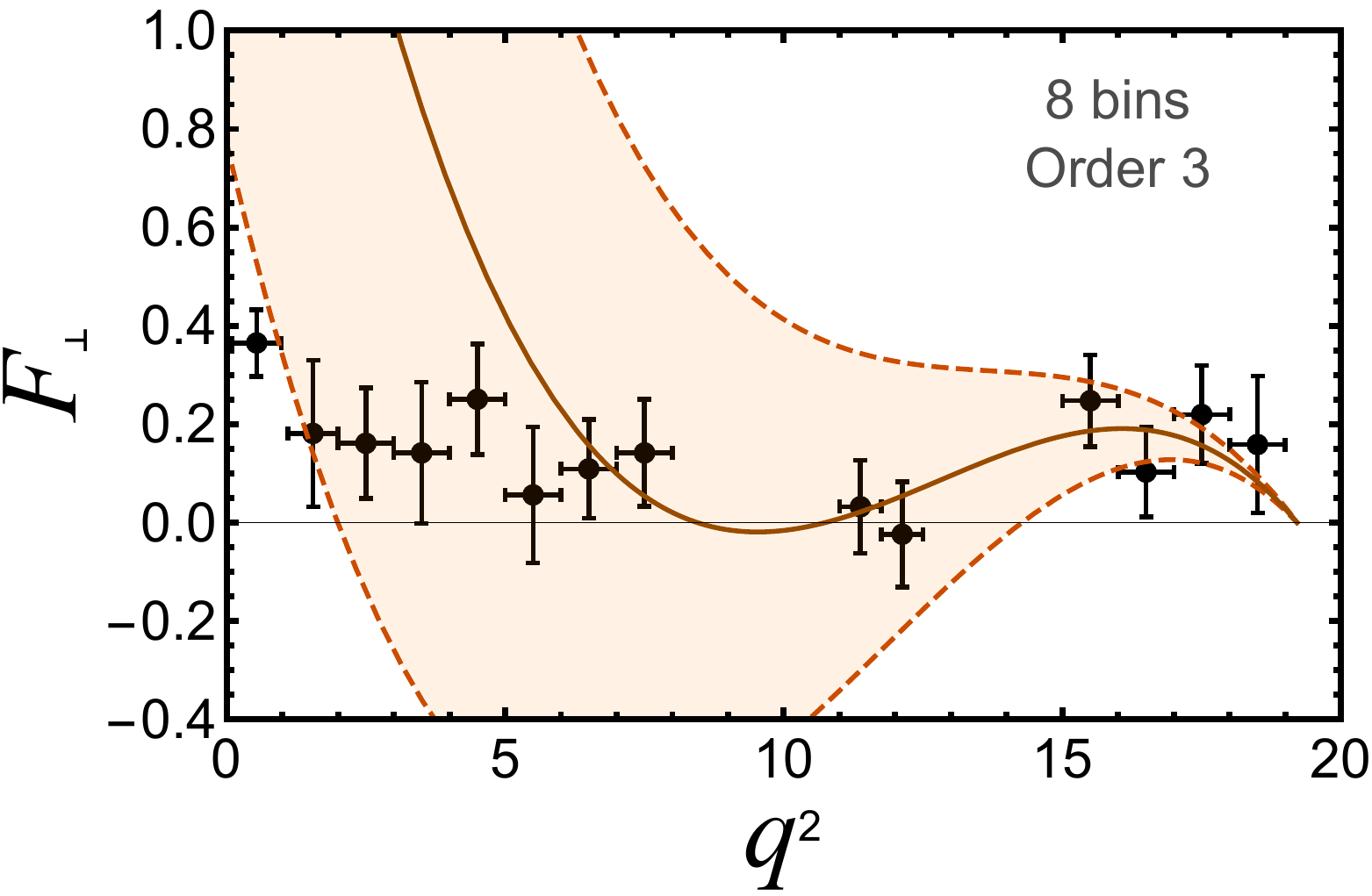}\\%
				\includegraphics*[width=1.5in]{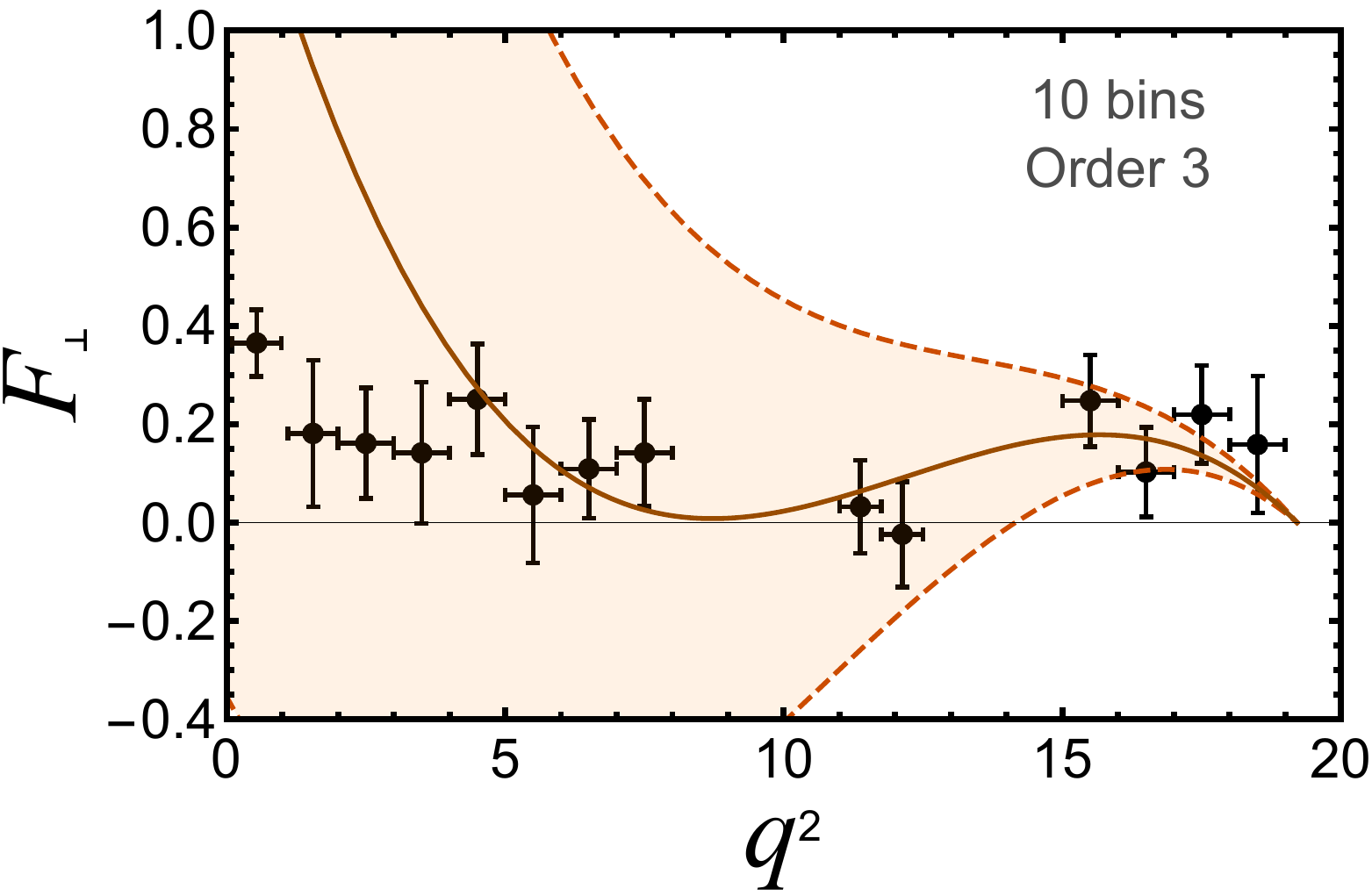}%
				\includegraphics*[width=1.5in]{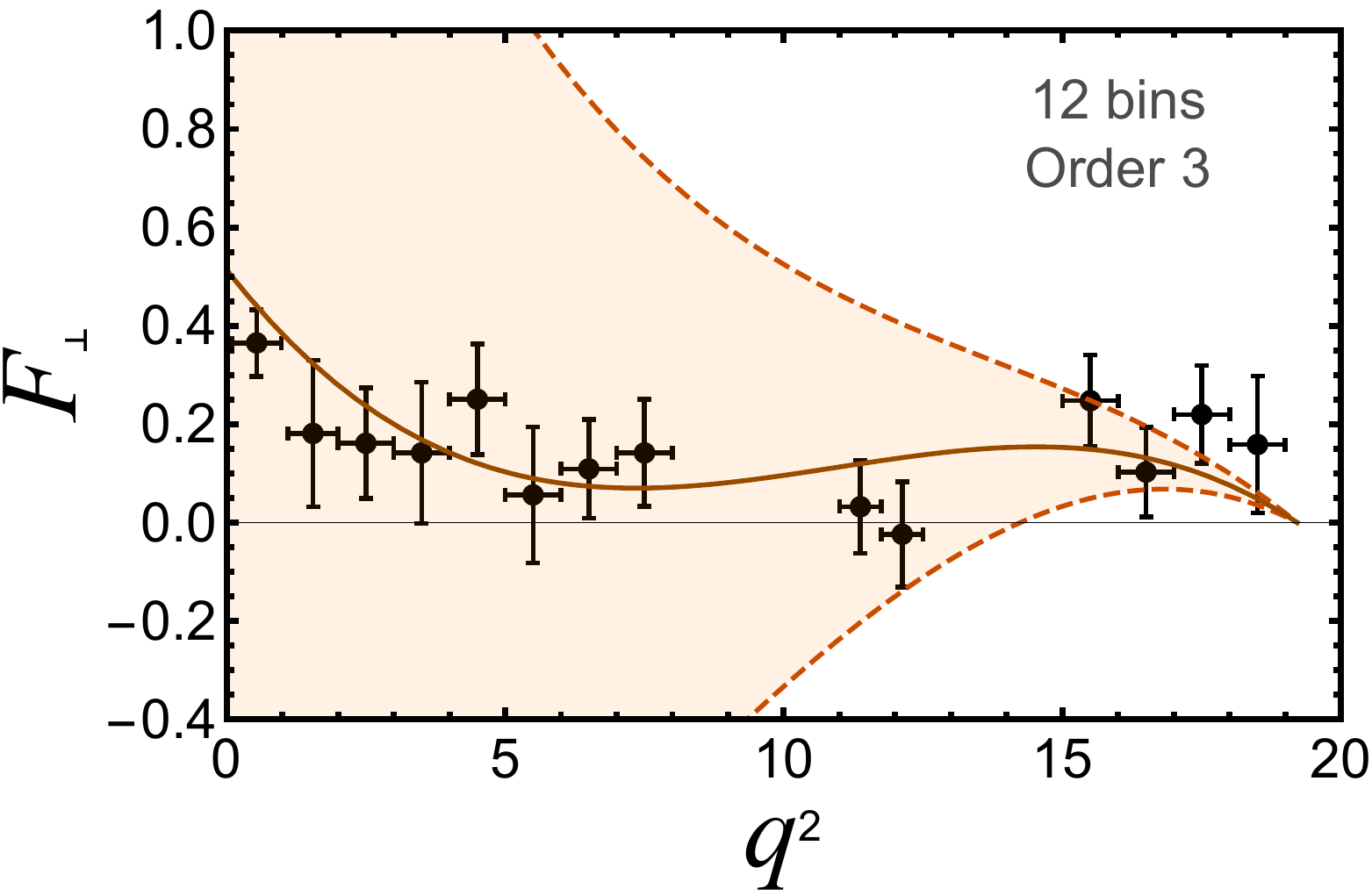}%
				\includegraphics*[width=1.5in]{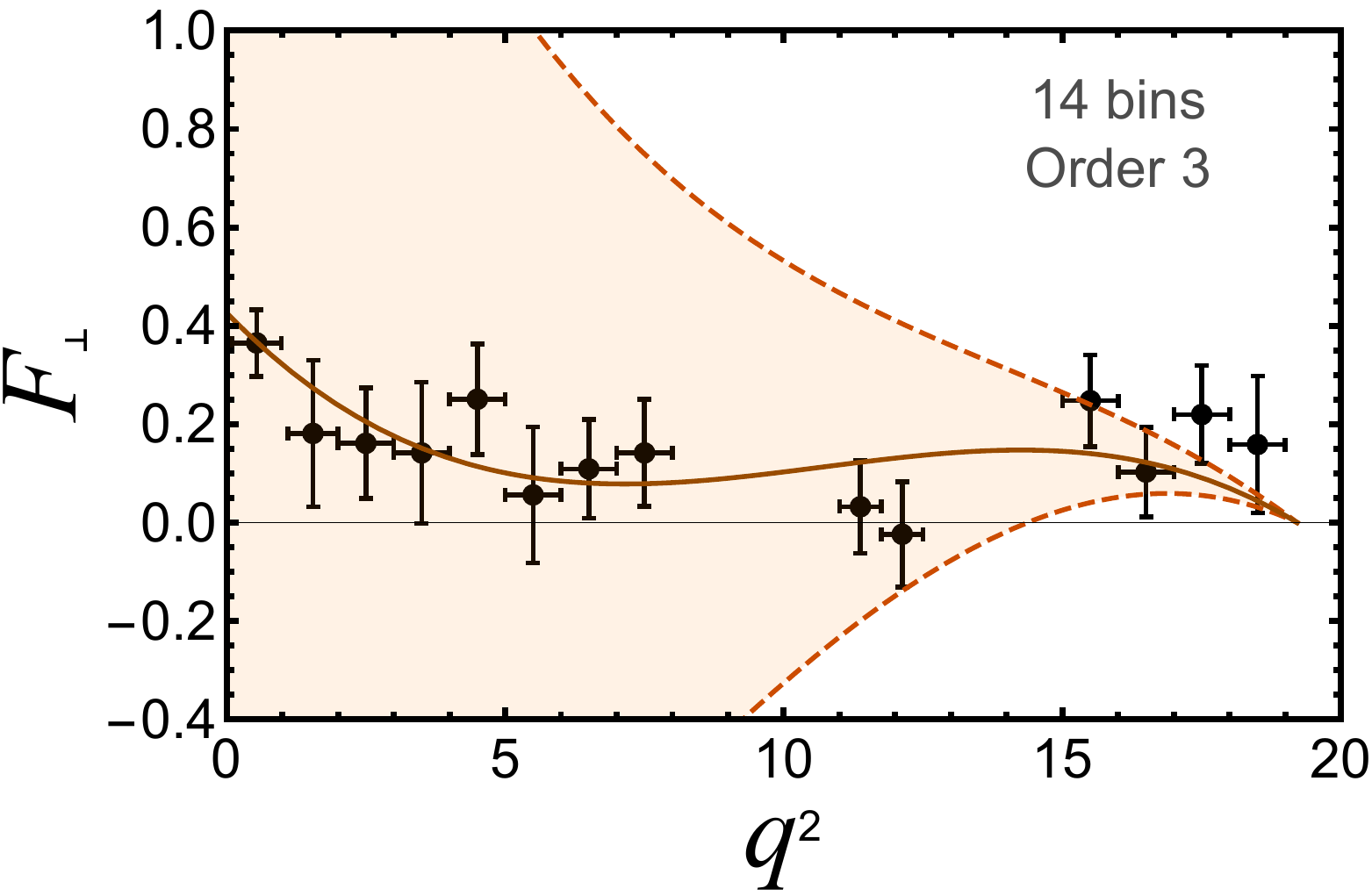}\\%
				\includegraphics*[width=1.5in]{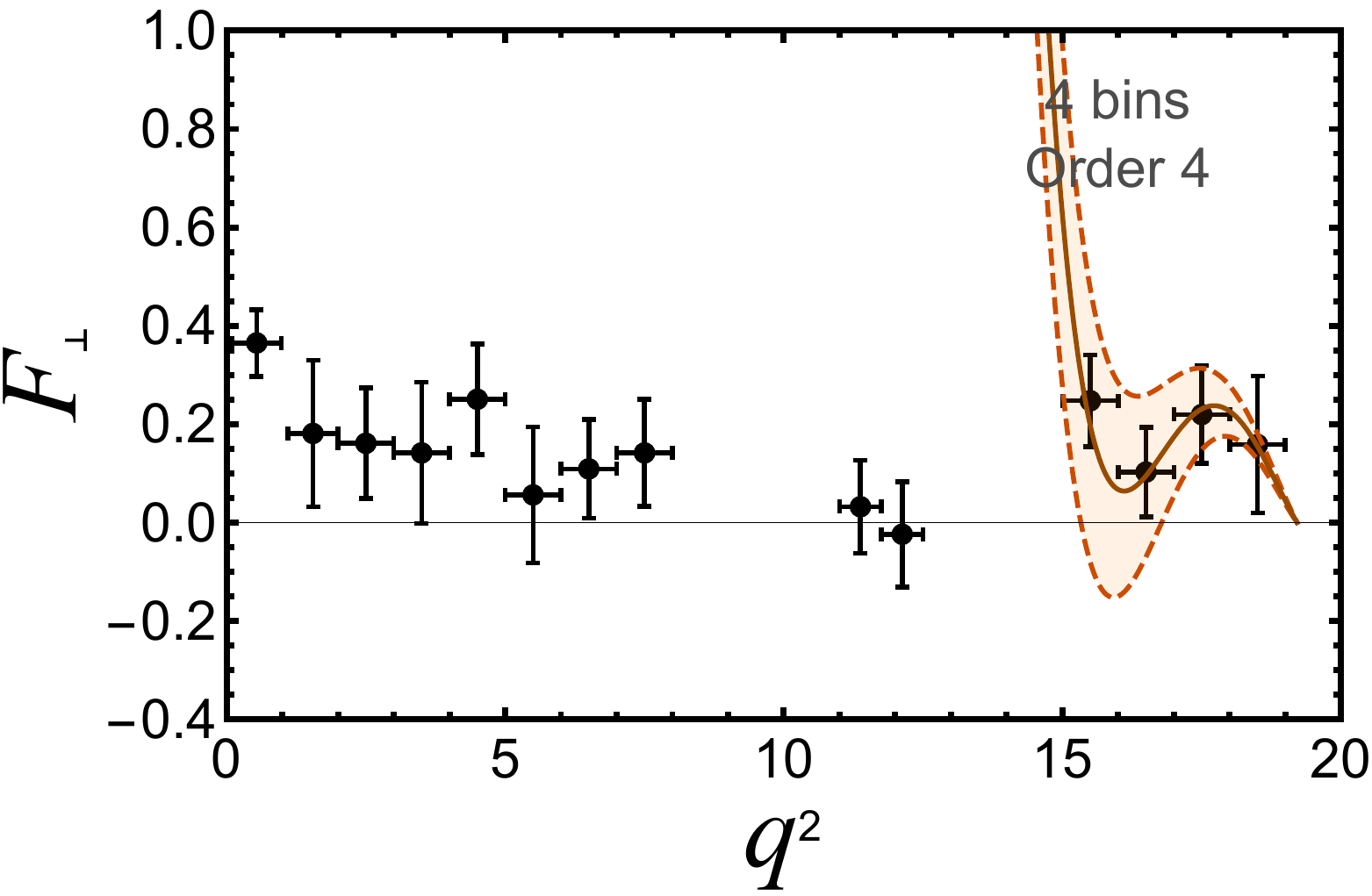}%
				\includegraphics*[width=1.5in]{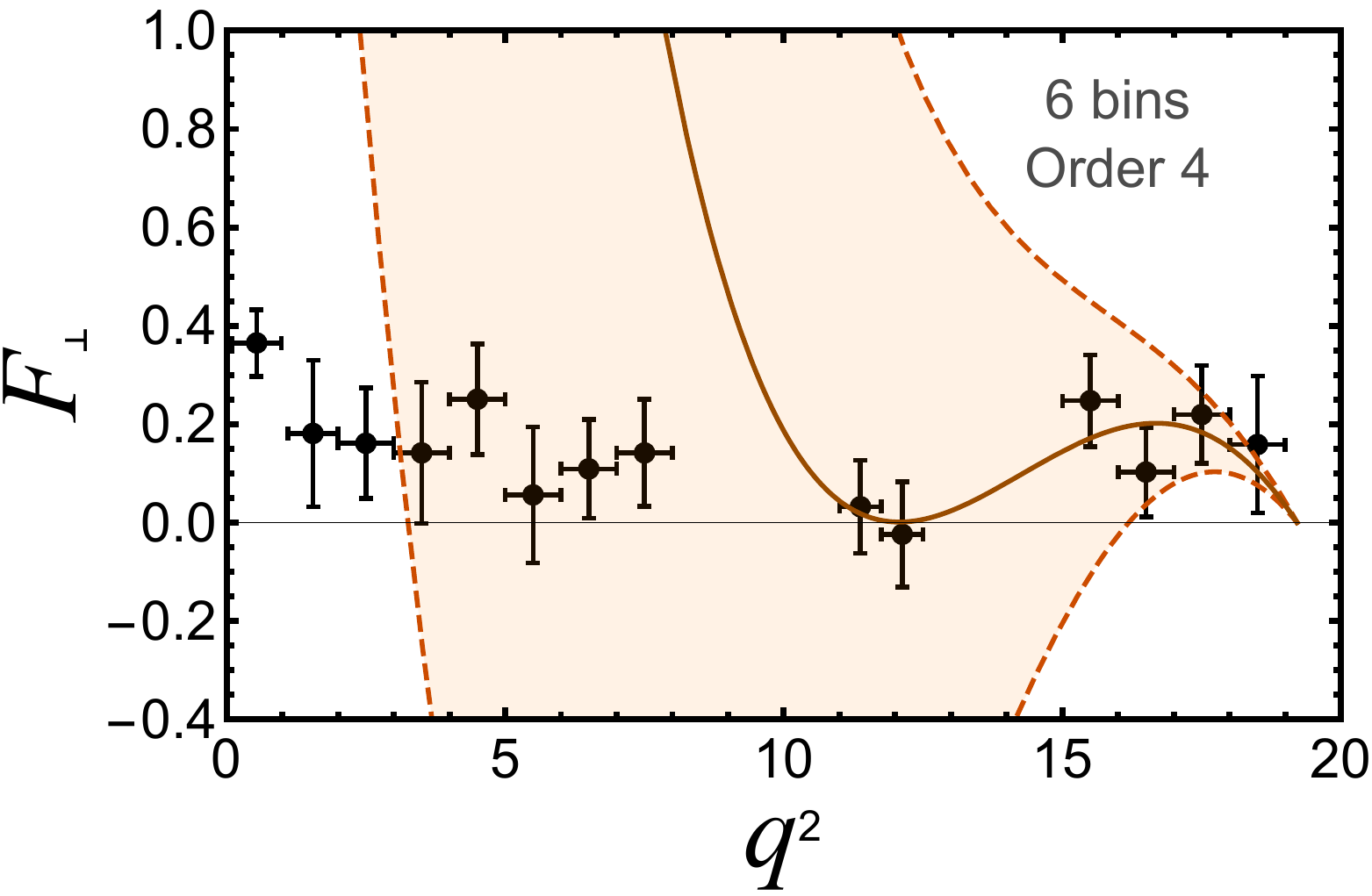}%
				\includegraphics*[width=1.5in]{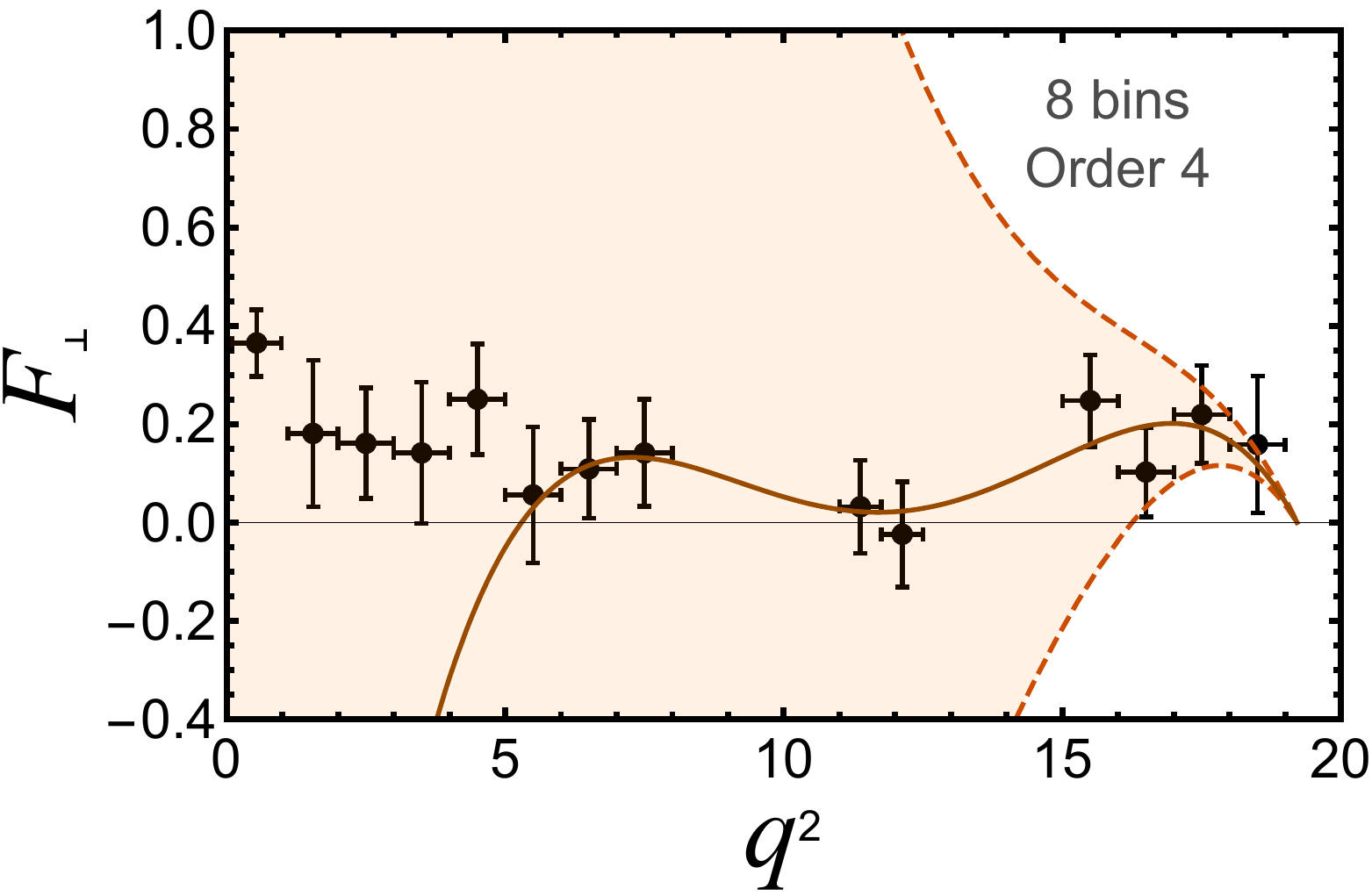}\\%
				\includegraphics*[width=1.5in]{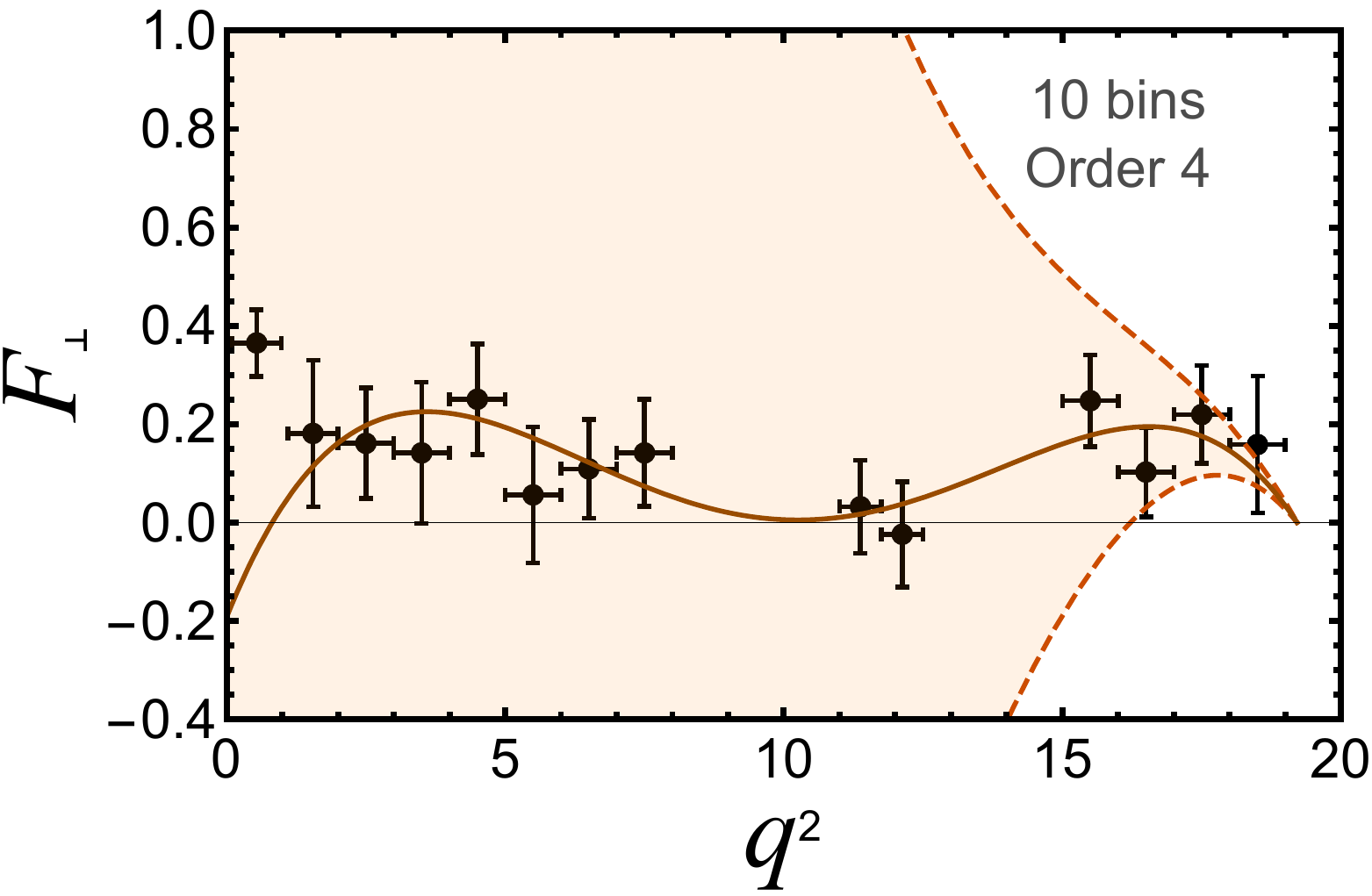}%
				\includegraphics*[width=1.5in]{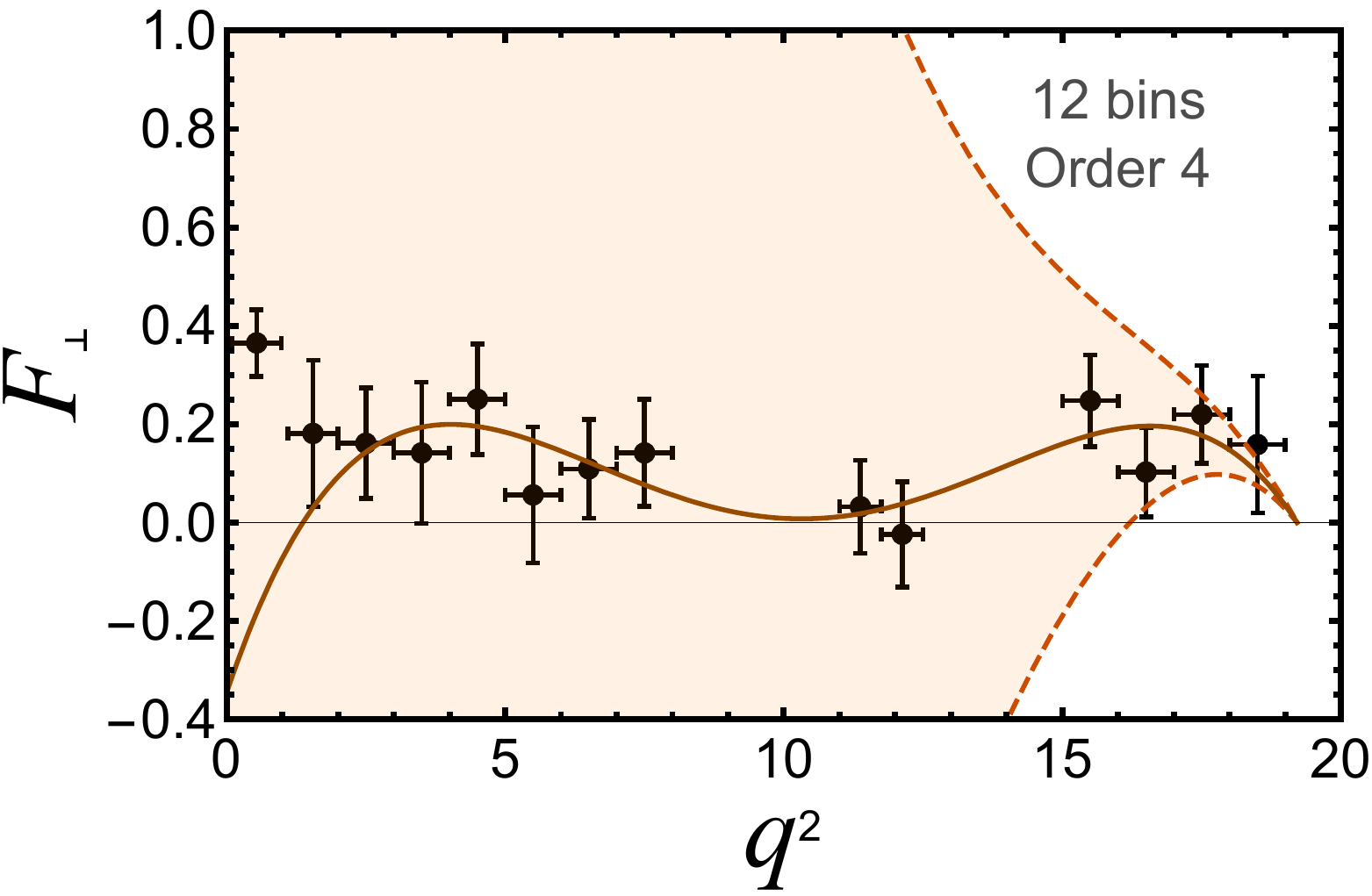}%
				\includegraphics*[width=1.5in]{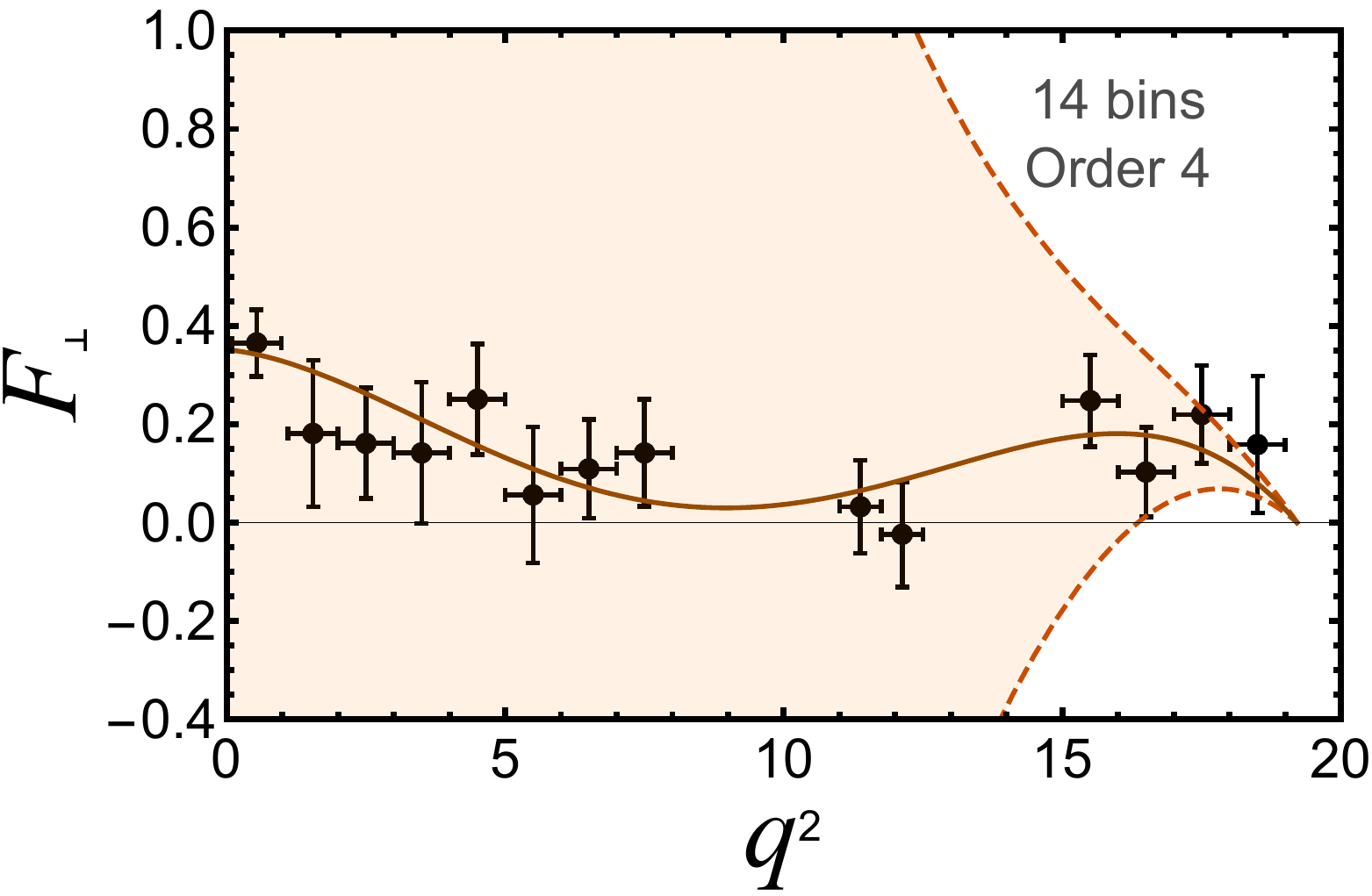}\\%
				\caption{Fits to $F_\perp$ various numbers of bins and polynomial parameterizations.  The color code is the same as in Fig.~\ref{fig:1}} 
				\label{fig:0}
			\end{center}
		\end{figure}
	\end{center}
	
\end{widetext}

\end{document}